\documentclass[iop]{emulateapj}

\newcommand{\htwo}{H$_2$}

\newcommand{\HI}{H$\;${\small\rm I}\relax}

\newcommand{\HII}{H$\;${\small\rm II}\relax}
\newcommand{\HeII}{He$\;${\small\rm II}\relax}
\newcommand{\NV}{N$\;${\small\rm V}\relax}

\newcommand{\SiII}{Si$\;${\small\rm II}\relax}

\newcommand{\NiII}{Ni$\;${\small\rm II}\relax}
\newcommand{\NI}{N$\;${\small\rm I}\relax}
\newcommand{\OI}{O$\;${\small\rm I}\relax}
\newcommand{\NHI}{$N$(\textsc{H~i})}
\newcommand{\NDI}{$N$(\textsc{D~i})}

\newcommand{\NH}{$N$(H)}
\newcommand{\nhtot}{$N$(H$_{tot})$}
\newcommand{\dhtot}{D/H$_{tot}$}
\newcommand{\Nhtwo}{$N$(H$_2$)}
\newcommand{\NHtwo}{$N$(H$_2$)}
\newcommand{\kms}{km~s$^{-1}$\relax}

\newcommand{\chisq}{$\chi^2$}

\newcommand{\fuse}{{\em FUSE}\relax}
\newcommand{\lya}{Ly$\alpha$}

\newcommand{\err}[2]{\ensuremath{^{+ #1}_{- #2}}}
\newcommand{\fD}{$f\rm_{D}$}

\shorttitle{D/H Ratio in the Nearby ISM}
\shortauthors{Friedman et al.}

\graphicspath{{./}{figures/}}

\begin{document}

\title{A High Precision Survey of the D/H Ratio in the Nearby Interstellar Medium}

\author{Scott D. Friedman\altaffilmark{1}}
\altaffiltext{1}{Space Telescope Science Institute, 3700 San Martin Drive, Baltimore, MD 21218, USA}

\author{Pierre Chayer\altaffilmark{1}}

\author{Edward B. Jenkins\altaffilmark{2}}
\altaffiltext{2}{Department of Astrophysical Sciences, Princeton University, Princeton, NJ 08544-1001, USA}

\author{Todd M. Tripp\altaffilmark{3}}
\altaffiltext{3}{Department of Astronomy, University of Massachusetts, Amherst, MA 01003, USA}

\author{Gerard M. Williger\altaffilmark{4,5,6}}
\altaffiltext{4}{Department of Physics \& Astronomy, University of Louisville, Louisville, KY 40292, USA}
\altaffiltext{5}{Jeremiah Horrocks Institute, University of Central Lancashire, Preston PR1 2HE, England}
\altaffiltext{6}{Institute for Astrophysics and Computational Sciences, Catholic U. of America, Washington DC 20064}

\author{Guillaume H\'ebrard\altaffilmark{7,8}}
\altaffiltext{7}{Institut d’Astrophysique de Paris, CNRS, UMR 7095, Sorbonne Universit\'{e}, F-75014 Paris, France}
\altaffiltext{8}{Observatoire de Haute Provence, CNRS,Universit\'{e} d’Aix-Marseille, F-04870 Saint-Michel-l’Observatoire, France}

\author{Paule Sonnentrucker\altaffilmark{9}}
\altaffiltext{9}{European Space Agency (ESA), ESA Office, Space Telescope Science Institute, 3700 San Martin Drive, Baltimore, MD 21218, USA}

\begin{abstract}

We present high S/N measurements of the \HI\ \lya\ absorption line toward
16 Galactic targets which are at distances between approximately 190 and 2200 pc,
all beyond the wall of the Local Bubble. We describe
the models used to remove stellar emission and absorption features and the methods
used to account for all known sources of error in order to compute high precision
values of the \HI\ column density with robust determinations of uncertainties. When
combined with \htwo\ column densities from other sources, we find  
total H column densities ranging from 10$^{20.01}$ to 10$^{21.25}$ cm$^{-2}.$ Using
deuterium column densities from \fuse\ observations we determine the D/H ratio
along the sight lines. We confirm and strengthen the conclusion that D/H is spatially variable
over these \HI\ column density and target distance regimes, which predominantly probe
the ISM outside the Local Bubble.
We discuss how these results affect models of Galactic chemical evolution. We also
present an analysis of metal lines along the five sight lines for which we have high
resolution spectra and, along with results reported in the literature,
discuss the corresponding column densities in the context of a
generalized depletion analysis. We find that D/H is only weakly correlated with metal
depletion and conclude that the spatial D/H variability is not solely due to dust depletion.
A bifurcation of \dhtot\ as a function of depletion at high depletion levels provides
modest support that deuterium-rich gas is infalling onto the Galactic plane.

\end{abstract}


\section{Introduction} \label{sec:intro}

The observed abundance of deuterium is one of the cornerstones of
modern cosmology.  Building on the idea that some elements more massive
than hydrogen could be synthesized in the first few minutes of the Big
Bang \citep[e.g.,][]{vonweiz38,gamow48} combined with the discovery of
the cosmic microwave background, detailed predictions of the
abundances of deuterium and helium from Big-Bang nucleosynthesis (BBN)
were developed in the 1960s \citep{peebles66,wagoner67}. Subsequently,
measurements of deuterium in the diffuse interstellar medium (ISM) of
the Milky Way \citep{rogerson73, yorkrogerson76} were found to be in good agreement with
the predictions of BBN, which provided a spectacular confirmation of
Big Bang theory and an estimate of the baryonic content of the
universe.

The utility of deuterium abundances in the ISM as a probe of the early
universe depends on the importance of subsequent processes that can
either destroy or produce this isotope as the Galaxy evolves.  On the
one hand, we know for certain that deuterium is destroyed in stars (astration).
The importance of this effect depends on how much of the stellar material
replenishes the gas in the ISM and how much this loss of deuterium is balanced by
contributions that comes from the infall of pristine gas from the intergalactic
medium.  A general consensus from modeling these processes in our Galaxy
is that D/H probably does not decrease from the primordial value by more
than a factor of about 2 \citep{steigman92, vangioni94, galli95, steigman95,
dearborn96, prantzos96, chiappini02, romano06, prodanovic08, leitner11, weinberg17}.

On the other hand, we also must be aware of processes after BBN that can
create new deuterium.This was first investigated by
\citet{epstein76} who considered both synthesis and spallation production
mechanisms including pregalactic cosmic rays, shock waves, hot
explosions, and the disruption of neutron stars by black holes.
They concluded that post Big-Bang deuterium production requires
extremely violent and exotic conditions for which there is little
supporting evidence and most processes would over- or under-produce
other light elements, which other observations have adequately constrained.
However, \citet{mullan98} pointed out that these
investigators neglected a potentially important process, the production
of neutrons in stellar flares which are then captured by protons to
form D, and this could be an important source of interstellar deuterium.
\citet{prodanovic03} subsequently examined
this hypothesis in more detail and ruled out this mechanism as a
significant source of D on the Galactic scale based on observed limits
of the 2.22 MeV $\gamma$-ray produced by this reaction. However, they
do agree that this process must occur at some level and the possibilty
of very local enrichment, while not likely, cannot be ruled out entirely.
A more exotic creation process is the proposal by \citet{gnedin92} that
deuterium could be produced by the photodisintgration of $^4$He by
gamma rays produced by the accretion of gas onto $10^6\,{\rm M}_\odot$
black holes.  Although a black hole with a mass comparable to this
exists at the center of the Milky Way, it has not been shown that it has
produced an appreciable amount of deuterium in the region of the
Galaxy that is the subject of our investigation, and in any case this probably would
not cause abundance variations over the scales we are probing.
{\citet{lubowich00} measured the distribution of 
DCN relative to HCN in a molecular cloud only 10 pc from the
Galactic center and conclude that D/H = $1.7\pm0.3$ ppm (parts per million), far below
any region in the local ISM. \citet{lubowich10} measured this
ratio in 16 molecular clouds at galactocentric distances ranging from 2 pc to
10 kpc. They find that D/H increases slightly with distance to a maximum
of 20.5 ppm. Both studies conclude that the observed deuterium is
cosmological and there are no other significant sources.
In this study we therefore assume, as most studies have since
\citet{epstein76}, that all observed deuterium is primordial in origin.
Consequently, the observed deuterium abundance may
provide an unambiguous probe of the chemical evolution of gas in
galaxies (i.e., the processing of gas due to cycling through stars).

However, as more interstellar D/H detections have accumulated, the
interpretation of the deuterium abundances has become less clear.  The
ensemble of D/H measurements from the \textit{Copernicus} and
\textit{International Ultraviolet Explorer (IUE)} satellites seemed to
indicate that D/H is spatially variable in the Milky Way, which caused
some tension between BBN and galactic evolution models
\citep{laurent79,madjar84,madjar98,hebrard99}.  The reality of the variability was
challenged based on uncertainties of the early measurements
\citep{mccullough92}, but subsequent (and more precise) D/H
determinations with the \textit{ORFEUS-SPAS II} Interstellar Medium
Absorption Profile Spectrograph (IMAPS) and with the \textit{Far
Ultraviolet Spectroscopic Explorer (FUSE)} have continued to provide
compelling evidence that D/H varies from place to place in our Galaxy
\citep{jenkins99, sonneborn00, moos02, wood04, linsky06} (hereafter L06).

Today, deuterium measurements in the lowest metallicity, high-redshift
QSO absorption systems are preferred for cosmological purposes in
order to measure a D/H abundance that is close to the primordial value
and minimally confused by astration
\citep[e.g.,][]{omeara06,pettini12, cooke18, zavarygin18}.  The metallicity of the Milky Way
ISM is 5-600 times higher than the metallicity of the QSO absorbers
typically used to constrain the primordial D/H and cosmological baryon
density \citep{cooke18}, but the Milky Way measurements are still
relevant for two reasons. First, it is important to understand the
origin of the spatial variability of D/H in the Galactic ISM in order
to ensure that the high-redshift measurements are also interpreted
correctly. After years of work high$-z$ D/H measurements have now
fairly well converged but do exhibit some scatter \citep[see,
e.g., Figure 7 in][]{cooke18}. We want to understand this deuterium variability to make
sure we are interpreting all of these results correctly.  The Milky Way ISM is likely the best
laboratory for probing the physical processes that affect D/H.
Second, by comparison with high-redshift measurements, Milky Way
deuterium abundances constrain models of the chemical evolution of our
Galaxy.

One possible explanation for the Galactic D/H variability is that the
Milky Way could still be accreting relatively pristine gas with a high
deuterium abundance and a low metallicity.  The Galactic high-velocity
cloud Complex C is an example of a sub-solar metallicity cloud with a
high deuterium abundance that appears to be falling into the Milky Way
\citep{sembach04}; if infalling clouds like Complex C have merged into
the Galactic ISM but are poorly mixed, this could lead to patchy
(spatially variable) deuterium abundances.  However, in this scenario
an anticorrelation between metallicity and D/H would be expected
because the processing inside stars that destroys D also creates
metals.  This anticorrelation does not appear to be present in the
data \citep{hebrard03} but it is important to confirm this result with
larger samples and precise measurements.

\begin{deluxetable*}{lrrrccccc}
\tablecaption{Stellar Information and STIS Observation Log\tablenotemark{a}\label{stis_obslog}}
\tablewidth{0pt}
\tablehead{
\colhead{Object} & \colhead{$l$} & \colhead{$b$} &
\colhead{$V$} & \colhead{Obs Date} & \colhead{Exp. Time} & \colhead{Grating ($\lambda_c$)} &
\colhead{Aperture} & \colhead{Dataset} \\
&
\multicolumn{1}{c}{($\deg$)} & 
\multicolumn{1}{c}{($\deg$)} &
\multicolumn{1}{c}{(mag)} & &
\multicolumn{1}{c}{(s)} &
\multicolumn{1}{c}{(\AA)} &
(arcsec) \\
}
\startdata
HD 191877 	&	61.6		&	$-6.45$	&	6.26	&	2011-06-17	&	1183	&	 E140H (1271)		&	0.1x0.03	&	OBIE08010	\\
BD+39 3226  	&	65.0		&	28.8		&	10.2	&	2011-07-24	&	1757	&	 E140H (1271)		&	0.2x0.09	&	OBIE09010	\\
Feige 110 	 	&	74.1		&	$-59.1$	&	11.4	&	2010-12-12	&	1734	&	 G140M (1218)		&	52x0.05	&	OBIE01010	\\
PG 0038+199 	&	119.8	&	$-42.7$	&	14.5	&	2010-12-31	&	1882	&	 G140M (1218)		&	52x0.05	&	OBIE10010	\\
HD 41161  	&	165.0	&	12.9		&	6.76	&	2010-12-12	&	1433	&	 E140H (1271)		&	0.1x0.03	&	OBIE11010	\\
HD 53975  	&	225.7	&	$-2.3$	&	6.48	&	2011-10-02	&	1161	&	 E140H (1271)		&	0.1x0.03	&	OBIE12010	\\
TD1 32709  	&	233.0	&	28.1		&	12	&	2011-05-06	&	1732	&	 G140M (1218)		&	52x0.05	&	OBIE13010	\\
WD 1034+001 	&	247.6	&	47.8		&	13.2	&	2011-04-20	&	1904	&	 G140M (1218)		&	52x0.05	&	OBIE14010	\\
LB 3241 	 	&	273.7	&	$-62.5$	&	12.7	&	2011-06-24	&	2070	&	 G140M (1218)		&	52x0.05	&	OBIE07010	\\
LSS 1274\tablenotemark{b}  	&	277.0	&	-5.3	&	12.9	&	2011-04-14	&	0	&	 G140M (1218)		&	52x0.05	&	\ldots \\
HD 90087  	&	285.2	&	$-2.1$	&	7.8	&	2011-09-27	&	2038	&	 E140H (1271)		&	0.2x0.09	&	OBIE15010	\\
CPD$-$71 172 	&	290.2	&	$-42.6	$	&	10.7	&	2011-08-21	&	1847	&	 G140M (1218)		&	52x0.05	&	OBIE05010	\\
LB 1566 	 	&	306.4	&	$-62.0$	&	13.1	&	2011-07-09	&	2185	&	 G140M (1218)		&	52x0.05	&	OBIE06010	\\
LSE 44 	 	&	313.4	&	13.5		&	12.5	&	2011-07-12	&	2071	&	 G140M (1218)		&	52x0.05	&	OBIE04010	\\
JL 9 		 	&	322.6	&	$-27.0$	&	13.2	&	2011-07-22	&	2301	&	 G140M (1218)		&	52x0.05	&	OBIE16010	\\
LSE 234 	 	&	329.4	&	$-20.5$	&	12.6	&	2011-02-06	&	2197	&	 G140M (1218)		&	52x0.05	&	OBIE17010	\\
LSE 263 	 	&	345.2	&	$-22.5	$	&	11.3	&	2011-05-26	&	1479	&	 G140M (1218)		&	52x0.05	&	OBIE03010	\\
\enddata
\tablenotetext{a}{HST data for this program can be obtained from the  {\it Mikulsky Archive for Space Telescopes\/} ({\it MAST\/}) at doi: {\doi{10.17909/j4tb-bk98}}.}
\tablenotetext{b}{Observation failed due to target coordinate error. No data were obtained.}
\end{deluxetable*}

A second hypothesis, originally proposed by \citet{jura82}, is that
the D/H spatial variability is due to depletion by dust grains, which
might more effectively remove D than H
\citep{tielens83,draine04,draine06}. In this case, a correlation
between D/H and abundances of depletable metals would be expected:
increased dust content would lead to a lower D/H ratio in the gas
phase (a higher portion of the D would be stuck on dust grains) as
well as lower gas-phase abundances of metals that tend to be
incorporated into dust (e.g., titanium, nickel, or iron).  The first
observational support for this hypothesis was provided by
\citet{prochaska05}, who found that D/H and Ti/H are correlated at
95\% confidence.  Further evidence supporting this explanation was
subsequently reported by L06, \citet{ellison07}, and
\citet{lallement08}. Some other aspects
of the observations do not entirely fit with the dust-depletion hypothesis.
For example, sight lines with low H$_{2}$ fractions and low values of the
$E_{B-V}$ color excess would be expected to have high D/H values because little
depletion should occur, but the opposite is observed -- some sight
lines with low $E_{B-V}$ and low H$_{2}$ fractions also have \textit{low}
D/H ratios.  Likewise, some of the directions with high H$_{2}$
fractions have the highest D/H ratios \citep{steigman07}.  These unexpected
behaviors implicitly assume that all species (D, metals, molecules, and dust)
share similar distribution with similar fractional abundances regardless
of the sight line considered, which  is not necessarily true \citep{welty20}
and is difficult to verify with the data at hand.  When combined with the presence
of significant outliers and the peculiar slopes in the relationships between D/H
and depleted metal abundances (Ellison et al. 2007), this potential degeneracy
also makes the dust-depletion hypothesis more problematic to probe.

This paper is organized as follows. In Section~\ref{sec:considerations} we describe some practical
considerations in the choice of observing strategy for this program and in Section~\ref{sec:targets} we describe
the targets and observing details. In Section~\ref{sec:stellar_model} we discuss the computation of the
stellar models used in the modeling of the interstellar \lya\ absorption line. In
Section~\ref{sec:NHI} we describe in detail the computation of the \HI\ column density and its
associated error.  Our objective is to determine D/H and we rely on published values
of \NDI. However, there are no published values of this for five of our targets, and in
Section~\ref{sec:NDI} we discuss our measurements of \NDI\ for these targets. In Section~\ref{sec:results}
we describe the principal results of this study, our new D/H results along 16
lines of sight, and how they compare with previous measurements in
the high \NHI\ regime. In Section~\ref{sec:metals} we
describe our abundance measurements of various metals in the five targets for
which we obtained high resolution spectra. The correlation of these abundances with
D/H and their interpretation in terms of a unified depletion analysis is given in Section~\ref{sec:depletions}.
In Section~\ref{sec:discussion} we discuss our results in the broader context of the distribution of D/H
measurements as a function of \NHI, the evidence for depletion of deuterium
onto dust grains and for infall of deuterium rich material, and how our results fit into models of Galactic chemical evolution.
We summarize the results of this study in Section~\ref{sec:summary}.

\section{Practical Considerations} \label{sec:considerations}

Ironically, in many interstellar sight lines, the column density of
the much rarer deuterium isotope is easier to measure than the
column density of the abundant \textsc{H~i}, and in many cases the
uncertainty in D/H is dominated by the uncertainty in
$N$(\textsc{H~i}) (L06).  When the \textsc{H~i} column is high enough so
that \textsc{D~i} can be detected, most of the higher \textsc{H~i}
Lyman series lines are strongly saturated but do not exhibit
well-developed damping wings; the \textsc{H~i} Ly$\alpha$ line must be
observed to tightly constrain $N$(\textsc{H~i}).  Consequently, many
\textit{FUSE} observations (which do not cover Ly$\alpha$) provide
precise measurements of $N$(\textsc{D~i}) but only very crude
constraints on $N$(\textsc{H~i}).  In some cases, \textit{FUSE}
results can be combined with \textit{Hubble Space Telescope (HST)}
spectra of Ly$\alpha$ resulting in exquisite D/H measurements
\citep[e.g.,][]{sonneborn02}.  Unfortunately, for some of the key
sight lines, only single, low-quality \textit{IUE} spectra of the
\textsc{H~i} Ly$\alpha$ line were available for the analyses above (if
any Ly$\alpha$ data were available at all), and the uncertainties in
$N$(\textsc{H~i}) were large and dominated by difficult to assess
systematic errors \citep{friedman02,friedman06}. For example, the
significance of the correlation between D/H and Ti/H observed by
\citet{prochaska05} hinges on the sight line to Feige 110, which
unfortunately has a very uncertain \textsc{H~i} column derived from
a single, poor-quality \textit{IUE} spectrum \citep{friedman02}.
Likewise, outliers in the various studies above could be spurious
measurements due to poor $N$(\textsc{H~i}) constraints, or they could
be real outliers that indicate that the spatial variability is not
necessarily due entirely to dust depletion.  Better measurements are
needed to understand the D/H variability and its implications.

To rectify the uncertainties in D/H measurements due to poor or
non-existent \textsc{H~i} Ly$\alpha$ data, we embarked on a program in
\textit{HST} Cycle 18 to obtain much better \textsc{H~i} Ly$\alpha$
spectra using the Space Telescope Imaging Spectrograph (STIS)
\citep{kimble98,woodgate98} on \textit{HST}.  This spectrograph
provides vastly better spectra of the Ly$\alpha$ line \citep[see,
  e.g., Figure 4 in][]{sonneborn02} than \textit{IUE} and enables
precise measurement of $N$(\textsc{H~i}) even in cases where the
interstellar Ly$\alpha$ is blanketed with narrow stellar lines
\citep{sonneborn02}.  Moreover, the new STIS spectra have high
spectral resolution and enable measurement of a variety of metal
column densities including species such as \textsc{O~i},
Mg~\textsc{ii}, \textsc{P~ii}, Cl~\textsc{i}, Mn~\textsc{ii},
Ni~\textsc{ii}, and Ge~\textsc{ii}, so the new STIS data also enable
improvements of the metal measurements.  In this paper we present the
findings of this study. 

\begin{deluxetable*}{lccccccc}
\tablecolumns{8}
\tablecaption{Stellar Parameters Used for the Model Atmospheres\label{tab:star_param}}
\tablewidth{0pt}
\tablehead{
\colhead{Star} & \colhead{Sp Type} & \colhead{$T_{\rm{eff}}$} & \colhead{$\log g$} &
\colhead{$\log N({\rm{He}})/N({\rm{H}})$} & \colhead{$v\sin i$} & 
    \colhead{$Distance\tablenotemark{a}$} & \colhead{Ref.}  \\
\colhead{} &
\colhead{} &
\colhead{(K)} &
\colhead{(cm s$^{-2}$)} &
\colhead{} &
\colhead{(km s$^{-1}$)} &
\colhead{(pc)} &
\colhead{} 
}
\startdata
\cutinhead{Hot Subdwarfs}
BD+39 3226 & He-sdO & $45970\pm1000$ & $6.05\pm0.10$ & $0.50\pm0.10$ & \nodata & $189\pm2$  & 1 \\
Feige 110 & sdOB & $44745\pm2000$ & $5.96\pm0.15$ & $-1.78\pm0.10$ & \nodata & $271\pm4$ & 2 \\
TD1 32709 & He-sdO & $46500\pm1000$ & $5.60\pm0.15$ & $2.0\pm0.3$ & $31\pm3$ & $522\pm16$ & 3,4,5 \\
LB 3241   & sdO & $42200\pm2000$ & $5.60\pm0.20$ & $< -3.3$ & \nodata & $653\pm18$ & 2 \\
CPD$-$71 172 & sdO    & $60000\pm5000$ & $5.4 \pm0.2 $ & $-1.0$ & \nodata & $328\pm2$ & 6 \\
LB 1566   & He-sdO & $49320\pm2000$ & $5.84\pm0.20$ & $> 1.8$ & \nodata & $823\pm26$ & 2 \\
LSE 44    & sdO & $39820\pm2000$ & $5.50\pm0.20$ & $-2.90\pm0.10$ & \nodata & $615\pm15$ & 2 \\
JL 9      & He-sdO & $75000\pm5000$ & $5.50 \pm0.25$ & $0.21\pm0.10$ & \nodata & $1592\pm78$ & 2 \\
LSE 234   & sdO    & $90000\pm5000$ & $6.0\pm0.3$ & $-1.0\pm0.1$ & \nodata & $601\pm16$ & 7 \\
LSE 263   & He-sdO & $70000\pm2500$ & $4.90\pm0.25$ & $> 1.0$ & \nodata & $719\pm59$ & 8 \\
\cutinhead{White Dwarfs}
PG 0038+199 & DO & $125000\pm5000$ & $7.0\pm0.5$ & $ 1.7$ & \nodata & $400\pm7$ & 9 \\
WD 1034+001 & DO & $115000\pm5000$ & $7.0\pm0.5$ & $> 1.9$ & \nodata & $193\pm2$ & 9 \\
\cutinhead{O and B Stars}
HD 191877   & B1.0 Ib & 21700 & 2.67 & \nodata & 152 & $1811\pm141$ & 10,11,12 \\
HD 41161    & O8.0 Vn  & 34877 & 3.92 & \nodata & 296 & $1489\pm134$ & 13,14,15 \\
HD 53975    & O7.5 Vz  & 35874 & 3.92 & \nodata & 163 & $1154\pm65$ & 13,14,12 \\        
HD 90087    & O9.0 II & 31607 & 3.38 & \nodata & 259 & $2193\pm126$ & 16,14,15 \\
\enddata
\tablenotetext{a}{Distances from ${\it Gaia}$ Data Release 3 (DR3).}

\tablerefs{
(1) \citet{chayer_etal2014}; 
(2) This study; 
(3) \citet{dreizler93}; 
(4) \citet{schindewolf_etal18};
(5) \citet{hirsch_09};
(6) \citet{deleuil_viton92}; 
(7) \citet{haas_etal95}; 
(8) \citet{husfeld_etal89};
(9) \citet{werner_etal17}; 
(10) \citet{lesh68}; 
(11) \citet{searle_etal08};
(12) \citet{howarth_etal97}; 
(13) \citet{sota_etal11}; 
(14) \citet{martins_etal05}; 
(15) \citet{penny96};
(16) \citet{garrison_etal77}. 
}
\end{deluxetable*}

\section{Target Selection and Observations} \label{sec:targets}

The Local Bubble (LB) is a volume of space that contains a mixture of low density ionized and neutral gas in which the Sun is embedded \citep{Breitschwerdt98} and has an irregular boundary about $100-300$ pc from the Sun \citep{pelgrims20}. This boundary consists of
neutral hydrogen, so sight lines with \NHI\ $\gtrsim 10^{19.2}$ cm$^{-2}$ usually extend beyond the LB into the more distant ISM. Previous work has shown that D/H is approximately constant within the LB but it exhibits considerable variability beyond it (see Figure 1 in L06). To investigate the
cause of this variability we selected targets that (1) lie outside the LB, (2) have \fuse\ spectra of sufficient quality to permit accurate computation of \NDI\ or that such published values already exist, and (3) that the stellar fluxes were appropriate to obtain excellent \lya\ profiles in a single HST orbit. The 17 targets selected as part of program ID 12287 are listed in Table~\ref{stis_obslog} along with various exposure parameters. Sixteen targets were successfully observed.

The targets of choice for studying sight lines just beyond the LB are hot subdwarf stars (spectal type sdB, sdOB, sdO, He-sdO). Hot stars are favored because their flux peaks in the ultraviolet where several interesting atomic transitions occur, and subluminous stars because their spatial density allows the observation of bright stars at distances of a hundred to several hundred parsecs. To the distances where we find these hot subdwarfs, we can add the extremely hot white dwarfs, although they are much less numerous than the hot subdwarfs.  At distances well beyond the LB, O and B stars are used to explore long sight lines. Although these faint and bright blue stars are suitable targets for studying the interstellar \ion{H}{1} \lya\ line, their stellar spectra present challenges for measuring accurate \ion{H}{1} column densities.

The spectra of 11 targets were obtained using the first order grating G140M on STIS. This setup provides a very clean spectrum spanning approximately $1192-1245$\,\AA\ at a velocity resolution of $\sim30$ \kms, which is adequate for measuring \NHI\ using the damping wings of the \lya\ line. Five targets were too bright to be observed with G140M and were instead observed in echelle mode with the E140H grating, spanning $1164-1356$\,\AA\ at a resolution of $\sim2.6$ \kms, which is sufficient for measuring metal abundances for these sight lines, as discussed in Section~\ref{sec:metals}. A potential disadvantage of this observing configuration is that echelle reductions can suffer from imperfect ripple corrections, causing the spectral orders to be improperly joined. We found that the IDL procedure hrs\_merge.pro produced excellent 1-d spectra and no further correction was required. The standard pipeline-reduced data show zero flux at the center of the saturated \lya\ cores (aside from minor geocoronal \lya\ emission) so no additional background correction was required for any spectrum. We note that the flux zero point was an especially troublesome issue for many previous analyses, especially those that relied on \textit{IUE} observations (see e.g., \citet{friedman06}).

\begin{figure*}
\epsscale{1.1}
\plotone{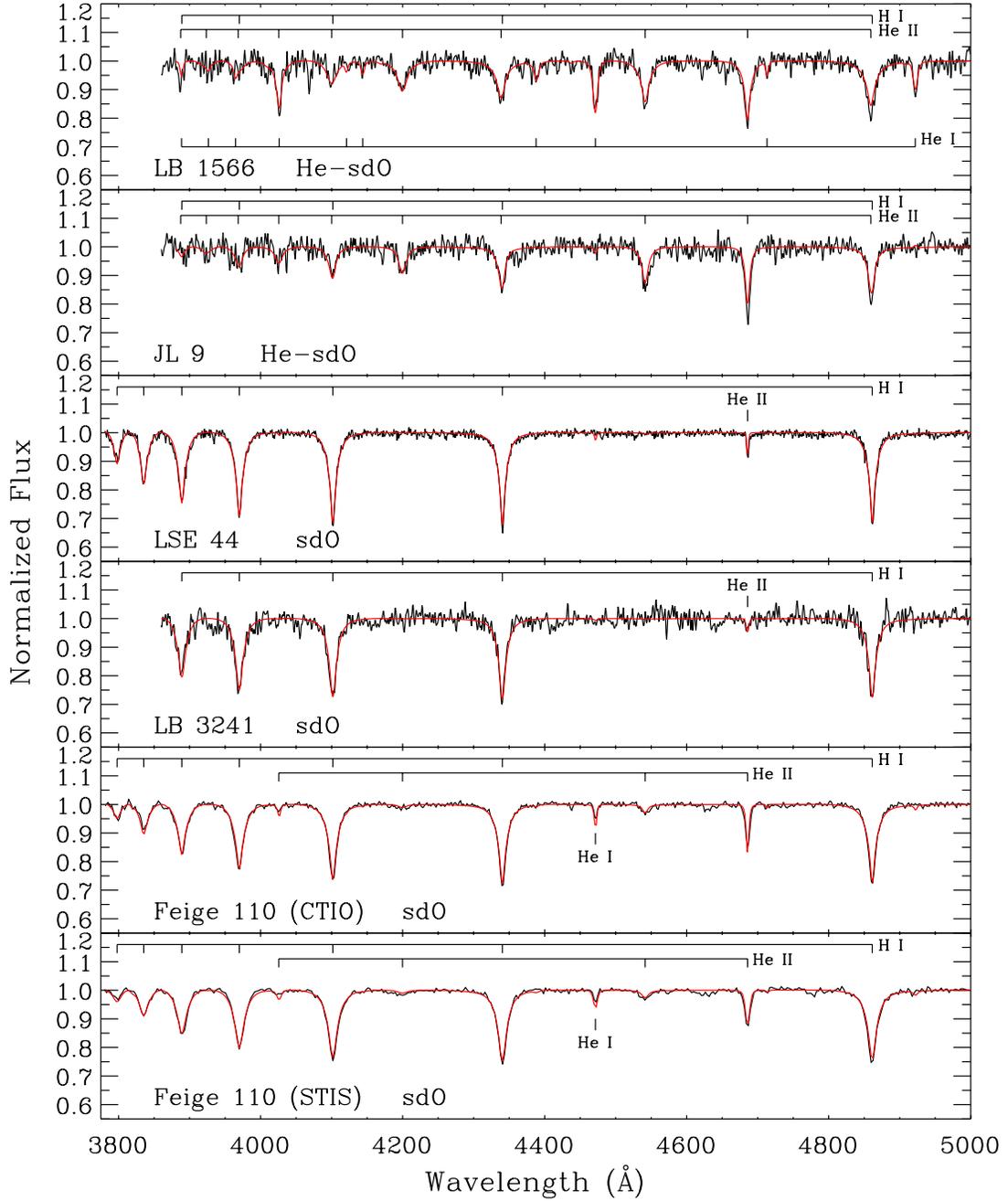}
\caption{Best model-atmosphere fits to the optical spectra of the hot subdwarf stars analyzed in this study. The names of the stars with their spectral types  are given in each panel along with identifications of \ion{H}{1}, \ion{He}{1}, and \ion{He}{2} lines. Two optical spectra are available for Feige~110 and both are used to estimate the uncertainties of the atmospheric parameters.}
\label{fig:optical_fits}
\end{figure*}

\section{Stellar Models}\label{sec:stellar_model}

In order to take into consideration the stellar contributions to the Ly$\alpha$ line profile and the placement of the continuum, we computed synthetic spectra using stellar atmosphere models. 
We used the stellar atmosphere and spectrum synthesis codes {\tt TLUSTY}\footnote{\url{http://tlusty.oca.eu}} \citep{hubany95}  and {\tt Synspec}\footnote{\url{http://tlusty.oca.eu/Synspec49/synspec.html}} to compute non-local thermodynamic equilibrium (NLTE) stellar atmosphere models and synthetic spectra.  Both codes were developed by I. Hubeny and T. Lanz.  A detailed description and use of the codes are presented in a series of three papers by \citet{hubeny_lanz2017a, hubeny_lanz2017b, hubeny_lanz2017c}. Synthetic spectra are calculated from models of stellar atmospheres that describe the surface properties of stars and are based on the results of spectroscopic data analysis. Table~\ref{tab:star_param} gives the atmospheric parameters of the 16 stars considered in this study. These parameters are the surface gravity, the effective temperature, and the number ratio of helium-to-hydrogen. We also take into account the chemical composition of the atmospheres which is important for hot stars and in particular hot subdwarfs and white dwarfs. These high-gravity stars show abundance anomalies that arise from the effects of diffusion. We determined the atmospheric parameters of five stars in this study, and collected the parameters of other stars from the literature.

The atmospheric parameters of these five stars were determined by fitting the H and He lines observed in optical spectra with two grids of NLTE atmosphere models.  LB~3241, LB~1566 and JL9 were observed with the CTIO\footnote{Cerro Tololo Inter-American Observatory, \url{http://www.ctio.noao.edu}} 1.5-m Cassegrain spectrograph by H.E. Bond in 2011. The spectrograph was configured to use the 26/Ia grating and a slit of 110~$\mu$m. This configuration produced wavelength coverage ranging from 3650 to 5425 \AA\ and spectral resolution of FWHM = 4.3 \AA. JL~9 and LB~1566 have exposure times of 400~s each while LB~3241 has an exposure time of 350~s. The signal-to-noise ratio from each exposure is about 50. The optical spectra of Feige~110 and LSE~44 are described in \citet{friedman02, friedman06}. As Feige~110, LB~3241, and LSE~44 are He-poor stars, we used the grid of NLTE models for extreme horizontal branch stars developed by \citet{brassard_etal2010}. This grid covers the ranges of 20,$000\le T_{\rm{eff}} \le 50$,000~K in steps of 2000~K, $4.6 \le \log g \le 6.4$ in steps of 0.2 dex, and $-4.0 \le \log(N(\rm{He})/N(\rm{H})) \le 0.0$ in steps of 0.5 dex. These models assume a metallicity of C = 0.1, N = 1.0, O = 0.1, S = 1.0, Si = 0.2, and Fe = 1.0 $\times$ solar values \citep{grevesse98}, which is typical of these H-rich stars. For the He-rich stars LB~1566 and JL~9, we used our own grid of NLTE H-He models that covers the ranges 30,$000\le T_{\rm{eff}} \le 98$,000~K in steps of 2000~K, $4.8 \le \log g \le 7.0$ in steps of 0.2 dex, and $0.0 \le \log(N(\rm{He})/N(\rm{H})) \le 3.0$ in steps of 0.5 dex. Figure~\ref{fig:optical_fits} shows our best fits to the optical spectra. The effective temperature of JL~9 was increased from 68,820~K obtained from the optical fit alone to 75,000~K, as shown in Table~\ref{tab:star_param}, in order to reproduce the ionization balance of the \ion{Fe}{6} and \ion{Fe}{7} ions which are observed in the \textit{FUSE} and STIS spectra. \citet{werner_etal2022} came to a similar conclusion although their temperature and gravity are somewhat different with $T_{\rm{eff}} = 80,000\pm5000$~K and $\log g = 5.2\pm0.3,$ but the error analysis described in Section~\ref{sec:NHIerror} shows that this has a negligible effect on the value of \NHI\ we obtain. 

\begin{figure*}
\epsscale{0.85}
\plotone{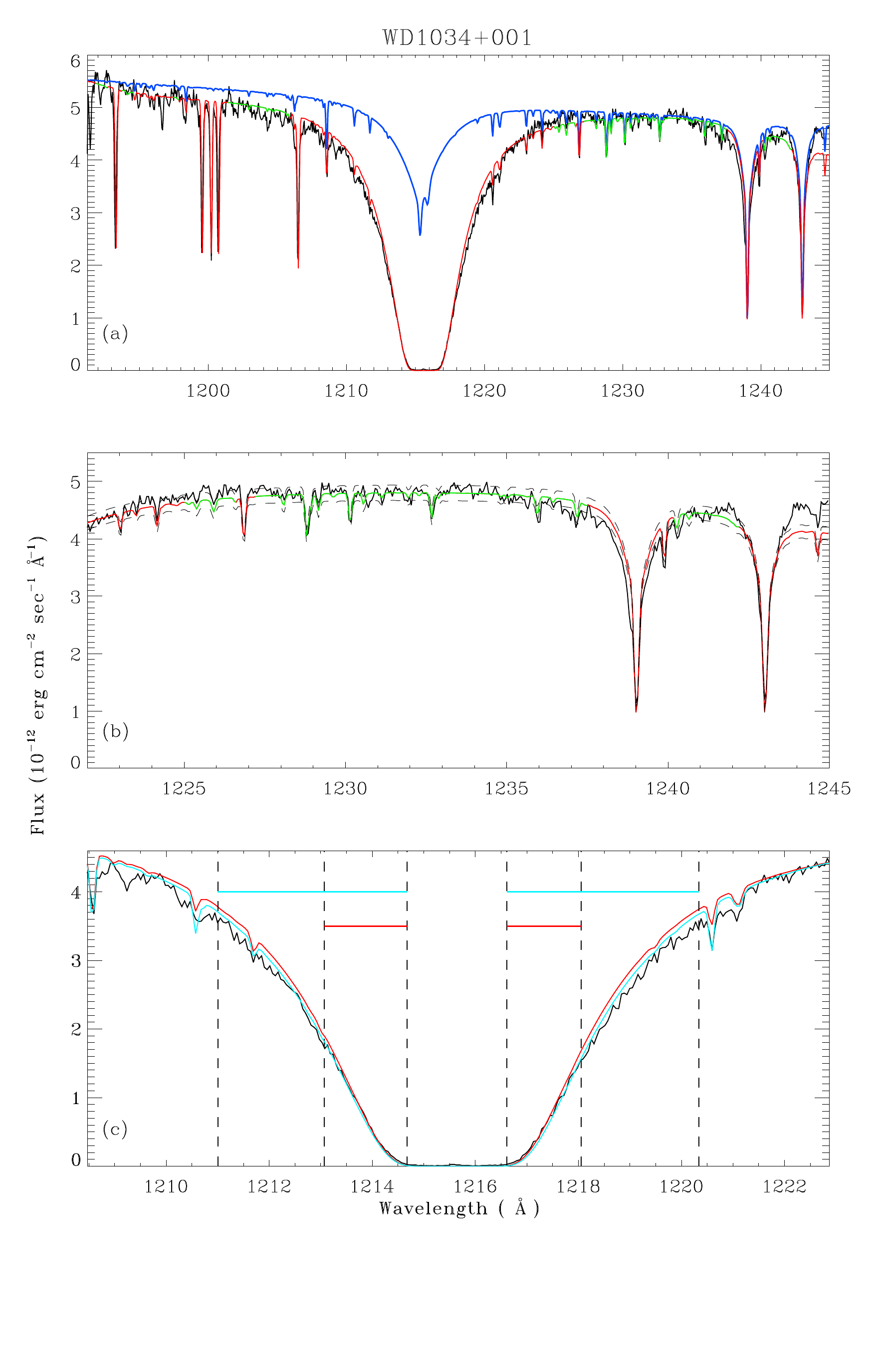}
\caption{The spectrum of WD1034+001 in detail. (a) The observed spectrum is shown in black and our best fit model spectrum in red. The spectral regions that are used to constrain the polynomial fit to the continuum are shown in green. The blue line is the model of the stellar atmosphere. (b) An expanded view of the long-wavelength portion of the spectrum showing a continuum region and the stellar absorption lines used to constrain the radial velocity of the stellar model. The dashed lines show the high and low continuum scalings used to determine the contribution of continuum placement errors to the error in \NHI. (c) An expanded view of the \lya\ line. \NHI\ and the radial velocity of the interstellar \HI\ are most strongly constrained in the spectral regions just outside the black core of the absorption line. The nominal spectral region used to constrain these parameters is shown by the red horizontal bars and corresponds to the red spectrum. To estimate the error associated with the choice of spectral region, we also calculate \NHI\ and the radial velocity based on the extended region indicated by the cyan horizontal bars and the corresponding cyan spectrum. The spectral coverage indicated by the red and cyan bars differs for each target. See text for a discussion of the extent of these bars and of the discrepancy between the black and red spectra in the upper wings of the damped \lya\ profile.
\label{wd1034}}
\end{figure*}

As Table~\ref{tab:star_param} indicates, we placed our targets into three classes: Hot subdwarfs, white dwarfs, and O and B stars. For each star, we calculated stellar atmosphere models which are based on the atmospheric parameters and their uncertainties, and the abundances of metals published in the literature (see references in Table~\ref{tab:star_param}).  For those stars that did not have metal abundances, we used \textit{FUSE} and STIS spectra to determine their abundances. In the case of O and B stars, we used the grids of model atmospheres that were calculated by \citet{lanz_hubeny2003, lanz_hubeny2007}. In order to take into account the effect of atmospheric parameter uncertainties on the stellar contribution to the determination of \ion{H}{1} column densities, we calculated models at the extremes of effective temperature and gravity. Models with $T_{\rm{eff}} - \Delta T_{\rm{eff}}$ and $\log g + \Delta \log g$ produce stronger Ly$\alpha$ and $\lambda$1215 \ion{He}{2} stellar lines, while models with $T_{\rm{eff}} + \Delta T_{\rm{eff}}$ and $\log g - \Delta \log g$ produce weaker lines (see Section~\ref{sec:NHIerror}). Synthetic spectra were calculated from these atmosphere models. They have been calculated to cover the wavelength ranges of the low and high resolution STIS spectra. They were convolved with Gaussians of FWHM = 0.1 and 0.03 \AA, respectively. Finally, a rotational convolution was performed on the spectra for stars with high $v \sin i$ values. In some cases the models were better constrained than previous ones in the literature due to accurate distances provided by the {\it Gaia} DR3 \citep{soszynski16, vallenari22}.

\section{Measurement of \NHI} \label{sec:NHI}

In this section we describe the method used to compute the interstellar \HI\ column density. 
There are 9 adjustable parameters in our fits to the observed spectra. They are the coefficients of the 6th order polynomial fit in clear portions of the continuum region adjacent to the \lya\ absorption line; the radial velocity of the modeled stellar spectrum with respect to the observed spectrum; the radial velocity of the modeled interstellar \lya\ absorption with respect to the observed spectrum; and, of course, the value of \NHI\ itself. The b-value of the absorbing gas is not important because the Gaussian part of the Voigt profile is buried deep inside the black core of the strong \lya\ line. 

Figure~\ref{wd1034} illustrates how we measure the \HI\ column density by examining the spectrum of WD1034+001 in detail. The first step is to select regions of the continuum which are relatively free of absorption lines, over which the polynomial will be fit. These are shown in green in panel (a). Note that we have used continuum regions on both the blue and red sides of \lya. (For HD90087 there is so much absorption on the blue side of \lya\ that the continuum never recovers, so only the red side was used to constrain the polynomial.) The purpose of this polynomial fit is to remove residual instrumental variations and the wavelength dependent reddening by dust. Next, the radial velocity of the stellar model is adjusted based on selected stellar absorption lines, such as those shown in the expended red spectral region in panel (b). We avoided lines that might have a significant interstellar contribution. For WD1034+001 we used the strong \NV\ $\lambda\lambda 1238, 1242$ doublet. The remaining weak stellar lines do not significantly improve the constraint on the radial velocity in this case, but they match the model well. We then fix the stellar velocity so the automated fitting program does not try to assign the stellar lines to one of the many interstellar features in the spectrum. The final value of \NHI\ is insensitive
to this velocity since the stellar \HI\ and \HeII\ absorptions (the most prominent absorption lines in the stellar model, shown in blue in panel (a)) are almost completely contained within the black core of the interstellar \HI\ absorption.

Next we do a simultaneous fit of the remaining 8 parameters listed in the previous paragraph by  minimizing \chisq\ between the model and observed spectra in the green continuum regions shown in panel (a) and in the damping wings of the \lya\ profile as shown in panel (c). For this task we used the {\tt amoeba} software routine in {\tt IDL}. In the \chisq\ calculation that minimizes the outcome for \NHI\ we assigned greater weight to the \lya\ region than to the continuum regions because we did not want  difficulties in the continuum fit to compromise the important \lya\ fit, particularly in continuum regions far from the \lya\ line which are unimportant for the determination of the \HI\ column density.

\begin{figure}
\epsscale{1.25}
\plotone{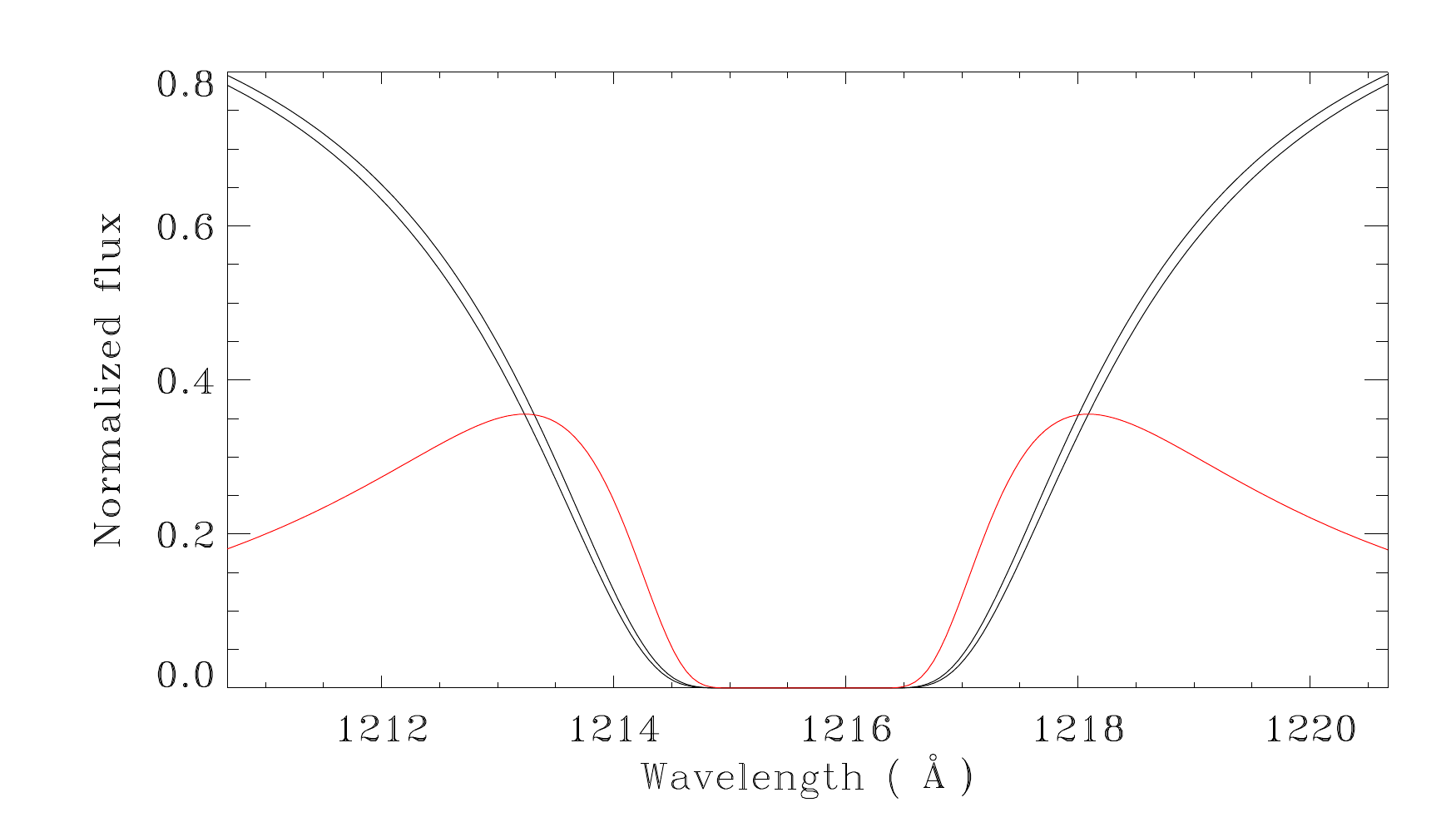}
\caption{The top black curve shows the continuum-normalized \lya\ profile with no other interstellar or stellar absorption lines for log(\NHI) = 20.12, corresponding to the column density of WD1034+001. The bottom black curve is for log(\NHI) greater by 0.03 dex. The red curve is the difference between these profiles multiplied by 15 to more clearly show the wavelengths where the two damped profiles differ the most. The peak of this curve guided our selection of the \lya\ fitting region described in Section~\ref{sec:NHIerror}. Note that our total error on the logarithmic \HI\ column density for every sight line in this study is between 0.01 and 0.02 dex, which is considerably less than the difference displayed in this illustration.
\label{profile_demo}}
\end{figure}

\begin{figure*}[t!]
\epsscale{0.87}
\plotone{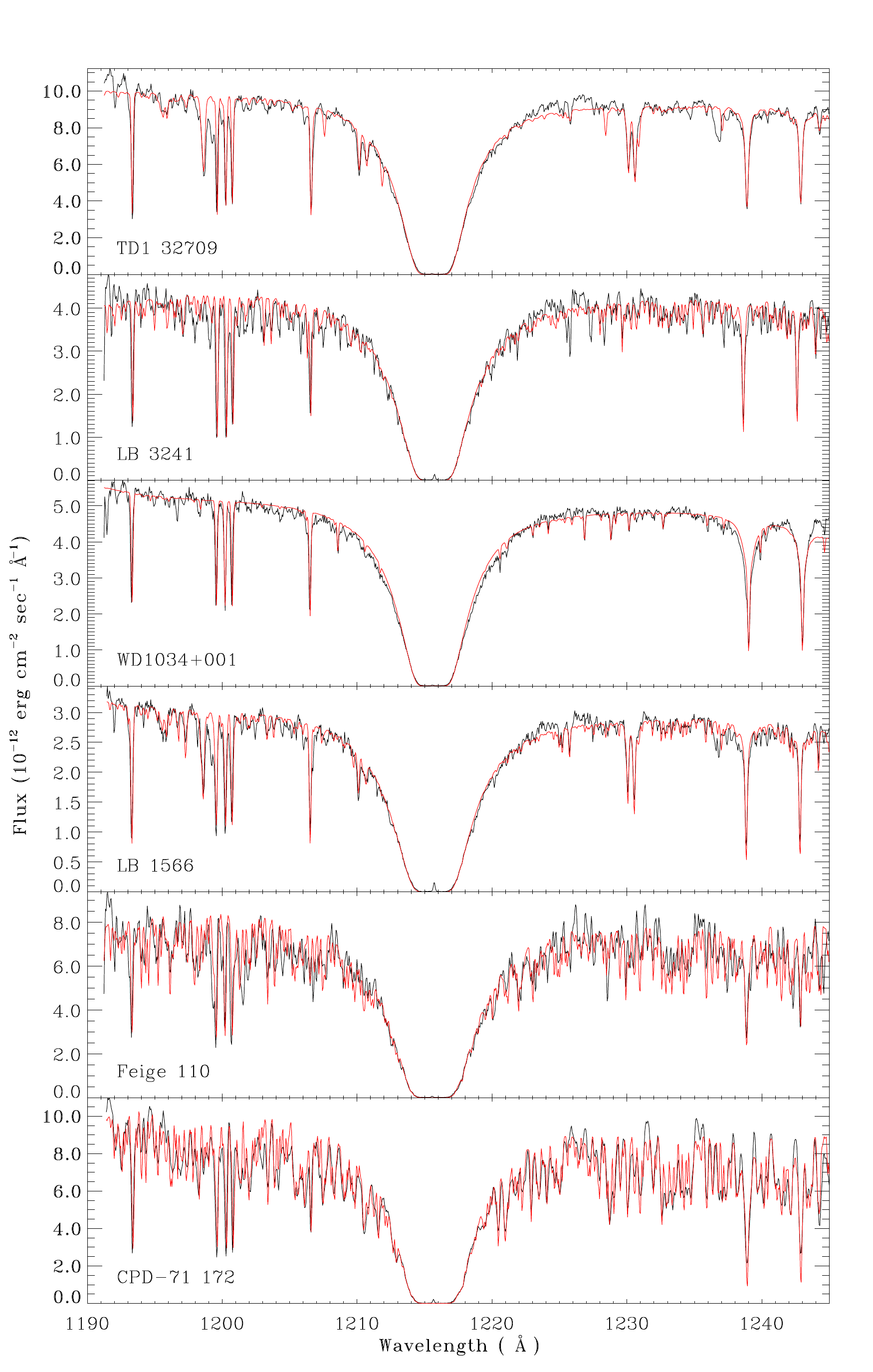}
\caption{Observed STIS spectra (in black) obtained with the STIS medium resolution G140M grating, plotted
with our model spectra (in red). The objects are labelled in each panel and are shown in order from lowest (top) to highest (bottom) values of \NHI.
The small peak in the core of the \lya\ line in some spectra is due to geocoronal \lya\ emission.
\label{spectra1}}
\end{figure*}

\setcounter{figure}{3}
\begin{figure*}[t!]
\epsscale{0.87}
\plotone{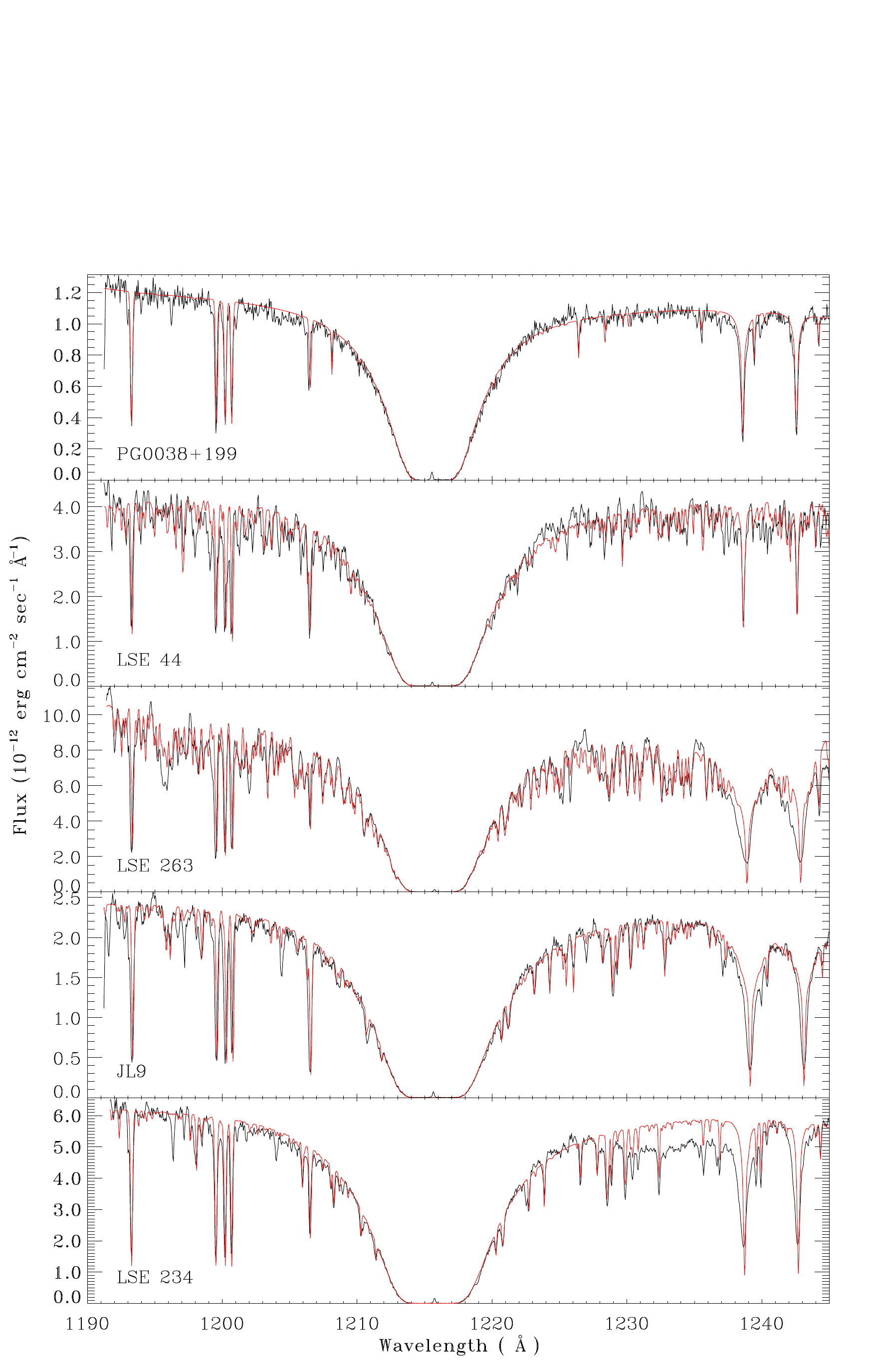}
\caption{{\it (Continued)}
\label{spectra2}}
\end{figure*}

\begin{figure*}[t!]
\epsscale{1.0}
\plotone{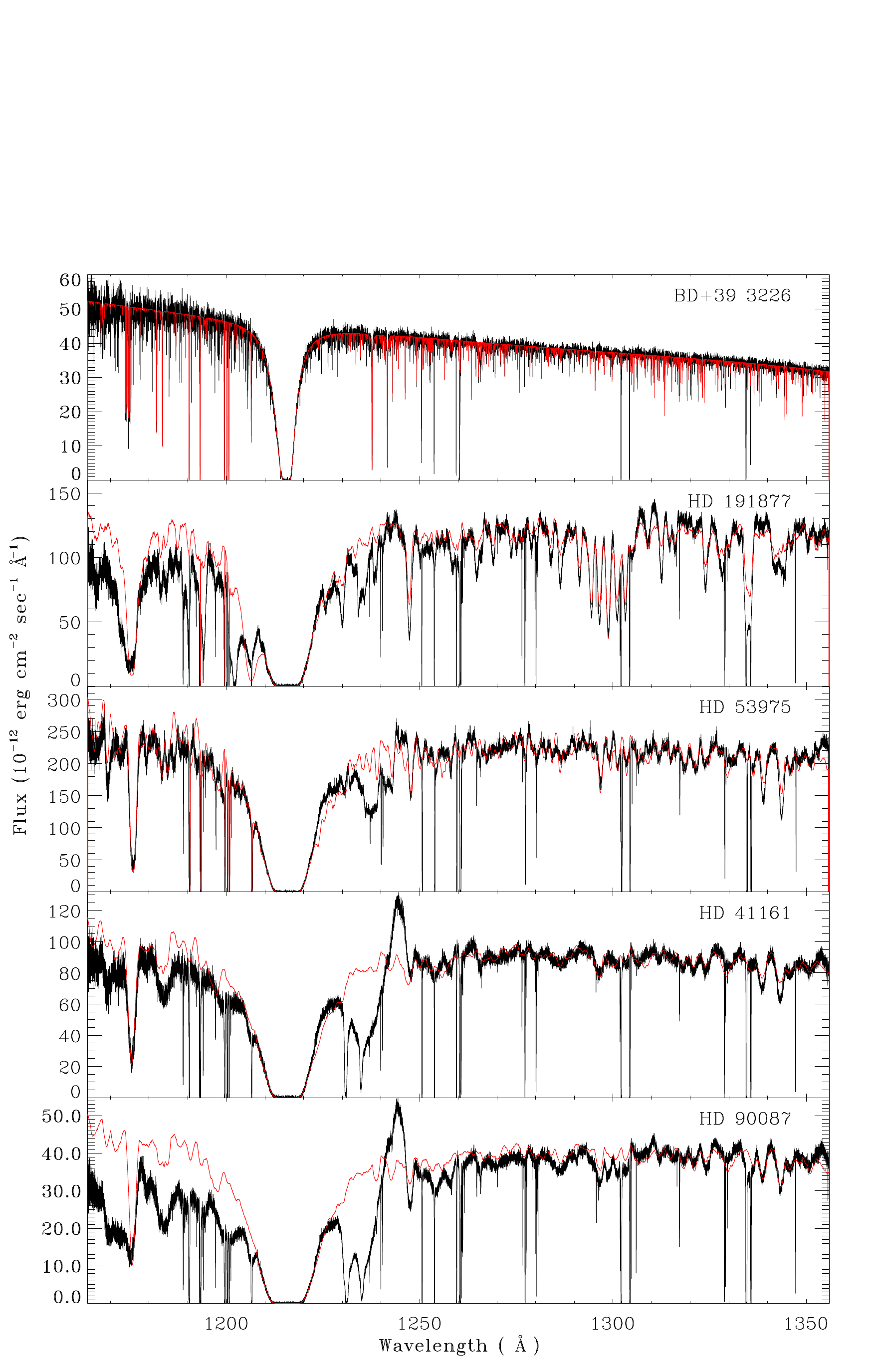}
\caption{Observed STIS spectra (in black) obtained with the STIS high resolution E140H echelle grating, plotted
with our model spectra (in red). The objects are shown in order from lowest (top) to highest (bottom) values of \NHI.
\label{spectra3}}
\end{figure*}

We now consider the proper spectral region of the damped absorption wings used to determine \NH. We want to select regions where the model spectrum most sensitively deviates from the observed spectrum due to errors in the modeled value of \NHI. Figure~\ref{profile_demo} shows a closeup of the continuum-normalized \lya\ region of WD1034+001 with no other interstellar or stellar absorption lines. The upper black line is for log(\NHI) = 20.12, the estimated column density for this object as we discuss in Section~\ref{sec:results}, and the lower line is for log(\NHI) greater by 0.03 dex. This is considerably larger than our total error on log(\NHI) for any sight line in our study, which range from $0.01-0.02$ dex, but was selected to emphasize the effect of a small increase in column density. The red curve is the difference between the two damped profiles multiplied by a factor of 15 so that it crosses the profiles at its peak values. This shows that the spectral region most sensitive to errors in log(\NHI) is about $\onethird$ of the way up to the full continuum at $y=1$ in the plot. This guided our selection of the \lya\ fitting region.

Figure~\ref{spectra1} shows the observed spectra (in black) for the 11 targets obtained with the STIS medium resolution G140M grating. Our modeled spectra are shown in red. Figure~\ref{spectra3} shows spectra of the 5 targets obtained with the high resolution E140H echelle grating, which covers a much wider wavelength interval than G140M. For the bottom 4 targets, all O and B stars (see Table~\ref{tab:star_param}), we did not attempt to model the \ion{N}{5} $\lambda\lambda$1240 P Cygni profiles, but this will not have a significant effect on the \NHI\ estimates which are primarily constrained by the spectral regions near the core of the \lya\ lines, as we just described.

\subsection{$N$(H {\small I)} error analysis} \label{sec:NHIerror}

The errors associated with determining \NHI\ are almost completely systematic in nature. Virtually every feature visible in the spectra of Figures~\ref{spectra1} and \ref{spectra3} is real but many of the lines are unidentified or are not fit well by stellar models due to unknown or inaccurate atomic physics data, such as oscillator strengths. It is this mismatch, along with uncertainties on the stellar model parameters, such as metal abundances, effective temperature, and gravity, that are responsible for the majority of the error in \NHI. 

Six possible sources of error contributed to the final error estimated for each value of \NHI. These were combined in quadrature to determine the final error. In this section we describe each of these contributions. All errors quoted in this paper are $1\sigma.$

\textbf{Statistical error.}  This error is computed based on a formal $\chi^2$ computation using the statistical error reported by the \textit{CalSTIS} pipeline. It is computed in the two regions between the vertical dashed lines around the red bars as shown, for example,  in Figure~\ref{wd1034} for WD1034. As noted above, this error is small, ranging from 0.001 to 0.004 dex for HD5 53975 and Feige 110, respectively.

\begin{figure*}
\epsscale{0.9}
\plotone{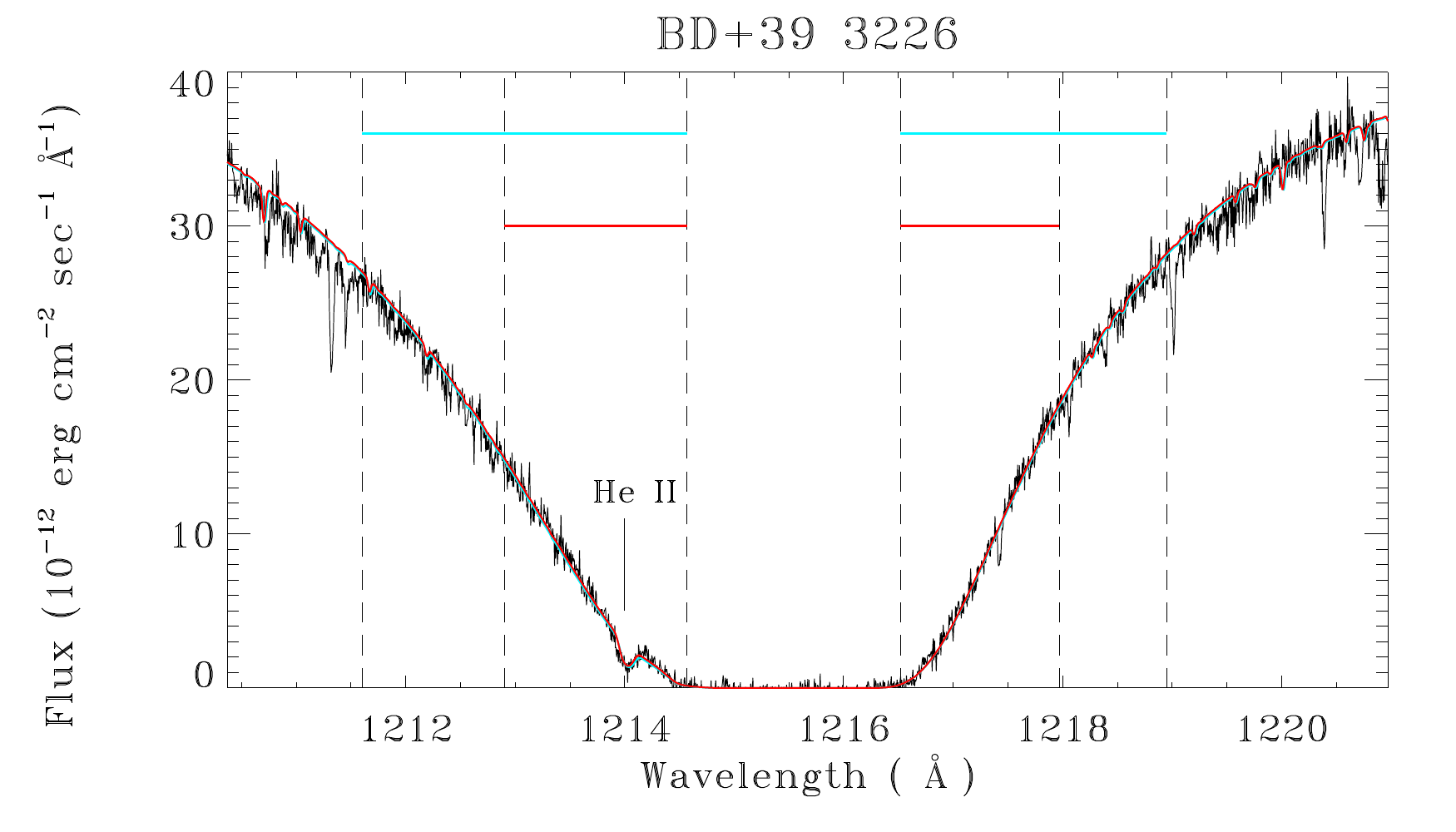}
\caption{A detailed view of the \lya\ region of BD+39 3226. The meanings of the colors and dashed lines are the same as in Figure~\ref{wd1034}c. The red spectrum is almost indistinguishable from the cyan spectrum because \NHI\ differs by only 0.006 dex between the two fits.
\label{bd39}}
\end{figure*}

\textbf{Continuum placement errors.} The process for determining the best value of \NHI\ minimizes the residual between the observed spectrum and the model spectrum. In the continuum region this is done by computing appropriate values of the coefficients of a 6th order polynomial. To estimate the contribution of continuum placement errors we compute the RMS deviation between the model and the observed spectrum in the continuum fitting regions and scale the model by this relative factor. We compute a new value of \NHI\ with this fixed, high continuum, and then do the same with the similarly scaled low continuum placement (Figure~\ref{wd1034}b). The mean of the differences of these two values of \NHI\ establishes the continuum placement error. This error ranges from 0.002 to 0.014 dex for LSE 234 and Feige 110 respectively. It is the largest source of error in \NHI\ for five stars in our sample: Feige 110, CPD$-$71 172, LSE 263, BD+39 3226, and JL9. It may be surprising that the continuum contribution to the error is so small for LSE 234 (finale panel of Figure~\ref{spectra2}) when there is such an enormous deviation between the model and the spectrum on the red side of \lya. However, our continuum scaling region excludes 1226.2 to 1240.6\AA\ for this object. More importantly, having one of the largest value of \NHI\ in our target sample, the \lya\ line has very well-developed damping wings, which renders our estimate of \NHI\ quite insensitive to continuum placement errors. In other words, far out on the wings of the interstellar absorption profile there is a large covariance in the errors of the continuum level and the amount of \HI. Near the core this is much less important. Thus, it is the core region which most strongly constrains the \HI\ column density estimate.

\textbf{Interstellar absorption velocity errors.} We assume the interstellar absorption is from a single component at a single velocity. This velocity is common to \HI\ and to low-ionization metals in the interstellar cloud, and is one of the free parameters determined as part of the \chisq\ minimization procedure. We determined the uncertainty in this velocity by measuring the velocities of metal lines such as \SiII\ $\lambda\lambda 1193.3, 1250.6$, \NI\ $\lambda\lambda 1199.5, 1200.2, 1200.7$, and \OI\ $\lambda\lambda 1302.2, 1355.6$, when these lines are present and not too saturated. The RMS dispersion in the velocities of these lines provides a measure of the uncertainty in the velocity of the absorbing cloud. This velocity error was added to the best-fit interstellar velocity and then fixed during a new computation of \NHI. The RMS error estimate was then subtracted and the same procedure followed. The mean of the differences of the resulting values provided the error contribution to \NHI. It ranged from 0.0002 to 0.003 dex for JL9 and HD 41161, respectively, and is not the largest source of error for any target in our sample.

\begin{deluxetable*}{ccccrc}
\tablecaption{\fuse\ Observation Log\tablenotemark{a}\label{newfuse_obslog}}
\tablewidth{0pt}
\tablehead{
\colhead{Target} & \colhead{Obs Date} & \colhead{Data ID} & \colhead{Exp. Time\tablenotemark{b}} &
\colhead{$N_{\rm exp}$\tablenotemark{c}} & \colhead{Aperture\tablenotemark{d}} \\
}
\startdata
LB\,1566		&	2003-07-15	&	P3020801		&	5.8	&	3	&	LWRS	\\
			&	2003-09-11	&	P3020802		&	21.8	&	21	&	MDRS	\\
LB\,3241		&	2002-09-21	&	M1050301	&	9.1	&	17	&	LWRS	\\
			&	2002-11-14	&	M1050302	&	5.6	&	11	&	LWRS	\\
			&	2002-11-16	&	M1050303	&	11.1	&	22	&	LWRS	\\
			&	2002-11-18	&	M1050304	&	9.7	&	19	&	LWRS	\\
			&	2003-09-10	&	Z9040501		&	3.2	&	7	&	LWRS	\\
LSE\,263		&	2003-05-30	&	D0660401		&	4.0	&	8	&	MDRS	\\
			&	2004-09-14	&	E0450201		&	23.1	&	32	&	MDRS	\\
LSE\,234		&	2003-04-07 	&	P2051801		&	9.8	&	20	&	LWRS	\\
			&	2003-05-30	&	P3021101		&	10.6	&	20	&	LWRS	\\
			&	2006-04-22 	&	U1093901		&	8.6	&	17	&	LWRS	\\
CPD$-$71\,172	&	2003-07-13	&	P3020201		&	11.4	&	25	&	MDRS	\\
\enddata
\tablenotetext{a}{FUSE data used in this program can be obtained from ({\it MAST\/}) at doi: {\doi{10.17909/wna5-2a75}}.}
\tablenotetext{b}{Total exposure time of the observation in ks}
\tablenotetext{c}{Number of individual exposures during the observation}
\tablenotetext{d}{LWRS and MDRS are low and medium resolution \fuse\ slits, respectively}
\end{deluxetable*}

We note that the width of the \lya\ line is more than 1300 \kms\ (FWHM) for all sightlines, so large that any reasonable {\it b}-value has no discernible effect on the computed value of \NHI, and is therefore not included in our fit and does not materially contribute to the error.

\textbf{Stellar model velocity errors.} We bound the velocity uncertainty based on the width of the stellar lines. Typically, we find that shifts of approximately half the width of most stellar lines is the upper bound on the stellar velocity error. This was $\le \pm$ 30 \kms\ for the four OB stars and $\le \pm$ 16 \kms\ for the remaining stars. However, the error in \NHI\ is highly insensitive to the stellar model velocity, and ranges from 0.0002 to 0.003 dex for HD 41161 and BD+39 3226, respectively, and is not the largest source of error for any target in our sample. This is expected since wings of the stellar \HI\ absorption profile are so much narrower than the well developed damping wings of the interstellar \HI\ profile.

\textbf{\lya\ fitting region errors.} We showed in the previous section that the spectral location that most sensitively constrains $N$(\ion{H}{1}) is about $\onethird$ of the way up from zero flux to the full continuum level, but it is broad so it is best to use this spectral location plus some region on each side of it. However, we cannot always use this full range due to the presence of underlying stellar features that are not included in our model. The detailed selection of spectral region differs for each target. As an example, Figure~\ref{wd1034}(c) shows this region for WD1034+001. The red bars indicate the region used to compute $N$(\ion{H}{1}). There is important information even just outside the core region so the inner extent of the red bars begins there. The bars extend to approximately 1213\AA\ and 1218\AA, corresponding to the peak sensitivity shown in Figure~\ref{profile_demo}. In this case we do not extend them further due to some obvious absorption features in the spectrum.

To test the sensitivity to this choice, we performed the complete multi-parameter fit of \NHI\ using a wider \lya\ fitting region, shown by the horizontal cyan bars in Figure~\ref{wd1034}(c), with the best fit model spectrum also shown in cyan. For some targets this larger width included obvious discrepancies between the observed spectrum and the model but we accepted this to avoid underestimating the error associated with our nominal choice of width. A close examination of Figure~\ref{wd1034}(c) shows that the cyan spectrum lies largely below the observed (black) spectrum in the red bar region but the fit is better than the red one further from the core. This is because \NHI\ is forced to increase in order to fit the wings of the \lya\ profile over the wide (cyan) region. The red and cyan spectra have \HI\ column densities of 20.119 and 20.142 dex, a difference of only 0.023 dex. Another example is shown in Figure~\ref{bd39}, a detailed view of the \lya\ region for BD39+3226, one of five targets observed at high resolution. In this case the red and cyan spectra are nearly indistinguishable, and correspond to \NHI\ column densities of 20.011 and 20.017 dex respectively, a difference of only 0.006 dex. These examples demonstrate the exquisite quality of the data and the great sensitivity of \NHI\ to the fit in the region just outside the \lya\ core. The error associated with the selection of the width of the fitting region ranges from 0.00 (indicating that the standard and wide selections give the same value of \NHI) to 0.016 dex for CPD$-$71 172 and LB 1566, respectively. The fitting region width is the largest source of error for PG0038+199, TD1 32709, WD1034+001, and LB 1566.

The discrepancy between the observed spectrum (black) and the best fit spectrum (red) for some targets shown in Figure~\ref{spectra1} is forced by the need to not underestimate the continuum farther from the line core. The better fit in the upper wing region shown by the cyan spectrum comes at the penalty of a poorer fit near the bottom of the \lya\ profile. Examination of Figure~\ref{spectra1} shows that several other targets such as TD1 32709 and LB 1566, exhibit similar behavior to WD1034+001 to some degree. This effect has also been seen in the published profiles for PG0038+199 and WD1034+001 \citep{werner_etal17} and JL9 \citep{werner_etal2022}, so it is not unique to our fitting procedure.

\textbf{Stellar model errors.} In section~\ref{sec:stellar_model} we described the stellar models we used to reproduce the observed spectra. Uncertainties in the stellar models can contribute to errors in the determination of \NHI. To assess the magnitude of this error for most targets we computed two extreme stellar models in which we changed the stellar atmospheric temperature and surface gravity to their maximum or minimum plausible values. We used our standard, 9 parameter fit to compute the best value of \NHI\ for the pair of extreme cases, one leading to a low value and one leading to a high value of \NHI. The average of the difference between these values and the best value of \NHI\ was the error associated with the stellar models. This error ranged from 0.001 to 0.015 dex for JL9 and LB 3241, respectively. The stellar model error is the largest contributor to the error in \NHI\ for LSE 44 and LB 3241.

For the OB stars HD~41161, HD~90087, HD~53975, HD~191877 we did not compute extreme stellar models, because our static models do not take into account the strong P Cygni profiles in the \ion{N}{5} doublet, which are observed on the red side of the \lya\ line profile. For these 4 cases, to be conservative we adopted an error of {\it twice} the mean of the corresponding errors of the remaining 12 targets, or 0.012 dex.

\section{New Measurements of $N$(D~I)} \label{sec:NDI}

\begin{figure*}
\plotone{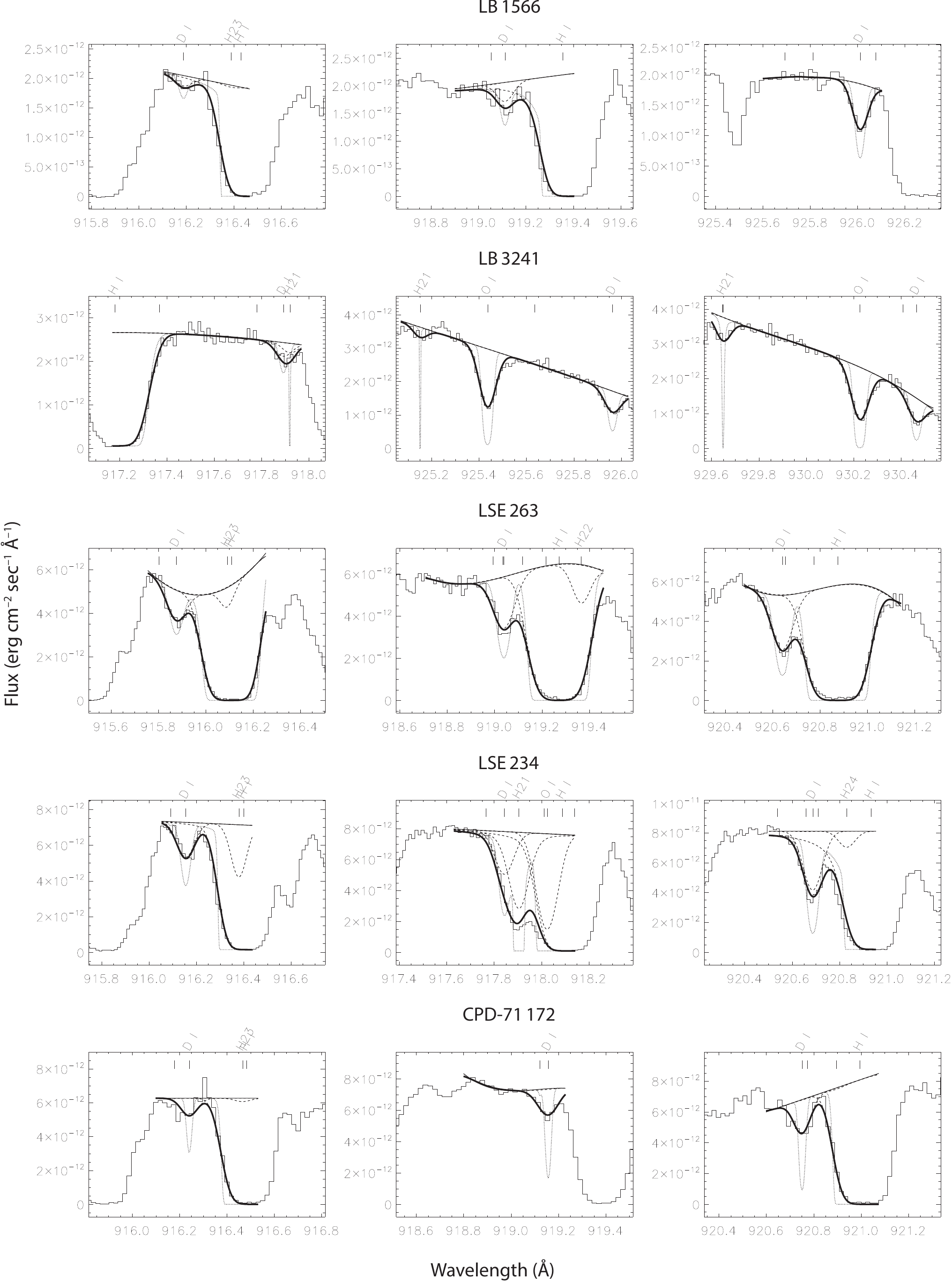}
\caption{Examples of \fuse\ spectral windows showing deuterium lines toward the
five targets in this study without previously published values of \NDI.  Histogram
lines are the data, and the solid lines are
continua and fits broadened by convolution with the \fuse\ LSF.
The dashed  lines are the fits for each species. The dotted lines are the model profiles
prior to convolution with the LSF.  The H$_2$ lines of the levels $J=1$ to $J=4$ are
denoted as H21 to H24. 
\label{fig_fit_fuse}}
\end{figure*}

\subsection{Observations and data processing} \label{sec:NDIobs}

Five of our targets have no published \NDI\ measurements: 
LB\,1566, LB\,3241, LSE\,263, LSE\,234, and CPD$-$71\,172.
All have archival \fuse\ data, with observations 
secured between 2002 and 2006, using the low or medium resolution slits,
LWRS and MDRS, respectively (see Table~\ref{newfuse_obslog}).
They were all obtained in histogram mode, except LB\,3241 which was 
observed in time-tagged mode. We obtained from the \fuse\ archive 
the one-dimensional spectra, which were extracted from the 
two-dimensional detector images and calibrated using the CalFUSE pipeline \citep{dixon07}.  
The data from each channel and segment (SiC1A, SiC2B, etc.) 
were co-added separately for each of the two
slits, after wavelength shift corrections of the individual calibrated
exposures. Wavelength shifts between exposures were typically a few
pixels.  In the case of LB\,1566 which was observed using both
slits, the LWRS and MDRS data were co-added separately. The 
line spread function (LSF) and dispersions are different 
depending on the segments and the slits, thus requiring this separate 
treatment. These different datasets for a given target 
are used simultaneously but separately in the analysis reported below.
The spectral resolution in the final spectra ranges between $\sim13000$ 
and $\sim18500$, depending on detector segment and wavelength. 
Clear  \ion{D}{1} Lyman series absorption lines are detected for all the targets.

\subsection{Data analysis} \label{sec:NDIanalysis}

The deuterium column densities \NDI\ on the five lines of sight were 
measured by Voigt profiles fits of the interstellar spectral absorption lines.  
We used the profile fitting method presented in detail by \citet{hebrard02},
which is based on the procedure Owens.f,
developed by Martin Lemoine and the French \fuse\ Team \citep{lemoine02}.
We split each spectrum into a series of small
sub-spectra centered on absorption lines, and fitted them
simultaneously with Voigt profiles using
\chisq\ minimization. Each fit includes \ion{D}{1} lines, as well 
as those of other species blended with them.
Due to the redundancy of \fuse\ spectral coverage, a given transition 
might be observed in several segments and with one or two slits.
These different observations allow some instrumental
artifacts to be identified and possibly averaged out.
The laboratory wavelengths and oscillator strengths are
from \citet{abgrall93a, abgrall93b} for molecular hydrogen, and from
\citet{morton03} for atoms and ions.

Several parameters are free to vary during the fitting procedure,
including the column densities, the radial velocities of the
interstellar clouds, their temperatures and turbulent velocities, and
the shapes of the stellar continua, which are modeled by low order
polynomials. Owens.f produces solutions that are coherent between all the
fitted lines, assuming for each sightline one absorption component with a single radial velocity,
temperature, and turbulence.  Some instrumental parameters are also
free to vary, including the flux background, the spectral shifts
between the different spectral windows, or the widths of the Gaussian
line spread functions used to convolve with the Voigt profiles.
The simultaneous fit of numerous lines allows statistical and
systematic errors to be reduced, especially those due to continuum
placements, line spread function uncertainties, line blending, flux and 
wavelength calibrations, and atomic data uncertainties. 

In Section~\ref{sec:metals} we present the complexity of the sight lines that we observed at high spectral resolution, particularly toward the most distant targets with numerous components present at different radial velocities. 
However, the velocity structure along these five lines of sight is not known and therefore we
assumed a single interstellar component for each line. As discussed and tested in \citet{hebrard02}, our measured column densities and their associated uncertainties are reliable with respect to this assumption, and they are also reliable considering typical  temperature and turbulence of interstellar clouds, as well as the shape and width of the line spread function (LSF) of the observing instrument. Thus we report total deuterium column densities, integrated along each line of sight. The error bars were obtained using the $\Delta$\chisq\ method presented by \citet{hebrard02}. The measured \ion{D}{1} column densities are given in Table~\ref{tab:d2h}. Examples of the fits are shown in Figure~\ref{fig_fit_fuse}.}

Our measurements of $N$(\ion{D}{1}) were derived from unsaturated lines. Saturated lines on the flat part of the curve of
growth were excluded from consideration. We thus only kept the \ion{D}{1} lines for which the model profiles prior to convolution with the LSF do not reach the zero flux level (see Figure~\ref{fig_fit_fuse}).  Indeed, saturated lines can introduce systematic errors on column density measurement \citep{hebrard02, hebrard03}. Issues related to saturation and other systematic effects were discussed extensively by \citet{hebrard02} and the method used here is exactly the same. In particular, Section 4.2 of that study discusses and tests the reliability of reported column densities and their uncertainties with respect to the number of interstellar clouds on the line of sight, their temperature and turbulence, and the shape and width of the LSF. The tests reported by \citet{hebrard02} show that the uncertainties on column densities are reliable when they are derived from the fit of unsaturated lines.

To exclude the saturated \ion{D}{1} lines from our fits, we checked that their profiles prior to convolution with the LSF (shown as dotted lines in Figure~\ref{fig_fit_fuse}) do not reach the zero flux level. The unconvolved profiles are constrained as numerous lines of several species are fitted simultaneously. For example, in the fit around 918\AA\ of LSE 234 (fourth line, middle panel in Figure~\ref{fig_fit_fuse}), the unconvolved profile appears to reach zero flux level but this is actually due to blending with a $J = 1$ transition of molecular hydrogen. The \ion{D}{1} transition is located $\sim$0.05\AA\ to the blue side of this line. It does not reach the zero flux level and is not saturated, thus providing a reliable column density.

\section{\NHI\ and D/H Results} \label{sec:results}

\begin{figure*}
\epsscale{1.0}
\plotone{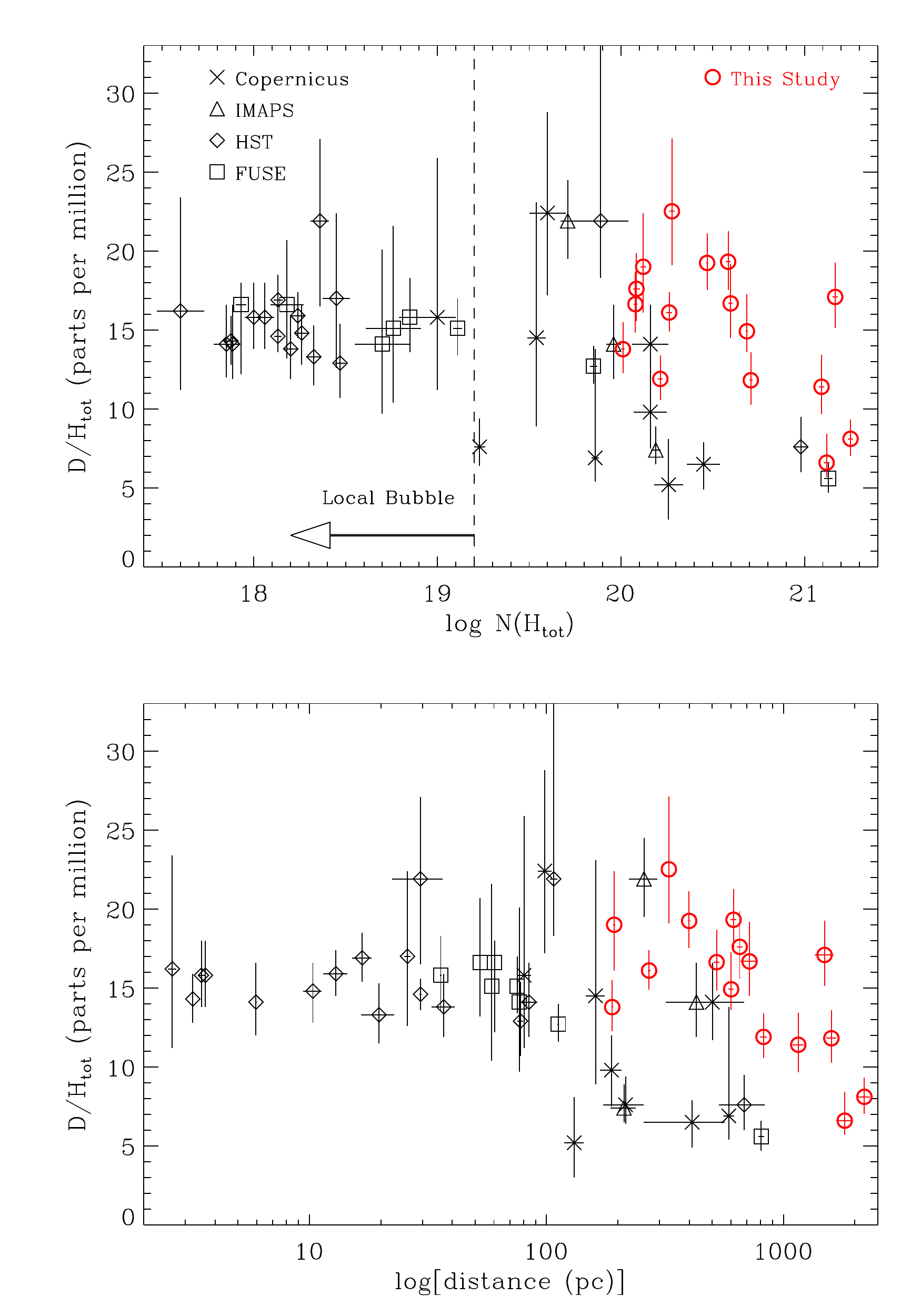}
\caption{Top: \dhtot\ vs.~log(\nhtot), where \nhtot = \NHI\ + 2\Nhtwo\ is the total neutral and molecular
hydrogen column density. \NHI\ is used if \nhtot\ is not available. The
red symbols are from this study. The symbols for the other data points designate the spacecraft
that observed the line of sight. The boundary of the Local Bubble, taken to be at log(\nhtot) = 19.2,
is shown as the vertical dashed line. Bottom: \dhtot\ vs. distance. Distances are from {\it Gaia} and
{\it Hipparcos}.
\label{dh_vs_h}}
\end{figure*}

The principal results of this study are shown in Figure~\ref{dh_vs_h} and Table~\ref{tab:d2h} where our new
measurements of \NHI\ are given in column 2. Column 3 lists
the column densities of \NHtwo. For the 7 targets with the TP code this was calculated using the method
described in Section 3.2 of \citet{jenkins19} who used the optical depth profiles created by
\citet{mccandliss03}. Column 4 shows
our five new results of \NDI\ presented in section~\ref{sec:NDI} and for the remaining targets, previously published
values. All measurements of \NDI\ come from \fuse\ spectra. Column 5 gives our resulting values
of D/H$_{tot}$, where \nhtot\ = \NHI+2\NHtwo. 2\Nhtwo\ is a relatively minor constituent
of \nhtot, ranging from 15.9\%\ down to 7.0\% for HD 191877, PG 0038+199, HD 41161, HD 90087,
JL 9, and less than 3.5\% for the remaining 11 stars in our sample. We can ignore
the presence  of HD in our assessment of the deuterium abundance, since $N({\rm HD})$ is 
generally of order $3\times 10^{-7}N({\rm H_{tot}})$ (Snow et al. 2008).

\begin{deluxetable*}{llllll}
\tablecaption{The column densities of neutral hydrogen, molecular hydrogen, neutral deuterium, and D/H$_{tot}^{\rm a}$\label{tab:d2h}}
\tablewidth{0pt}
\tablehead{
\colhead{Target} & \colhead{log(\NHI)} & \colhead{log(\NHtwo)} & \colhead{log(\NDI)} & \colhead{D/H$_{tot}$} & \colhead{References$^{\rm b}$} \\
}
\startdata
BD+39 3226	&	$	20.01	\pm	0.01	$  &  $	15.65	\err{	0.06	}{	0.07	}  $  &  $	15.15	\pm	0.05			$  &  $	13.78	\err{	1.71	}{	1.53	}  $  &	O++06, O++06	\\
TD1 32709	&	$	20.08	\pm	0.01	$  &  $	14.48	\err{	0.12	}{	0.11	}  $  &  $	15.30	\pm	0.05			$  &  $	16.63	\err{	2.08	}{	1.86	}  $  &	O++06, O++06	\\
LB 3241      	&	$	20.08	\pm	0.02	$  &  $	14.50	\err{	0.30	}{	0.50	}  $  &  $	15.33	\pm	0.05			$  &  $	17.60	\err{	2.27	}{	2.04	}  $  &	TP, TP	\\
WD 1034+001	&	$	20.12	\pm	0.02	$  &  $	15.72	\err{	0.13	}{	0.12	}  $  &  $	15.40	\pm	0.07			$  &  $	19.00	\err{	3.39	}{	2.91	}  $  &	O++06, O++06	\\
LB 1566      	&	$	20.21	\pm	0.01	$  &  $	15.48	\err{	0.18	}{	0.18	}  $  &  $	15.29	\pm	0.05			$  &  $	11.90	\err{	1.49	}{	1.33	}  $  &	TP, TP	\\
Feige 110     	&	$	20.26	\pm	0.02	$  &  $	15.20	\err{	0.30	}{	0.40	}  $  &  $	15.47	\pm	0.03			$  &  $	16.10	\err{	1.28	}{	1.20	}  $  &	TP, F++02	\\
CPD$-$71 172	&	$	20.28	\pm	0.01	$  &  $	15.60	\err{	1.10	}{	0.35	}  $  &  $	15.63	\err{	0.08	}{	0.07	}  $  &  $	22.51	\err{	4.61	}{	3.43	}  $  &	TP, TP	\\
PG 0038+199	&	$	20.40	\pm	0.01	$  &  $	19.33	\err{	0.02	}{	0.02	}  $  &  $	15.75	\pm	0.04			$  &  $	19.24	\err{	1.89	}{	1.73	}  $  &	W++05, W++05	\\
LSE 44         	&	$	20.57	\pm	0.01	$  &  $	18.82	\err{	0.10	}{	0.10	}  $  &  $	15.87	\pm	0.04			$  &  $	19.31	\err{	1.94	}{	1.78	}  $  &	TP, F++06	\\
LSE 263         	&	$	20.60	\pm	0.01	$  &  $	16.40	\err{	0.40	}{	0.50	}  $  &  $	15.82	\pm	0.06			$  &  $	16.68	\err{	2.51	}{	2.20	}  $  &	TP, TP	\\
JL 9                	&	$	20.68	\pm	0.01	$  &  $	19.25	\err{	0.02	}{	0.02	}  $  &  $	15.78	\pm	0.06			$  &  $	11.81	\err{	1.77	}{	1.54	}  $  &	W++04, W++04	\\
LSE 234        	&	$	20.69	\pm	0.01	$  &  $	16.35	\err{	0.25	}{	0.35	}  $  &  $	15.86	\err{	0.06	}{	0.04	}  $  &  $	14.91	\err{	2.35	}{	1.33	}  $  &	TP, TP	\\
HD 191877     	&	$	21.05	\pm	0.02	$  &  $	20.02	\err{	0.05	}{	0.05	}  $  &  $	15.94	\err{	0.11	}{	0.06	}  $  &  $	6.60  	\err{	1.82	}{	0.88	}  $  &	SDA21, H++03 \\
HD 53975     	&	$	21.08	\pm	0.02	$  &  $	19.18	\err{	0.04	}{	0.04	}  $  &  $	16.15	\err{	0.07	}{	0.07	}  $  &  $	11.40	\err{	2.04	}{	1.74	}  $  &	OH06, OH06	\\
HD 41161      	&	$	21.10	\pm	0.02	$  &  $	20.02	\pm{	0.03	}                   $  &  $	16.40	\err{	0.05	}{	0.05	}  $  &  $	17.30	\err{	2.24	}{	2.01	}  $  &	SDA21, OH06	\\
HD 90087       	&	$	21.21	\pm	0.02	$  &  $	19.91	\pm{	0.03	}                   $  &  $	16.16	\err{	0.06	}{	0.06	}  $  &  $	8.09		\err{	1.23	}{	1.08	}  $  &	SDA21, H++05	\\
\enddata
\tablenotetext{a}{All values of \NHI\ were determined in this study. D/H$_{tot}$ is given in parts per million.}
\tablenotetext{b}{The first source listed is for the determination of \NHtwo\ and the second for \NDI.  The keys to references are explained in Table~\ref{refs} of Appendix B. 
The code TP means the value was determined in this paper. See text for explanation of \Nhtwo\ with the TP code.}
\end{deluxetable*}

Our values of \dhtot\ as a function of \nhtot\ are plotted in Figure~\ref{dh_vs_h} together with
previously reported measurements
made with data from  \textit{Copernicus},  \textit{IMAPS},  \textit{HST}, and  \textit{FUSE}. 
This figure may be compared directly to Figure 1 in L06. Note that the figures
differ slightly in that we plot D/H$_{tot}$ vs. H$_{tot}$ while they plot D/\HI\ vs. \HI.
We also plot \dhtot\ vs distance. Distances are taken from {\it Gaia} DR3
\citep{soszynski16, vallenari22} when available (38 stars) and {\it Hipparcos} \citep{perryman97} when not (15 stars).

The primary question we sought to answer in this study is whether the previous determinations of \dhtot\ seen in
targets beyond the Local Bubble are due to errors in the measured values of \NHI. It is now clear that
this is not the case. The 16 new values of \dhtot\ are not consistent with a single value of the
deuterium abundance. In fact, the best straight line fit through these points without constraining
the slope yields \chisq\ = 57 and the probability that a linear fit would give this value or greater is
$4 \times 10^{-7}.$  However, the scatter of the points is now substantially reduced compared to previous
determinations. The standard deviation of D/H$_{tot}$ for the 16 targets in our study is 4.3 ppm, compared
to 6.0 ppm in L06 for the 9 targets that are common to both studies, despite
the fact that the means of the distributions are almost unchanged at 15.2 and 15.7 ppm, respectively.
Including \NHI\ values from \citet{diplas94} we find the interesting result that 
11 of the 13 points with both old and new published values of \dhtot\ moved closer to the mean.
That is to say, the high points moved lower and the low points moved higher. The typical change is
$1-2\sigma$ which for an individual point would not be noteworthy but perhaps is for such a large
majority of points. The exceptions are HD191877 and HD90087, which moved 0.9 $\sigma$ and 1.0 $\sigma$
away from the mean, respectively. We have identified no systematic effect which may be responsible
for this general trend toward the mean.

With the recent {\it Gaia} DR3 data release most of the distances to our targets are now known to high
accuracy and in the bottom of Figure~\ref{dh_vs_h} we plot \dhtot\ vs. distance. Since we found greater scatter in
\dhtot\ at large values of \nhtot, we expected to see a similar scatter at large distances. The plot shows
exactly this result with no particular trend with distance other than approximate constant \dhtot\ within
$\sim100$ pc, as was previously known. This is consistent with estimates of the distance to the wall of the Local Bubble ranging
from $65-250$ pc, depending on direction \citep{sfeir99}.

\section{Measurement of Metal Abundances} \label{sec:metals}

Our observations permit a detailed analysis of metal abundances toward the five
targets for which we obtained high resolution echelle data. We discuss the analysis
and results in this section.

The metal lines for the stars observed at high resolution,
HD~191877, HD~41161, HD~53975, HD~90087 and BD+39~3226,
were fitted with {\sc vpfit}~9.5~\citep{carswell14}.  
Table~\ref{tab-fvalues} in Appendix A shows the sources of the {\it f}-values we used.
In addition to the basic wavelength and flux vectors from the datasets listed in Table~\ref{stis_obslog},
we used the associated 1$\sigma$ error arrays, and created continua using
a semi-manual method.
We first employed a 15 pixel median filter for a rough continuum estimate, then
fitted around the lines using either a series of linear interpolations, 
selected to connect the median-filtered curves over absorption lines, or
the {\sc IRAF} {\it continuum} package \citep{tody86, tody93} around more complex regions.

The $1 \sigma$ error arrays were then verified for consistency with the
root mean square (rms) variations for the normalized (flux/continuum) vectors.
The check was done for all bin values from 5 to 1000 pixels. 
Deviations from rms values were typically on the order of 
10-20\%.  We made a correction
in {\sc vpfit} parameter files to employ this correction, though in
certain wavelength intervals, particularly in line troughs,
larger correction factors sometimes had to be employed.  This reflected
a combination of systematic errors which may not have been completely
accounted for in the HST pipeline reductions, and also under-sampling of
the LSF when using grating 
E140H with the Jenkins slit ($0.1''\times 0.03''$).  (We did not
request any special detector half-pixel sampling with the observations, which would be
necessary to exploit the full resolving power of this slit.)

For the profile fits with {\sc vpfit}, we employed the library STScI LSF for the given grating and slit combination.
We generally required a probability of the fitted profiles being consistent with the data of at least $p=0.01$,
as a goodness of fit threshold.  In some cases, we tied the radial velocities of
several ions together, to make multiple simultaneous fits.  Also in some cases,
we allowed a linear offset of the continuum level as a free parameter, which
effectively compensates for unidentified line blends with other ions, though
this was in a small minority of cases.  Finally, due either to the under-sampled
LSF, noise spikes, or potential artifacts in the reduced spectra, we occasionally increased
the error values by factors up to 2-3 to reduce the effects of individual pixels
on the fits and obtain acceptable statistical fits.

Detailed notes on individual objects are presented in Appendix A along with column
density and other observational data for each sight line.

\section{Correlation with Depletions of Heavy Elements} \label{sec:depletions}

There have been numerous studies that compared D/H measurements with the 
relative abundances of other elements (Prochaska et al. 2005; Linsky et al. 2006; 
Oliveira et al. 2006; Ellison et al. 2007; Lallement et al. 2008), in order to investigate 
the hypothesis that D more easily binds to dust grains than H, as suggested by Jura 
(1982), Draine (2004, 2006), and Chaabouni et al. (2012).  All have shown that there 
are correlations between D/H and gas-phase abundances of certain elements that 
exhibit measurable depletions onto dust grains in the interstellar 
medium. While these correlations are statistically significant and
reinforce the picture that the more dust-rich regions have lower deuterium 
abundances, the scatters about the trend lines are larger than what one could expect 
from observational errors.

In this section, we investigate this issue once again, including not only our own data 
but also results reported elsewhere.  However, our analysis here incorporates two 
important differences in approach from the earlier studies.  First, we characterize 
the depletions of heavy elements in terms of a generalized depletion parameter 
$F_*$ developed by \citep{jenkins09,jenkins13}.  The use of $F_*$ instead of the depletion of a 
specific element allows us to include in a single correlation analysis the data for 
cases where the depletion of {\it any\/} element is available instead of just one 
specific element.  Moreover, if for any given case more than one element has had its 
column density measured, the results will effectively be averaged, yielding a more 
accurate evaluation of the strength of depletion by dust formation.  A second 
important aspect of our study is that we limit the results to cases where $\log 
N({\rm H_{tot}})=\log [N({\rm H~I})+2N({\rm H}_2)]>19.5$, following a criterion 
defined by Jenkins (2004, 2009), so that we can reduce the chance that the 
abundance measurements are distorted by ionizations caused by energetic starlight 
photons that can penetrate part or much of the H~I region(s) (Howk \& Sembach 
1999; Izotov et al. 2001).

Much of the information about heavy element column densities is taken from the 
compilation of Jenkins (2009), with its specific standards for quality control and 
adjustments for revised transition $f$-values.  We have added a few new 
determinations that came out later in the literature.  An evaluation of $F_*$ for any 
individual element $X$ is given by the relation,
\begin{equation}\label{F*}
F_*(X)={\log N(X)-\log N({\rm H_{tot}})- (X/{\rm H})_\odot-B_X\over A_X}+z_X~,
\end{equation}
where the constants $(X/{\rm H})_\odot$, $A_X$, $B_X$, and $z_X$ are specified for 
each element in Table~4 of Jenkins (2009).  We can arrive at an error in $F_*(X)$ by 
using Geary's (1930) prescription\footnote{A simplified description of Geary's 
(1930) scheme is described in Appendix~A of Jenkins (2009).} for the error of a 
quotient for an expression $N/D$ (numerator over denominator),
\begin{equation}\label{sigmaQ}
\sigma(Q)=\sigma \left[ {N\pm \sigma(N)\over D\pm \sigma(D)}\right]
\end{equation}
with
\begin{equation}
\sigma(N)=\{ \sigma[\log N(X)]^2 + \sigma[\log N({\rm H_{tot}})]^2 + 
\sigma[B_{\rm red}]^2 \}^{1/2}
\end{equation}
and
\begin{equation}
\sigma(D)=\sigma(A_X)~.
\end{equation}

The error in the $B_X$ term in Eq.~\ref{F*} is a reduced form
\begin{equation}
\sigma[B_{\rm red}]=\{ \sigma(B_X)^2 - \sigma[(X/{\rm H})_\odot]^2 \}^{1/2}
\end{equation}
because any uncertainty in the solar abundance $(X/{\rm H})_\odot$ has no effect 
on the outcome for $F_*$; $B_X$ would change by an equal amount in the opposite 
direction.  Put differently, $\sigma[B_{\rm red}]$ represents just the uncertainty in 
the original fit without the systematic error from $(X/{\rm H})_\odot$. There is no 
error in $z_x$; this constant is used to insure that the error in $A_X$ is uncorrelated 
with that of $B_X$.  Ultimately, we use $\sigma(Q)$ as the value for the uncertainty 
in $F_*(X)$.

Table~\ref{basic_data} lists the data that were assembled for constructing the 
correlation, and Table~\ref{refs} indicates the sources in the literature that led to 
the values shown by the codes listed in Table~\ref{basic_data}.  For each sight line, 
a weighted average for $F_*$ over all elements $X$ was determined\footnote{Since we
made our computation of $F_*$ several new  {\it f}-values have been published and
are listed in Table~\ref{tab-fvalues}. This will result in modifications to
log N, as reflected in the remaining tables of the Appendix, but the $F_*$ numbers will not change.
This is because adjustments in the $B_X$ parameters
were implemented to reflect the changes prior to deriving the $F_*$ values.}  from
\begin{equation}
\langle F_*\rangle=
\sum_X  F_*(X)\sigma[F_*(X)]^{-2} \Bigg/ \sum_X \sigma[(F_*(X)]^{-2 } ~,
\end{equation}
where the error in this quantity is given by
\begin{equation}
\sigma[\langle F_*\rangle]=\left[ \sum_X \sigma[(F_*(X)]^{-2} \right]^{-1/2}~.
\end{equation}

For the elements C, N and Kr, $\sigma(A_X) > A_X/3$, which makes  $\sigma(Q)$ in 
Eq.~\ref{sigmaQ} untrustworthy (and the errors large).  Results for these three 
elements were ignored and not included in Table~\ref{basic_data}.  
Figure~9 shows $\log[N({\rm D~I})/N({\rm H_{tot}})]$ 
as a function of $\langle F_*\rangle$.  As noted above, we can
ignore the presence of HD in our assessment of the deuterium abundance.  $\langle 
F_*\rangle$ for LSE~44 was determined from only the abundance of oxygen; the 
error here is so large that this case was not included in the analysis or the plot. 

\begin{figure*}
\plotone{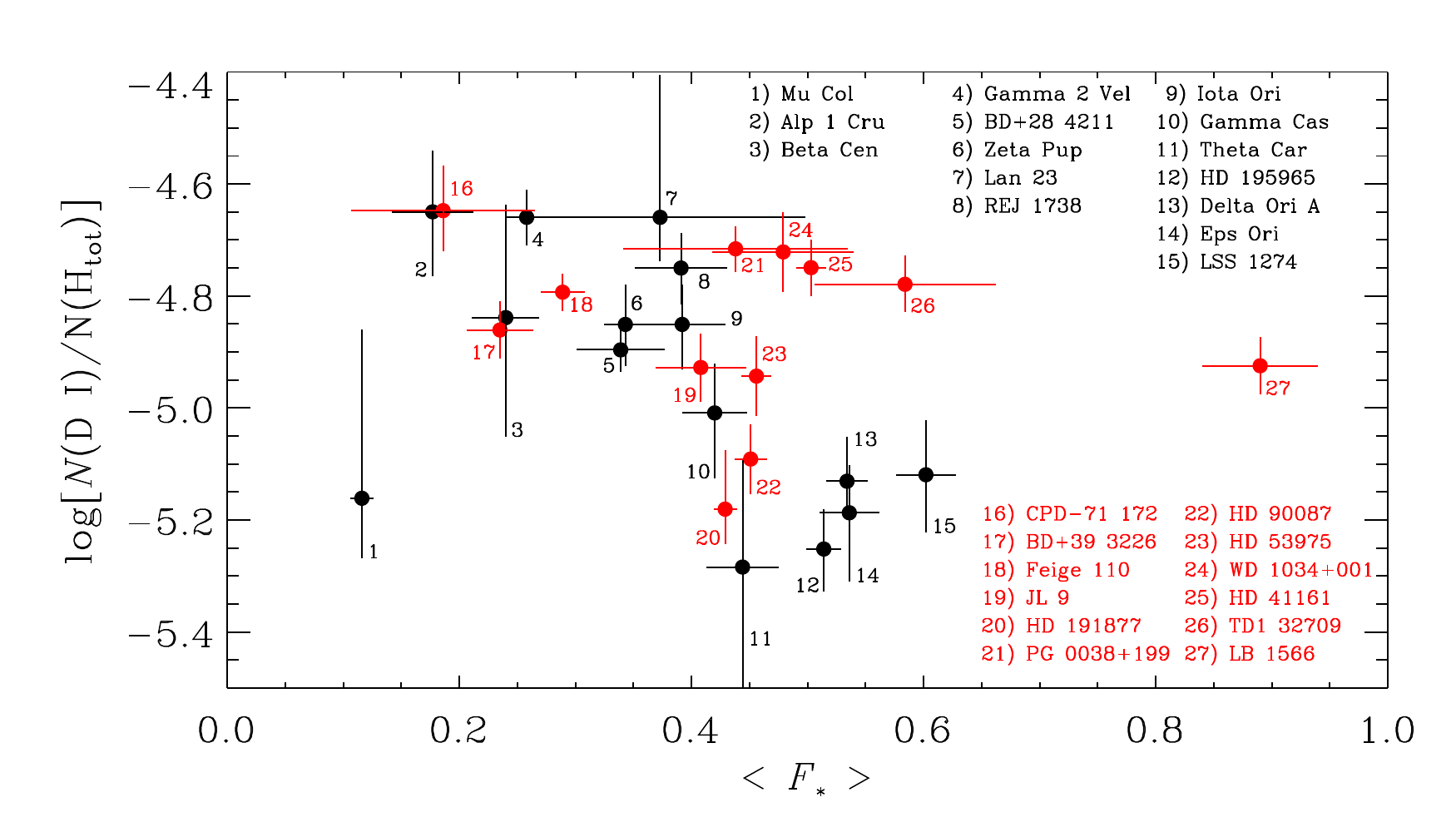}
\caption{The relationship between $\log [N({\rm D~I})/N({\rm H_{tot}})]$ and the generalized
depletion parameter $\langle F_*\rangle$ for our determinations (in red) and others from data in the literature (in black), as
listed in Table~\protect\ref{basic_data}.
\label{fig_depl}}
\end{figure*}

To assess whether or not the D/H and metal depletion measurements in Figure~\ref{fig_depl} are anticorrelated (note that as depletion
becomes more severe it becomes more negative), we begin with nonparametric Spearman
and Kendall $\tau$ correlation tests.  Using all of the data in Figure~\ref{fig_depl}, we find a Spearman correlation coefficient $r_{s} = -0.42$ with a
$p-$value of 0.028, and we obtain a Kendall $\tau = -0.32$ with $p-$value = 0.021.  Both of these tests indicate that the data are weakly correlated at 
slightly better than 2$\sigma$ significance. This is similar to results obtained in previous studies, although we note that L06 did not find a significant 
correlation in their sample with log $N$(H~I) $>$ 19.2 (i.e., their sample that most closely matches the criteria we have used to select our sample).  We 
have reduced the uncertainties of some of the measurements in L06 and we have added new sightlines; evidently these improvements have revealed a 
weak correlation even in this higher-$N$(H~I) sample.

This correlation may be slightly misleading, since both variables in Figure~\ref{fig_depl} have experimental errors that are partly composed of errors of a 
single quantity, $\log N({\rm H_{tot}})$. In our comparison shown in Figure~\ref{fig_depl}, the $y$ values are driven in a negative direction by positive errors 
in $\log N({\rm H_{tot}})$, while the reverse is true for the $x$ values (see Eq.~\ref{F*}), since for most elements $A_X\approx -1$. Hence, measurement 
errors in $\log N({\rm H_{tot}})$ will artificially enhance the magnitude of the negative correlation over its true value in the absence of such errors. To 
investigate this concern, we have carried out two additional tests that are less vulnerable to this problem.  As we discuss in the next two paragraphs, these 
two additional tests further support the finding that D/H is weakly correlated with metal depletion.

First, we have examined whether $N$(D~I)/$N$(Fe~II) is correlated with $\log N({\rm H_{tot}})$.  Iron depletion is typically strongly correlated with $\log 
N({\rm H_{tot}})$ because sightlines with higher $\log N({\rm H_{tot}})$ tend to have higher gas densities, higher molecular-hydrogen fractions, and physical 
conditions that are more conducive to elemental depletion by dust.  Using the data in Figure~\ref{fig_depl}, the Spearman test comparing iron depletion vs. 
$\log N({\rm H_{tot}})$ yields $r_{s} = -0.67$ with $p-$value = 0.0001, which confirms that Fe depletion is correlated with the hydrogen column in these 
data.  Therefore if the deuterium abundance is \textit{not} correlated with iron depletion, then $N$(D~I)/$N$(Fe~II) vs. $\log N({\rm H_{tot}})$ should be 
correlated with a positive slope --- as $\log N({\rm H_{tot}})$ increases and the relative iron abundance decreases due to depletion, $N$(D~I)/$N$(Fe~II) 
should go up.  This is not what we observe.  Instead, we find no correlation between $N$(D~I)/$N$(Fe~II) and $\log N({\rm H_{tot}})$ (Spearman $r_{s} = 
0.11$ with $p-$value = 0.58), which suggests that as the relative abundance of Fe decreases, the deuterium abundance decreases accordingly so that $N$
(D~I)/$N$(Fe~II) stays more or less the same.  A linear fit to $N$(D~I)/$N$(Fe~II) vs. $\log N({\rm H_{tot}})$ has a slope consistent with zero within the 
errors ($m = 2.1\pm 3.6$).  We note that there is substantial scatter in $N$(D~I)/$N$(Fe~II) vs. $\log N({\rm H_{tot}})$, just as there is substantial scatter in 
Figure~\ref{fig_depl}.

\begin{figure}
\epsscale{1.25}
\plotone{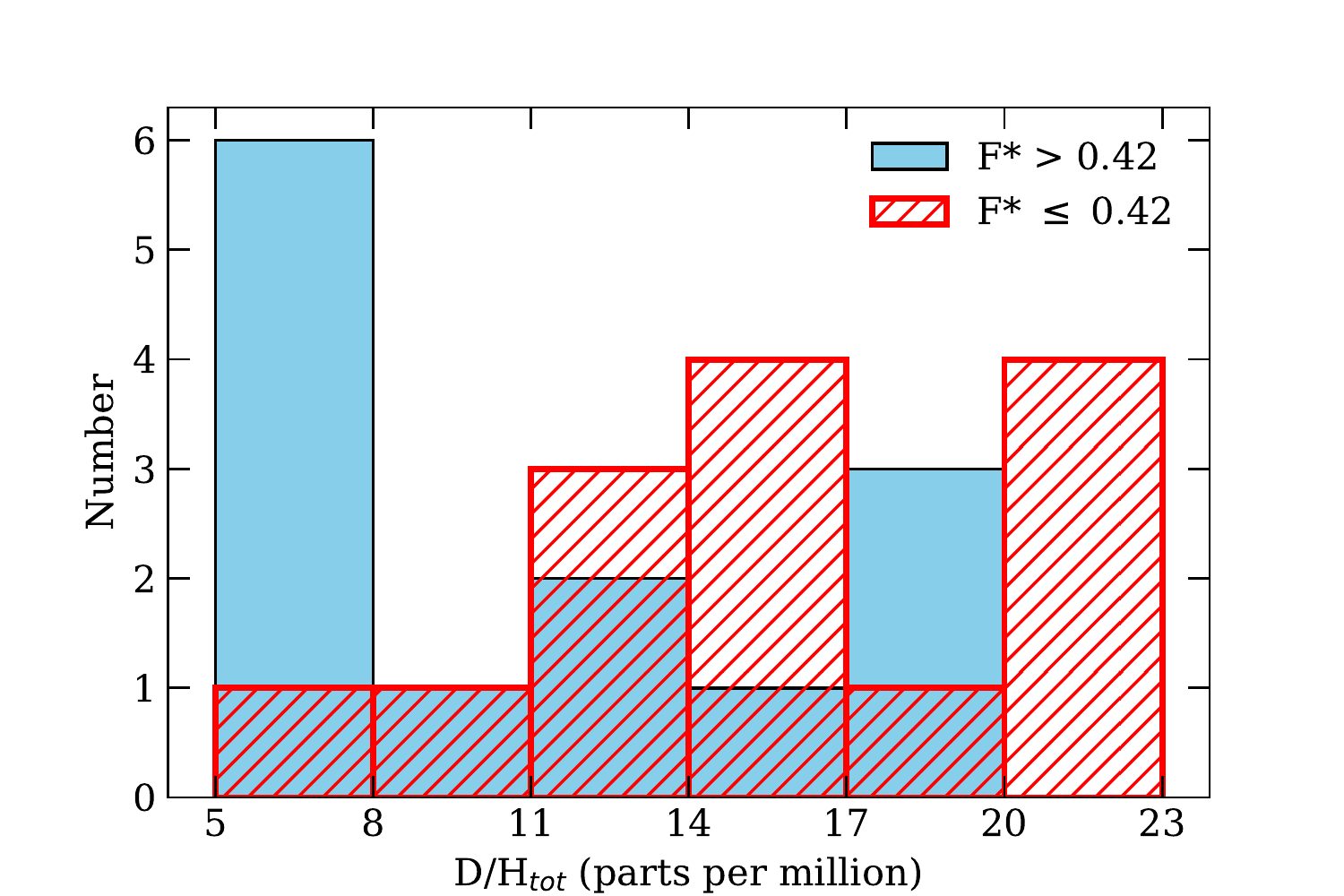}
\caption{Comparison of \dhtot\ distributions from the sightlines studied here, split into two samples: the sightlines with low metal depletion ($\langle F_* \rangle \leq 0.42$, red-hatched histogram) and the sightlines with higher metal depletion ($\langle F_* \rangle > 0.42$, solid-blue histogram).
\label{fig_highdp_vs_lowdp}}
\end{figure}

Second, we have split the data in Figure~\ref{fig_depl} into two equal-sized bins, one with lower amounts of metal depletion and one with higher depletions, 
and we have compared the D/H distributions in each bin.  Figure~\ref{fig_highdp_vs_lowdp} overplots the resulting D/H distributions for the data with $
\langle F_*\rangle \leq 0.42$ (lower metal depletion) vs. the data with  $\langle F_* \rangle > 0.42$ (greater metal depletion). 
Applying a Kolmogorov-Smirnov (KS) test to the 
two samples shown in Figure~\ref{fig_highdp_vs_lowdp}, we find the KS statistic $D = $ 0.48 with $p-$value = 0.062. This only tentatively rejects the null 
hypothesis (that the distributions are drawn from the same parent distribution) at slightly less than 2$\sigma$ confidence. However, the only criterion used 
to choose $\langle F_* \rangle$ = 0.42 to delineate the ``low-depletion'' and ``high-depletion'' samples is that it divides the data into two (almost) equal 
halves, and this results in 14 data points in the low-depletion bin and 13 points in the high-depletion bin. One of the measurements is right on the $\langle 
F_* \rangle$ = 0.42 boundary and has a low D/H ratio; if we change the definition slightly by placing all points with $\langle F_* \rangle \geq 0.42$ in 
the high-depletion group (resulting in 13 points in the low-depletion bin and 14 in the high-depletion bin), the KS test changes to $D$ = 0.57 with $p-$value 
= 0.015.  Clearly more D/H and depletion measurements would be helpful. The Anderson-Darling (AD) two-sample test, which can be applied in the same way 
as the KS test but may be more effective in some situations \citep{engmann11}, returns $p-$value = 0.023 and 0.012 in comparisons of the samples with 
number of low/high depletion points = 14/13 and 13/14, respectively.  The AD test therefore indicates that the low-depletion and high-
depletion samples are different at a slightly better significance, but nevertheless all of these tests provide weak indications that the distributions of D/H 
ratios are different when the metal-depletion level is low or high.

We have conservatively required our D/H sample to have log $N$(H~I) $\geq$ 19.5 to avoid systematic confusion from ionization effects.  This is a 
reasonable threshold for distant sightlines that may probe regions with high starlight intensities and high ionization parameters, which can elevate the 
contribution of ionized gas along a sightline.  However, inside the Local Bubble, the ionizing radiation field and ISM gas physics have been studied in detail 
\citep[e.g.,][]{redfield08,frisch11}, and while there are uncertainties, inside the Local Bubble the ionization parameter is likely quite low \citep{slavin02}, and 
Local Bubble sightlines with log $N$(H~I) $\geq$ 18.0 will have iron ionization corrections less than 0.15 dex \citep[see Fig.6 in][]{lehner03}.  Therefore we 
can add Local Bubble sightlines from L06 with log $N$(H~I) $\geq$ 18.0 without introducing appreciable error from ionization corrections.  If we combine 
our data with the 13 Local Bubble sightlines from L06 with log $N$(H~I) $\geq$ 18.0 that have Fe depletion measurements, we find a similar result with 
somewhat better significance: a Spearman test for D/H vs. [Fe/H] including the Local Bubble gives $r_{s}$ = 0.39 with $p-$value = 0.014.  Of course, this 
gives the Local Bubble, where D/H is fairly uniform and the metal depletion is relatively low, considerable weight, but it is interesting that even with this 
significant increase in the overall sample size, the resulting correlation is not very strong.

In Figure~\ref{fig_depl} there appears to be a bifurcation of D/${\rm H_{tot}}$ values for $\langle 
F_*\rangle > 0.4$.  L06 noted a similar separation, but with a slightly different selection of 
target stars and using the depletion of Fe instead of $\langle F_*\rangle$ as a discriminant.  If 
this effect is real and not a product of random processes, can we devise a possible explanation?  
L06 proposed that differences in grain properties could explain this phenomenon.  We think 
that an alternate interpretation is possible.  As we mentioned in Section~\ref{sec:intro},
there is evidence that low-metallicity gas in the Galactic halo has a higher than usual deuterium 
fraction.  When this gas mixes with material at the upper or lower boundaries of the Galactic 
plane gas, it might modify the D/H to higher values without appreciably changing the apparent 
values of $F_*$.  We could test this proposition by examining whether or not the high and low 
branches of the D/${\rm H_{tot}}$ trends shown in Figure~\ref{fig_depl} have significant differences 
in the distances $|z|$ of the target stars from the Galactic plane.  Table~\ref{zvalues} shows 
the $|z|$ values for stars in the two groups.

While the high group has an average $|z|$ equal to 343~pc and the low group has an average
$|z|$ equal to 160 pc, it is not clear that these differences are significant. In order to test the 
proposition that these outcomes represent separate populations in $|z|$, we performed a
KS test, and it revealed that there was a 5\% probability that the two
populations were drawn from a single parent distribution. We also performed an AD test
which gave a p-value for the null hypothesis of 0.026, corresponding to a $2\sigma-3\sigma$
significance. Thus, at only a modest significance level, we suggest that $|z|$ is a possible
discriminant for the two branches in \dhtot\ for sight lines that exhibit moderate to high depletions
of heavy elements. We propose that less dust in the infalling gas means that the freeze out of
deuterium would be reduced, which may add to the effect of this gas having had less destruction
of deuterium by astration. 

\begin{deluxetable}{
l	
c	
c	
l	
c	
}
\tablewidth{0pt}
\tablecaption{Distances from the Galactic Plane for Stars with $\langle F_*\rangle > 
0.4$\label{zvalues}}
\tablecolumns{5}
\tablehead{
\multicolumn{2}{c}{Low D/${\rm H_{tot}}$} & \colhead{~~~} & \multicolumn{2}{c}{High 
D/${\rm H_{tot}}$}  \\
\cline{1-2} \cline{4-5}
\colhead{Star} & \colhead{$|z|$} & \colhead{} & \colhead{Star} & \colhead{$|z|$}\\
\colhead{} & \colhead{(pc)} & \colhead{} & \colhead{} & \colhead{(pc)} }
\startdata
HD~191877&203&&PG0038~+199&271\\
HD~90087&80&&WD1034~+001&142\\
HD~53975&46&&HD~41161&332\\
JL9&722&&TD1~32709&245\\
HD~36486&64&&LB~1566&726\\
HD~37128&179&&&\\
HD~93030&12&&&\\
HD~195965&72&&&\\
LSS~1274&63&&&\\
\enddata
\end{deluxetable}

\section{Discussion} \label{sec:discussion}

While it has long been known that the measured values of D/H along many sightlines
within the Local Bubble are consistent with a single value \citep{linsky98, moos02, hebrard03},
beyond this structure the measurements show variability. The primary
goal of this study is to determine whether this variability was the result of
errors in the relatively poorly measured values of \HI\ column density. With this study we have
more firmly established that the variability is real but slightly smaller than previously estimated.

L06 suggested there are three separate regimes of D/H values each defined by
a range of \HI\ column densities (see their Figure 1). The first spans log \NH\ $< 19.2$, which is approximately
the range within the Local Bubble and was chosen because D/H is constant within this limit.
Here they find (D/H)$_{LB} = 15.6 \pm 0.4$ ppm for 23 sight lines where the uncertainty is the standard
deviation in the mean. The highest column density range corresponds to log N(\HI) $> 20.7$ where again they note
that D/H is approximately constant
and, notably, lower than (D/H)$_{LB}.$ In this regime they found D/H$_{dist} = 8.6 \pm 0.8$ ppm
(standard deviation in the mean) for 5 sight lines toward the most distant targets (HD 90087, HD 191877,
LSS 1274, HD 195965, and JL 9). The standard deviation of the D/H values is 0.95 ppm. In the intermediate regime,
19.2 $\leq$ \NH\ $\leq$ 20.7, D/H is highly variable spanning a range from $5.0\err{2.9}{1.4}$ for $\theta$ Car
to $22.4\err{11.7}{6.2}$ for LSE 44 or, selecting a target with much smaller errors, $21.8\pm2.1$ for $\gamma^2$ Vel.

Our study did not include any targets in the first regime. However, our new results do not support the idea the D/H is
constant in the most distant regime. We have computed revised values of \NHI\ for HD 90087, HD 191877, and JL 9; see
Table~\ref{tab:d2h}. Combining our new values of D/H with those in L06 for LSS 1274 and HD 195965
gives D/H$_{dist} = 8.3 \pm 0.7$ (standard deviation in the mean) with a standard deviation of D/H of 2.0 ppm.
This dispersion is more than twice
L06 value. The targets most responsible for this increase in scatter are HD 41161 and HD 53975, neither
of which was in the L06 study. And for JL 9, due to the decrease in our estimate of log \NHI\ from $20.78 \pm 0.05$
to $20.68 \pm 0.009$, this star would not formally be included in the 3rd (highest \HI\ column density) regime.

If we continue with the same \NH\ criteria for the 3rd regime, we now have 6 stars that qualify: LSS 1274, HD 191877,
HD 53975, HD 41161, HD 195965, and HD 90087, four of which have been revised or are newly determined in this study.
These have a mean of 7.9 ppm and a standard deviation of 4.8 ppm. In the intermediate region our study has
25 sight lines with a mean of 13.0 ppm and a standard deviation of 5.3 ppm. There is no statistical distinction between
the intermediate and distant regions, suggesting that similar physical processes are responsible for the distribution
of D/H values in both regimes.

There is another simple way to compare the gas inside and outside the LB. For the group of 22 target stars shown in
Figure~\ref{dh_vs_h} that lie within the LB we compute the sum of all \NDI\ values and the sum of all \nhtot\ values, and take the
ratio of these group sums. We compute the identical sums and ratio for the 31 points outside the LB. The results are
15.4 ppm and 11.3 ppm, respectively, which are consistent with idea that high \nhtot\ sight lines have higher  $F_*$
 and that D/H$_{tot}$ decreases as  $F_*$ increases. Based on the D abundance, this shows in a general way that
 the material in the LB is not simply a homogenized sample of material found at greater distances.

Comparisons of present-day Milky Way abundances to observations of three primordial species
can be used to constrain models of Galactic chemical evolution.
First, emission lines from \HII\ regions in low metallicity star-forming galaxies yield the mass fraction of $^4$He \citep{izotov14, aver15}.
Second, the $^7$Li abundance has been measured in the atmospheres of metal-poor stars \citep{spordone10}. Third, and
most relevant to this study, quasar absorption line observations of clouds with extremely low metallicity give the D/H ratio
in gas that is as close to pristine as possible \citep{burles98a, burles98b, kirkman03, cooke14, cooke16, cooke18}.  \citet{zavarygin18}
computed a weighted average of 13 high quality D/H measurements in QSO absorption line systems as D/H$_{prim} =
25.45\pm0.25$ ppm. The highest precision
measurement is (D/H)$_{prim} = 25.27\pm0.30$ ppm \citep{cooke18} in a system with an oxygen abundance
[O/H] = $-2.769\pm0.028$, or about 1/600 of the solar abundance. Approaching this in a different way, using improved
experimental reaction rates of  $d(p,\gamma)^3$He, $d(d,n)^3$He, and $d(d,p)^3$H, \citet{pitrou21} theoretically
calculate the primordial ratio as (D/H)$_{prim} = 24.39\pm0.37,$ about 2.1$\sigma$ below the quasar measurements.
Similarly, \citet{pisanti21} find (D/H)$_{prim} = 25.1\pm0.6\pm0.3$, where the two errors are due to uncertainties in the
nuclear rates and baryon density, respectively. This agrees very well with the measured value. For the purpose of
comparing to the local values of D/H in our study we prefer to be guided by the experimental values of \citet{zavarygin18} and
 \citet{cooke18} and take (D/H)$_{prim} = 25.4\pm0.3$ ppm.

In order to constrain models of Galactic chemical evolution, we want to compare (D/H)$_{prim}$ to the total deuterium abundance in the Galaxy.
As noted in section~\ref{sec:intro} there are two effects that could complicate assessing the total deuterium abundance. First, low metallicity
gas may still be accreting onto the disk of the Milky Way. This gas may have a higher D/H ratio than gas in the ISM that
has been polluted by material processed in stellar interiors and expelled via stellar winds and supernovae. \citet{sembach04}
showed that the high velocity cloud Complex C is falling into the Galaxy and has D/H = $22\pm7$ ppm. \citet{savage07}
measured the deuterium abundance in the warm neutral medium of the lower Galactic halo and found D/H = $22\err{8}{6}$ ppm,
virtually the same as in Complex C. Some of our sight lines have D/H values even greater than this, but note that the
Complex C and the neutral medium measurements have large error bars. Our results
provide some support for the infall hypothesis as the possible cause of the bifurcation of points in Figure~\ref{fig_depl}
at higher levels of depletion, $\langle F_*\rangle > 0.4.$ This D-rich material likely has low dust content reducing available
sites for deuterium depletion. More recently, by stacking spectra of many background QSOs to increase
the sensitivity to high velocity clouds, \citet{clark22} show that infalling gas tends to be in small, well-defined structures
with angular scales $\theta < 40\arcdeg.$ Observed metallicities range from 0.1 solar \citep{wakker99} to solar \citep{richter01, fox16}. This
patchiness may well be responsible for some of the observed variability in D/H reported here.

As noted earlier, one would expect an anticorrelation between gas-phase metal abundance and D/H which does not
appear to be the case \citep{hebrard03}. The second effect is while we assume that hydrogen depletion onto dust
grains is negligible \citep{prodanovic10}, the depletion of D onto the surfaces of dust grains \citep{draine04, draine06}
removes a fraction of the D from the gas phase that is measured in absorption line studies. In this case we expect a
correlation between metal abundance and D/H \citep{prochaska05, ellison07, lallement08}. \citet{prodanovic10} note that
while strong shocks would liberate both D and Fe, weaker shocks would liberate D only since it is weakly bound to
dust mantles while Fe is locked in grain cores. Our results only show a potentially weak anti-correlation
between the D/H and $\langle F_*\rangle$, adding weight to the conclusion that depletion onto dust grains is not always
the dominant factor and that the local sight line history needs be considered. This may be responsible for some of the
scatter in this correlation.

On the basis of several arguments including the metal abundance correlation, high D/H ratios observed in interplanetary dust
particles believed to originate in the ISM, and the effects of unresolved but saturated D lines, L06 conclude that the
large variation of D/H values beyond the Local Bubble are due to variable D depletion along different lines of sight.
They called attention to 5 stars ($\gamma^2$ Vel, Lan 23, WD1034+001, Feige 110, and LSE 44) outside the Local
Bubble that had high D/H values, ranging from 21.4 to 22.4 ppm. They stated that the total local Galactic D/H must be $\approx 22$
ppm or slightly greater.

In this study we have reevaluated \NH\ of three of these stars resulting in improved estimates of D/H,
all to lower values: WD1034+001 from $21.4\pm5.3$ to $19.00\err{3.39}{2.91}$; Feige 110 from $21.4\err{5.7}{3.8}$ to
$16.10\err{1.28}{1.20}$; and LSE 44 from $22.4\err{11.7}{6.6}$ to $19.31\err{1.94}{1.78}$. Ignoring Lan 23 due to its large
errors, there are now three stars with high D/H: $\gamma^2$ Vel at $21.9\err{2.6}{2.4}$, $\alpha$ Cru at $22.4\err{6.4}{5.2}$,
and a new one from this study, CPD$-$71 172 at $22.51\err{4.61}{3.43}$ the average of which, $\approx 22.1$, is almost the same as
L06 estimated. However, \citet{prodanovic10} point out the potential bias introduced when selecting only a small
number of high D/H values to consider when there are many other lower deuterium abundances that are consistent with
these within the errors. They used a more sophisticated Bayesian approach to estimate the undepleted abundance in
the local ISM using the 49 lines of sight in L06 and concluded that (D/H)$_{undepleted} = 20\pm1$ ppm.
In their analysis they used a ``top-hat" shaped prior for D/H, which is the least model-dependent of those they considered, although they also
modeled 4 others including positive and negative biased priors. Our new results as shown in Figure~\ref{dh_vs_h}
actually correspond to this unbiased prior better than the L06 data because the values of D/H in our study are more
uniformly distributed. For example, for log \NH $\ge 20.7$ we have 6 targets with D/H ranging from $5.6 - 17.3$ ppm.
L06 have 5 targets ranging from $7.6 - 10$ ppm. Thus, we adopt the \citet{prodanovic10} value of
(D/H)$_{undepleted}$ and using the current value of (D/H)$_{prim}$
discussed above we find an astration factor $f\rm_{D} = (25.4\pm0.3)/(20\pm1) = 1.27\pm0.07.$ This may be compared
to the values reported by L06 of $f\rm_{D} \le 1.19\err{0.16}{0.15}$ and $f\rm_{D} \le 1.12\pm0.14$,
depending on which value of (D/H)$_{prim}$ they used. Thus, while marginally higher, our astration factor does not significantly
differ from either of the L06 estimates.

We remind the reader that this may not represent the D/H value throughout the Galaxy \citep{lubowich10,leitner11,lagarde12}.

We now briefly consider this astration result in the context of models of Galactic chemical evolution in the Milky Way.
As previously noted, the gas-phase deuterium abundance can be enhanced by the local infall onto the Galactic disk
of primordial or at least less processed gas having low metallicity. Several investigators conclude that this and
other mechanisms are necessary to account for the low value of \fD\ found by L06.  \citet{tsujimoto11} argues
that this result is due to the decline in the star formation rate in the last several Gyr which has suppressed astration
over this same period.  \citet{prodanovic08} make a strong case for Galactic infall (the very title of their paper)
by showing that such small \fD\ requires both high infall rates and a low gas fraction, where
gas fraction is the present day ratio of gas to total mass. As the fraction of baryons that are returned to the
ISM by stars increases, even higher infall rates are required. These constraints are somewhat eased by higher \fD\ found
in our study. \citet{vandevoort18} confirm the importance of the return fraction in affecting the local deuterium
abundance but they also require patchy infall of intermediate metallicity material. Their models more easily
accommodate lower values of \fD. Their simulations also show that the deuterium fraction is lower at smaller
Galactic radii, which has been previously discussed \citep{lubowich10, lagarde12, leitner11}. In support of the concept of
a patchy distribution of infalling material \citet{decia21} note that such pristine gas can lead to chemical inhomogeneities
on scale sizes of tens of parsecs and that this gas is not efficiently mixed into the interstellar medium.

Other investigators come to the opposite conclusion. \citet{oliveira05} argue that fraction of infalling gas deposited within a
mixing time must be $\la$ 15\% based on the uniformity of O/H in the Local Bubble and along more distant sight lines.
Since the median hydrogen volume density $n_H$ in
the long sight lines is more than an order of magnitude greater than $n_H$ in the LB, greater levels of infall
would cause more variability in O/H than is observed. \citet{weinberg17}
notes that D/H is tightly coupled to the abundance of elements produced in core collapse supernovae, including
oxygen, and the baryon return fraction. He finds that producing variations in D/H of even a
factor of two, which is considerably less than we observe,
would give rise to large variations in O/H if they were caused by differential astration. His models
are consistent with the observed D/H variations if instead they are caused by variable depletion with D rather than
H occupying a large fraction of sites on polycyclic aromatic hydrocarbons.

We agree with previous investigators that the importance of deuterium depletion compared to infall will require improved
understanding of the properties and composition of dust grains and a greater understanding of some of the puzzling
relationships of D/H vs.~gas-phase metal abundance and reddening. Reducing the errors on D/H measurements
and observations of additional target stars would also help
to constrain the models but this is unlikely until high spectral resolution measurements of deuterium in the far-ultraviolet can once
again be obtained from space.

Finally, we call attention to an unusual result previously noted for Feige 110. D/H and O/H were first presented by
\citet{friedman02}. \citet{hebrard05} revisited this sightline and noted that D/H, O/H and N/H were all approximately
$2-3$ times larger than the values usually measured in the distant interstellar medium. This suggested that \NHI\
might be underestimated. We know from the current study that the value of D/H for Feige 110 is not at all unusual.
Furthermore, while the newly determined value log(\NHI) = $20.26 \pm{0.02}$ is slightly greater than the old
value, $20.14\err{0.13}{0.20}$ \citep{friedman02}, they agree within the errors. Our improved value of \NHI\ therfore
does not resolve the unexpectedly large values of O/H and N/H toward this target.

\section{Summary} \label{sec:summary}

In this investigation we observed 11 targets at medium spectral resolution 
($\sim30$ \kms) and 5 more at high resolution ( $\sim2.6$ \kms) in order to obtain high signal-to-noise absorption spectra
of the \HI\ \lya\ absorption line arising in the nearby interstellar medium. These targets range in distance from
189 to 2200 pc. With these data we reach the following conclusions.

\begin{enumerate}

\item We computed an atmospheric model for each star in our program. These models include temperature, gravity, and
a large number of metal lines of various ionization states. In some cases the models were better constrained than previous
ones in the literature due to accurate distances provided by the \textit{GAIA} DR3.

\item We fit the \lya\ absorption profile against the stellar flux model in order to compute \NHI. We demonstrated that
the most sensitive spectral region for constraining \NHI\ is where the damped profile lifts up from the saturated core region.
By carefully considering  statistical errors, continuum placement errors, stellar model errors, and others we arrive at robust
estimates of the total error in our measurement of \NHI.

\item We computed \NDI\ for the 5 sight lines that did not have previously published values. All estimates of \NDI\ come
from \textit{FUSE} observations.

\item With previously published estimates of \NHtwo\ we computed D/H$_{tot}$ for the 16 sight lines. We compared this to
similar previous studies, L06 in particular, and confirmed and strengthened the conclusion that D/H is variable over this range of \NHI\ values.
We also find the same range of D/H as was previously reported but we do not observe systematically low values of D/H at
the largest values of  \nhtot. Our results support a Bayesian analysis \citep{prodanovic10} that yields (D/H)$_{undepleted} = 20\pm1$ ppm.
When combined with the most modern estimates of the primordial D/H ratio this yields an astration factor of
 $f\rm_{D} = 1.27\pm0.07,$ a value marginally greater than those in the L06 study. This is more easily accommodated
 by many models of Galactic chemical evolution and reduces the need to invoke high levels of infall of deuterium-rich gas
 \citep{vandevoort18}.
 
 \item For the 5 sight lines observed at high resolution we did an analysis to compute the gas-phase column densities
 of a variety of metal species. These were used to supplement a previous generalized depletion analysis \citep{jenkins09}.
 We find only a weak correlation between D/H and depletion with considerable scatter. This implies that processes
 other than depletion are likely contributors to the observed variability in D/H. The bifurcation of D/${\rm H_{tot}}$ values for $\langle 
F_*\rangle > 0.4$ provides some evidence that infalling material onto the Galactic plane contributes to the variability.

\end{enumerate}

\begin{acknowledgments}
We thank Derck Massa for useful discussions about the profiles of damped absorption lines, Howard Bond for providing
the optical spectra obtained at CTIO, and the anonymous referee for several useful suggestions which improved the quality of this paper. Support for Program number 12287 was
provided by NASA through a grant from the Space Telescope Science Institute, which is operated by the Association of Universities
for Research in Astronomy, Incorporated, under NASA contract NAS5-26555. This research made use of NASA’s Astrophysics
Data System Bibliographic Services and of several PYTHON packages: NUMPY \citep{harris20}, MATPLOTLIB  \citep{hunter07}, SCIPY \citep{virtanen20}, and ASTROPY \citep{astropy18}. This work has made use of data from the European
Space Agency (ESA) mission {\it Gaia} (\url{https://www.cosmos.esa.int/gaia}), processed by the {\it Gaia}
Data Processing and Analysis Consortium (DPAC, \url{https://www.cosmos.esa.int/web/gaia/dpac/consortium}). Funding for the DPAC
has been provided by national institutions, in particular the institutions participating in the {\it Gaia} Multilateral Agreement.
\end{acknowledgments}

\vspace{5mm}
\noindent \textit{Facilities:} \textit{HST}~(STIS), \textit{FUSE}

\noindent \textit{Software:} TLUSTY, Synspec, owens.f, VPFIT, IRAF continuum package, NUMPY, MATPLOTLIB, SCIPY, ASTROPY


\appendix


\section{A. Notes and Data on the Metal Line Analysis}

We present notes on the metal line analysis of the five objects for which we obtained high spectral resolution data.
Table~\ref{tab-fvalues} gives the wavelengths, {\it f-}values, and references for the spectral lines used in the metal abundance analysis. In Tables~\ref{table_BD39}$-$\ref{table_HD191877} for each ion group, each table row represents one component along the line of sight, with the column density sum (if there is more than one component) shown in the row below the last component. The next row shows the wavelength intervals used in the fit, which was done simultaneously over all intervals. The last row shows
the reduced $\chi^2$ value, the number of degrees of freedom (dof) and
the probability of the fit.  The probability $p$ is the likelihood of
obtaining a $\chi^2$ residual at least as large as what was obtained
from the data and the fit.  For profile-fitting with {\sc vpfit},
$p\geq 0.01$ is considered acceptable.  Unless noted, a fit is done
for a complex in velocity space (covering one or more wavelength
intervals) for one ion. Otherwise, the statistics for cases in which a
fit is done for several ions simultaneously are indicated.

\underline{\it BD+39~3226:} An empirical comparison of the errors
and rms of the flux showed consistency, and in most cases, no adjustment was made to the error array.
There are a number of weak transitions, some which could only be satisfactorily fitted with one
component  of a multiplet e.g. Mn~II 1197. The S~II 1259 fitting region (four components)  required a
1.64~km~s$^{-1}$ offset to the red. It was not originally  possible to obtain a statistically acceptable
fit, even by including multiple components narrower than the LSF.  The summed column density was
consistent with a measurement using the apparent optical depth method (AOD, {\sc imnorm}, \citet{sembach92})
for the 1250~\AA\ transition, which has the lowest {\it f-}value. To obtain a statistically
acceptable fit, we therefore increased the error array by factors of 2.9-3.3 per region to compensate,
perhaps due to narrow unresolved components, still recovering a summed column density consistent with
the AOD method. The O~I~1355 region may be affected by a repeller wire artifact, and is not fitted.
Results are shown in Table~\ref{table_BD39}.

\underline{\it HD~41161:} We increased the errors by a factor of 1.1, based on global rms measurements.  
The Ge~II~1237 fit could be improved with the addition of a third component at the expense
of increased absorber parameter errors, however without a significant change in the column density sum.
We therefore leave it at two components.  The error arrays had to be increased by a factor of 1.7 
over rms for the 
Mn~II~1201 region, and by 1.4-1.7 over rms for the Cl~I~1347 region, possibly due to the under-sampling
of the LSF for the Jenkins slit. The component structure is
complex, with five components for Mn~II and Ni~II, 
and seven for Mg~II and Cl~I. Results are shown in Table~\ref{table_HD41161}.

\underline{\it HD~53975:} We increased the errors by a factor of 1.3 based on global rms measurements.  Around
the Mn~II triplet and P~II 1301 line this was increased to 1.3, around Cl~I~1347
by 1.5-2.5, and around the Mg~II~1240 line it was
doubled.  These adjustments were necessary for statistically acceptable profile
fits, and  are likely at least in part needed due to under-sampling of the
LSF for the Jenkins slit.  The Mg~II doublet is situated in a local flux maximum, and
we allowed linear offsets to the continuum there as a free parameter for each
member of the doublet to compensate
for continuum uncertainty. The component structure shows
some complexity, with five for Mn~II, seven for Cl~I P~II, and
eight for Mg~II.  The latter has one broad component which may
be suspect and due to unresolved blends or continuum issues.
Results are shown in Table~\ref{table_HD53975}.

\underline{\it HD~90087:}  
We increased the errors by a factor of 1.15 in our program data ($\sim 1190-1360$~\AA ),
and decreased them by a multiplicative factor of 0.7 for the 1390-1590~\AA\ archival
E140H data (Program 9434, PI J. Lauroesch).  We made adjustments of a factor 1.2
around P~II~1301, 3.3 around Ni~II~1317 and 1.3-3.0 around Cl~I~1347. We could not
make a satisfactory simultaneous fit for the P~II 1301 and 1532 transitions, therefore
we only use the 1301~\AA\ region.  We cannot identify a reason for the problem.  However, 
we measure the maximum optical depth for the 1301~\AA\ component to be $\tau \approx 1.4$.
The optical depth ratio of the 1301~\AA\ to 1526~\AA\ 
components is $\sim 1.9-2.1$ (with a possible small
unidentified blend in the 1526~\AA\ component), whereas the Morton (2003) {\it f-}value
for the 1526~\AA\ component 
of 0.00303 would imply $f_{1301}\lambda_{1301}/f_{1526}\lambda_{1526}\sim 3.6$.
A number of different ion components can have their
radial velocities tied to each other while still yielding statistically acceptable fits, which is
reassuring.  Ni~II, Ge~II and Cl~I (five components each) and Mg~II (six components) show
particularly complex structure.  We determined the fine structure absorption from O~I$^*$
and O~I$^{**}$ to be telluric. Results are shown in Table~\ref{table_HD90087}.

\underline{\it HD~191877:} We made no global change to the error arrays,
but increased them by a factor of 1.1 around Ni~II~1317,
and by a factor of 3 around Cl~I~1347 (in the echelle overlap region).
We find no evidence for general zero point problems.  However, we observe the flux for Cl~I  in
the line trough to drop to 2\% of the continuum, with a
signal to noise ratio of 2.7 (before the error array
correction), which we were unable to fit, possibly due
to undersampling of the Jenkins slit
LSF and unresolved components.  Mg~II (five components)
shows complex structure. The O~I~1355 transition may be affected by mild, narrow
artifacts, perhaps from the repeller wire.
Results are shown in Table~\ref{table_HD191877}.

\renewcommand{\thetable}{A\arabic{table}}

\setcounter{table}{0}

\begin{deluxetable}{llll}
\tablecaption{Wavelengths and {\it f}-values used for metal ions. The values adopted are from \citet{cashman17}, \citet{morton03},
and \citet{morton00}. The references shown refer to the original sources used to determine these values.\label{tab-fvalues}}
\tablewidth{0pt}
\tablehead{
\colhead{Ion} & \colhead{$\lambda$ (\AA)} & \colhead{{\it f}-value} & \colhead{References} \\
}
\startdata
O~I    & 1355.5977 & $1.16\times 10^{-6}$ & \citet{WFD96} \\
Mg~II & 1239.9253 & $6.32\times 10^{-4}$ & \citet{ThFe99}, \citet{F97}, \citet{FHBV98}, \\
&   &   & \citet{GF99}, \citet{SFH00}, \citet{MMGDMM02} \\
Mg~II & 1240.3947 & $3.56\times 10^{-4}$ & \citet{ThFe99}, \citet{F97}, \citet{FHBV98}, \\
&   &  & \citet{GF99}, \citet{SFH00}, \citet{MMGDMM02} \\
P~II   & 1301.8743 & 0.0196 & \citet{BAI18}\\  			
S~II  & 1250.578  & 0.00543 & \citet{OH89}, \citet{L69}, \citet{N97}\\
S~II  & 1253.805  & 0.0109 & \citet{OH89}, \citet{L69}, \citet{N97}\\
S~II  & 1259.518  & 0.0166 & \citet{OH89}, \citet{L69}, \citet{N97}\\
Cl~I   & 1347.2396 & 0.145 & \citet{OH13}\\
Mn~II & 1197.184  & 0.217 & \citet{DMPY74}, \citet{LBYO82} \\
Mn~II & 1199.391  & 0.169 &  \citet{DMPY74}, \citet{LBYO82} \\
Mn~II & 1201.118  & 0.121 &  \citet{DMPY74}, \citet{LBYO82} \\
Ni~II  & 1317.217  & 0.0571 & \citet{jenkins06}\\
Ni~II  & 1393.324  & 0.0125 & \citet{BB19}\\
Ni~II  & 1454.842  & 0.022 & \citet{BB19}\\
Ni~II  & 1467.259  & 0.0040 & \citet{BB19}\\
Ni~II  & 1467.756  & 0.0067 & \citet{BB19}\\
Ga~II & 1414.402  & 1.7720 & \citet{FH95} \\
Ge~II & 1237.0591 & 1.230 & \citet{biemont98} \\
Kr~I   & 1235.8380 & 0.204 & \citet{chan92}, \citet{lang98} \\	
\enddata
\end{deluxetable}

\begin{deluxetable}{lrrclcc}
\tablecaption{Metal line data for BD+39 3226. The lower and upper case letters ($a,A$) in the ion column denote tied components
in terms of radial velocity. The lower case letter varies independently, and the upper case letter is constrained to follow it.
The fits for the Mg~II, Ni~II, Ge~II, P~II, and Cl~I ions were done simultaneously. The fits for the Mn~II and S~II ions were done individually.\label{table_BD39}}
\tablewidth{0pt}
\tablehead{
\colhead{ion} & \colhead{rad. vel. (km~s$^{-1}$)} & \colhead{$b$ (km~s$^{-1}$)} & \colhead{$\log N$ (cm$^{-2}$)} \\
}
\startdata
  Mn~II  & -22.8  $\pm$ 0.6   &  5.2 $\pm$0.7   &12.38$\pm$0.04       \\
  \multicolumn{4}{l}{fitting intervals (\AA): 1197.062-1197.124} \\
   \multicolumn{4}{l}{$\chi^2$, dof, prob.~of fit: 0.889, 9, 0.042} \\
        &                    &                  &                   &     &                  &       \\
  S~II  & -22.8  $\pm$ 0.3   &  3.8 $\pm$0.1   &14.44$\pm$0.07      \\
  S~II  & -20.1  $\pm$ 0.0   &  1.8 $\pm$0.2   &14.67$\pm$0.05      \\
  S~II  & -15.6  $\pm$ 0.6   &  4.2 $\pm$0.8   &14.41$\pm$0.05      \\
  S~II  & -13.2  $\pm$ 0.3   &  1.3 $\pm$0.2   &13.92$\pm$0.11      \\
  S~II sum  &                       &		             & 15.04$\pm$0.03   &	  &		     &       \\ 
 \multicolumn{4}{l}{fitting intervals (\AA): 1250.455-1250.540, 1253.680-1253.778, 1259.390-1259.470} \\
 \multicolumn{4}{l}{$\chi^2$, dof, prob.~of fit: 1.235, 34, 0.163} \\
        &                    &                  &                   &     &                  &       \\
 Mg~II  & -23.1  $\pm$ 0.3   &  4.9 $\pm$0.5   &14.86$\pm$0.03       \\
\multicolumn{4}{l}{fitting intervals (\AA): 1239.788-1239.852, 1240.258-1240.322} \\
\multicolumn{4}{l}{$\chi^2$, dof, prob.~of fit: 1.288, 21, 0.033}    \\
        &                    &               &                   &     &                  &       \\
 Ni~II$^a$  & -20.7  $\pm$ 0.3&  5.1 $\pm$0.5   &13.14 $\pm$0.07  \\
 Ni~II  & -14.7  $\pm$ 0.9   & 11.7 $\pm$0.8   &13.33 $\pm$0.05    \\
 Ni~II  &  16.5  $\pm$ 0.0   &  4.8 $\pm$0.2   &13.19 $\pm$0.01   &	    &		      &       \\
 Ni~II sum  &                 &		     & 13.70$\pm$0.03   &	  &		     &       \\
  \multicolumn{4}{l}{fitting intervals (\AA): 1317.058-1317.344} \\
  \multicolumn{4}{l}{$\chi^2$, dof, prob.~of fit: 0.983, 93, 0.528} \\
          &                    &                  &              &     &                  &       \\
 Ge~II$^A$  & -20.7  $\pm$ 0.0&  6.4 $\pm$2.0   &11.13 $\pm$0.12        \\
 Ge~II  & -10.8  $\pm$ 0.3   &  4.0 $\pm$0.6   &11.38 $\pm$0.05   &	    &		      &       \\
 Ge~II sum  &                 &		    & 11.58$\pm$0.05 	&	  &		    &	    \\
  \multicolumn{4}{l}{fitting intervals (\AA): 1236.914-1237.057} \\
        &                    &                  &                   &     &                  &       \\
 P~II  & -23.7  $\pm$ 0.9   &  4.0 $\pm$2.1   &12.85 $\pm$0.10       \\
   \multicolumn{4}{l}{fitting intervals (\AA): 1301.700-1301.850} \\
        &                    &               &                   &     &                  &       \\
 Cl~I   & -27.3  $\pm$ 2.1   &  3.7 $\pm$2.6   &11.68 $\pm$0.20  \\
 Cl~I   & -21.6  $\pm$ 0.3   &  1.0 $\pm$0.4   &11.94 $\pm$0.11   &	    &		      &       \\ 
 Cl~I sum   &                &                  & 12.13$\pm$0.10       &         &    	            &       \\ 
 \multicolumn{4}{l}{fitting intervals (\AA): 1347.090-1347.170} \\  \hline
\enddata
\end{deluxetable}

\begin{deluxetable}{lrrclcc}
\tablecaption{Metal line data for HD~41161. The fit for each ion was done individually.\label{table_HD41161}}
\tablewidth{0pt}
\tablehead{
\colhead{ion} & \colhead{rad. vel. (km~s$^{-1}$)} & \colhead{$b$ (km~s$^{-1}$)} & \colhead{$\log N$ (cm$^{-2}$)}  \\
}
\startdata
Mn~II & -14.1$\pm$ 4.5  & 1.5$\pm$ 0.6  &  11.87$\pm$0.75  \\
Mn~II &  -8.4$\pm$ 1.2  & 2.8$\pm$ 1.4  &  12.64$\pm$0.17 \\
Mn~II &  -1.5$\pm$ 1.5  & 3.5$\pm$ 1.6  &  12.62$\pm$0.31  \\
Mn~II &   7.2$\pm$ 1.2  & 5.0$\pm$ 2.9  &  12.89$\pm$0.26   \\
Mn~II &  12.0$\pm$ 1.8  & 2.7$\pm$ 2.7  &  12.09$\pm$0.94  &  &  & \\
Mn~II sum &             &               &  13.26$\pm$0.18 &  &  & \\
 \multicolumn{4}{l}{fitting intervals (\AA): 1197.107-1197.273, 1199.314-1199.370,  1201.041-1201.207}  \\
 \multicolumn{4}{l}{$\chi^2$, dof, prob.~of fit: 0.93, 60, 0.620} \\
      &                &               &                &  &  & \\               
 Ge~II &  -4.5$\pm$ 0.6  & 6.3$\pm$ 0.6  &  12.04$\pm$0.04 \\
 Ge~II &   9.3$\pm$ 0.6  & 5.4$\pm$ 1.0  &  11.84$\pm$0.05   \\
Ge~II sum &             &               &  12.24$\pm$0.03 &  &  & \\
 \multicolumn{4}{l}{fitting intervals (\AA): 1236.980-1237.130} \\
 \multicolumn{4}{l}{$\chi^2$, dof, prob.~of fit:  1.46, 23' 0.071} \\
      &                &               &                &  &  & \\               
Mg~II & -12.3$\pm$ 8.7  & 3.2$\pm$ 0.9  &  14.71$\pm$1.13  &  \\
Mg~II &  -7.8$\pm$ 0.3  & 2.3$\pm$ 0.8  &  15.46$\pm$0.28  \\
Mg~II &  -3.6$\pm$ 0.9  & 2.4$\pm$ 1.1  &  15.31$\pm$0.28  &  \\
Mg~II &   1.2$\pm$ 0.9  & 2.7$\pm$ 1.5  &  15.27$\pm$0.24  &  &  & \\
Mg~II &   6.6$\pm$ 0.6  & 2.7$\pm$ 1.0  &  15.34$\pm$0.12  &  &  & \\
Mg~II &  10.8$\pm$ 0.9  & 1.8$\pm$ 0.4  &  15.18$\pm$0.10  &  &  & \\
Mg~II &  14.1$\pm$ 0.3  & 1.8$\pm$ 0.4  &  14.98$\pm$0.12  &  &  & \\
Mg~II sum &             &               &  16.07$\pm$0.14 &  &  & \\
 \multicolumn{4}{l}{fitting intervals (\AA): 1239.853-1240.001, 1240.310-1240.481} \\
 \multicolumn{4}{l}{$\chi^2$, dof, prob.~of fit: 1.46, 39, 0.031} \\
      &                &               &                &  &  & \\               
 P~II &  -8.4$\pm$ 0.9  & 3.6$\pm$ 0.1  &  13.88$\pm$0.19  \\
 P~II &  -1.8$\pm$ 1.5  & 4.9$\pm$ 2.5  &  13.87$\pm$0.20   \\
 P~II &   5.4$\pm$ 1.2  & 2.3$\pm$ 0.8  &  13.45$\pm$0.26  &  &  & \\
 P~II &  10.5$\pm$ 0.9  & 3.2$\pm$ 1.0  &  13.74$\pm$0.09  &  &  & \\
 P~II sum &             &               &  14.37$\pm$0.10 &  &  & \\
 \multicolumn{4}{l}{fitting intervals (\AA): 1301.760-1301.960} \\
 \multicolumn{4}{l}{$\chi^2$, dof, prob.~of fit: 1.57, 23, 0.041} \\
      &                &               &                &  &  & \\               
Ni~II &  -9.9$\pm$ 5.7  & 4.9$\pm$ 1.4  &  12.86$\pm$0.57   \\
Ni~II &  -6.6$\pm$ 1.2  & 1.8$\pm$ 0.6  &  12.89$\pm$0.34   \\
Ni~II &   0.3$\pm$ 1.2  & 5.0$\pm$ 3.1  &  13.29$\pm$0.34  &  &  & \\
Ni~II &   9.3$\pm$ 1.8  & 5.4$\pm$ 2.0  &  13.35$\pm$0.16  &  &  & \\
Ni~II &  20.7$\pm$ 1.2  & 1.6$\pm$ 1.4  &  11.99$\pm$0.17  &  &  & \\
Ni~II sum &             &               &  13.76$\pm$0.16 &  &  & \\
 \multicolumn{4}{l}{fitting intervals (\AA): 1317.120-1317.330} \\
 \multicolumn{4}{l}{$\chi^2$, dof, prob.~of fit: 1.51, 20, 0.065} \\
      &                &               &                &  &  & \\               
  O~I & -20.1$\pm$ 1.5  & 3.3$\pm$ 1.8  &  16.86$\pm$0.18  \\
  O~I &  -4.5$\pm$ 1.2  & 7.5$\pm$ 1.4  &  17.69$\pm$0.09   \\
  O~I &   9.6$\pm$ 1.2  & 6.2$\pm$ 1.9  &  17.57$\pm$0.10  &  &  & \\
  O~I sum &             &               &  17.96$\pm$0.06 &  &  & \\
 \multicolumn{4}{l}{fitting intervals (\AA): 1355.480-1355.690} \\
 \multicolumn{4}{l}{$\chi^2$, dof, prob.~of fit: 0.813, 24, 0.725} \\
      &                &               &                &  &  & \\               
 Cl~I & -20.1$\pm$ 5.7  & 4.5$\pm$ 7.4  &  11.82$\pm$0.77   \\
 Cl~I & -15.3$\pm$ 1.2  & 1.4$\pm$ 0.7  &  12.11$\pm$0.31   \\
 Cl~I &  -7.5$\pm$ 0.3  & 3.0$\pm$ 0.2  &  13.49$\pm$0.10  &  &  & \\
 Cl~I &  -2.1$\pm$ 0.6  & 4.6$\pm$ 0.5  &  13.53$\pm$0.03  &  &  & \\
 Cl~I &   6.3$\pm$ 0.3  & 1.7$\pm$ 0.2  &  13.03$\pm$0.12  &  &  & \\
 Cl~I &  10.5$\pm$ 0.3  & 2.2$\pm$ 0.4  &  13.07$\pm$0.04  &  &  & \\
 Cl~I &  14.1$\pm$ 0.3  & 1.3$\pm$ 0.4  &  12.35$\pm$0.12  &  &  & \\
 Cl~I sum &             &               &  13.96$\pm$0.04 &  &  & \\
 \multicolumn{4}{l}{fitting intervals (\AA): 1347.129-1347.320} \\
 \multicolumn{4}{l}{$\chi^2$, dof, prob.~of fit: 2.16, 11, 0.014} \\ \hline
\enddata
\end{deluxetable}

\begin{deluxetable}{lrrclcc}
\tablecaption{Metal line data for HD~53975. The fit for each ion was done individually.\label{table_HD53975}}
\tablewidth{0pt}
\tablehead{
\colhead{ion} & \colhead{rad. vel. (km~s$^{-1}$)} & \colhead{$b$ (km~s$^{-1}$)} & \colhead{$\log N$ (cm$^{-2}$)}  \\
}
\startdata
Mn~II &   7.5$\pm$ 1.5  & 2.5$\pm$ 1.5  &  12.07$\pm$0.17   \\
Mn~II &  21.0$\pm$ 0.6  & 3.0$\pm$ 0.5  &  12.69$\pm$0.10   \\
Mn~II &  25.5$\pm$ 4.5  &11.3$\pm$ 6.4  &  12.85$\pm$0.29  \\
Mn~II &  32.4$\pm$ 0.6  & 1.7$\pm$ 0.2  &  12.73$\pm$0.21  \\
Mn~II &  37.2$\pm$ 3.3  & 3.3$\pm$ 3.6  &  12.26$\pm$0.33  &  &  & \\
Mn~II sum &             &               &  13.31$\pm$0.12 &  &  & \\
 \multicolumn{4}{l}{fitting intervals (\AA): 1197.200-1197.368, 1199.400-1199.500, 1201.120-1201.285} \\
 \multicolumn{4}{l}{$\chi^2$, dof, prob.~of fit: 1.34, 62, 0.039}  \\
      &                &               &                &  &  & \\               
Ge~II &  21.9$\pm$ 0.3  & 3.7$\pm$ 0.2  &  11.84$\pm$0.02  \\
Ge~II &  33.9$\pm$ 0.3  & 2.2$\pm$ 0.2  &  11.77$\pm$0.05   \\
Ge~II &  39.6$\pm$ 1.2  & 2.2$\pm$ 1.8  &  11.09$\pm$0.16  &  &  & \\
Ge~II sum &             &               &  12.15$\pm$0.03 &  &  & \\
 \multicolumn{4}{l}{fitting intervals (\AA): 1237.101-1237.250 } \\
 \multicolumn{4}{l}{$\chi^2$, dof, prob.~of fit: 1.38, 20, 0.118}  \\
      &                &               &                &  &  & \\               
Mg~II &   2.4$\pm$ 2.7  & 1.3$\pm$ 0.9  &  14.05$\pm$0.72   \\
Mg~II &   8.4$\pm$ 0.9  & 3.5$\pm$ 1.5  &  14.83$\pm$0.18  \\
Mg~II &  20.7$\pm$ 0.6  & 2.9$\pm$ 0.2  &  15.43$\pm$0.11  \\
Mg~II &  23.4$\pm$ 0.6  & 1.6$\pm$ 0.3  &  15.05$\pm$0.21  &  &  & \\
Mg~II &  24.3$\pm$ 4.5  &10.1$\pm$ 7.5  &  15.35$\pm$0.17  &  &  & \\
Mg~II &  33.6$\pm$ 0.0  & 2.5$\pm$ 0.2  &  15.55$\pm$0.08  &  &  & \\
Mg~II &  39.6$\pm$ 0.3  & 2.2$\pm$ 0.5  &  14.89$\pm$0.14  &  &  & \\
Mg~II sum &             &               &  16.05$\pm$0.06 &  &  & \\
 \multicolumn{4}{l}{fitting intervals (\AA): 1239.925-1240.121, 1240.387-1240.586}  \\
 \multicolumn{4}{l}{$\chi^2$, dof, prob.~of fit: 1.47, 50, 0.017}  \\
      &                &               &                &  &  & \\               
 P~II &   1.8$\pm$11.1  & 4.0$\pm$ 2.3  &  12.70$\pm$1.26   \\
 P~II &   8.4$\pm$ 3.3  & 3.4$\pm$ 4.2  &  13.06$\pm$0.50   \\
 P~II &  19.2$\pm$ 3.0  & 2.9$\pm$ 0.5  &  13.61$\pm$0.59  &  &  & \\
 P~II &  22.5$\pm$ 1.5  & 2.6$\pm$ 1.6  &  13.65$\pm$0.40  &  &  & \\
 P~II &  31.8$\pm$ 3.0  & 3.2$\pm$ 1.2  &  13.58$\pm$0.44  &  &  & \\
 P~II &  33.0$\pm$ 1.2  & 1.5$\pm$ 0.3  &  13.53$\pm$0.52  &  &  & \\
 P~II &  38.1$\pm$ 3.3  & 3.5$\pm$ 4.3  &  13.40$\pm$0.37  &  &  & \\
 P~II sum &             &               &  14.30$\pm$0.22 &  &  & \\
 \multicolumn{4}{l}{fitting intervals (\AA): 1301.848-1302.066}   \\
 \multicolumn{4}{l}{$\chi^2$, dof, prob.~of fit: 1.76, 17, 0.027}  \\
      &                &               &                &  &  & \\               
Ni~II &   9.0$\pm$ 0.3  & 8.2$\pm$ 0.4  &  13.28$\pm$0.02  \\
Ni~II &  23.1$\pm$ 0.0  & 4.2$\pm$ 0.2  &  13.37$\pm$0.01  \\
Ni~II &  34.2$\pm$ 0.3  & 3.5$\pm$ 0.3  &  13.07$\pm$0.03  &  &  & \\
Ni~II &  41.1$\pm$ 0.6  & 2.1$\pm$ 0.7  &  12.50$\pm$0.07  &  &  & \\
Ni~II sum &             &               &  13.76$\pm$0.01 &  &  & \\
 \multicolumn{4}{l}{fitting intervals (\AA): 1317.160-1317.420}   \\
 \multicolumn{4}{l}{$\chi^2$, dof, prob.~of fit: 1.55, 33, 0.022}  \\
      &                &               &                &  &  & \\               
 O~I &   4.5$\pm$ 0.9  & 6.8$\pm$ 1.2  &  17.09$\pm$0.06  \\
 O~I &  21.3$\pm$ 0.3  & 4.7$\pm$ 0.3  &  17.46$\pm$0.02  \\
 O~I &  33.3$\pm$ 0.9  & 2.4$\pm$ 0.3  &  17.38$\pm$0.27  &  &  & \\
 O~I &  40.2$\pm$ 8.1  & 5.1$\pm$11.2  &  16.98$\pm$0.66  &  &  & \\
 O~I sum &             &               &  17.87$\pm$0.02 &  &  & \\
 \multicolumn{4}{l}{fitting intervals (\AA):  1355.570-1355.820}  \\
 \multicolumn{4}{l}{$\chi^2$, dof, prob.~of fit: 0.376, 25, 0.998}  \\
      &                &               &                &  &  & \\               
 Cl~I &  -0.0$\pm$ 2.4  & 3.7$\pm$ 1.7  &  11.78$\pm$0.25  \\
 Cl~I &   8.4$\pm$ 1.8  & 3.5$\pm$ 2.8  &  11.88$\pm$0.19  \\
 Cl~I &  19.5$\pm$ 0.3  & 1.6$\pm$ 0.1  &  12.83$\pm$0.04  &  &  & \\
 Cl~I &  23.7$\pm$ 0.3  & 1.2$\pm$ 0.2  &  12.40$\pm$0.04  &  &  & \\
 Cl~I &  31.5$\pm$ 3.0  & 1.8$\pm$ 0.3  &  12.78$\pm$0.65  &  &  & \\
 Cl~I &  34.2$\pm$ 0.3  & 1.4$\pm$ 0.3  &  13.14$\pm$0.19  &  &  & \\
 Cl~I &  37.5$\pm$ 6.6  & 3.9$\pm$ 9.0  &  12.31$\pm$0.64  &  &  & \\
 Cl~I sum &             &               &  13.51$\pm$0.17 &  &  & \\
 \multicolumn{4}{l}{fitting intervals (\AA): 1347.208-1347.435}  \\
 \multicolumn{4}{l}{$\chi^2$, dof, prob.~of fit: 1.89, 17, 0.015}  \\ \hline
\enddata
\end{deluxetable}

\begin{deluxetable}{lrrclcc}
\tablecaption{Metal line data for HD~90087. Just as in Table~\ref{table_BD39}, the lower and upper case letters in the ion column denote tied
radial velocities. The fits for the Mn~II, Kr~I, Mg~II, Ni~II, Ga~II, Ge~II, O~I, and P~II ions were done simultaneously.
The fit for Cl~I was done individually.\label{table_HD90087}}
tablewidth{0pt}
\tablehead{
\colhead{ion} & \colhead{rad. vel. (km~s$^{-1}$)} & \colhead{$b$ (km~s$^{-1}$)} & \colhead{$\log N$ (cm$^{-2}$)} \\
}
\startdata
Mn~II$^A$ &  -2.7$\pm$ 0.0  & 4.3$\pm$ 0.2  &  12.88$\pm$0.07  \\
Mn~II$^B$ &   1.8$\pm$ 0.0  & 2.9$\pm$ 1.1  &  12.71$\pm$0.11  \\
Mn~II$^C$ &   9.6$\pm$ 0.0  & 3.9$\pm$ 0.3  &  13.29$\pm$0.04  \\
Mn~II$^E$ &  18.9$\pm$ 0.0  & 7.9$\pm$ 4.3  &  12.49$\pm$0.18  \\
Mn~II sum   &           &               &  13.55$\pm$0.04 &  &  & \\ 
 \multicolumn{4}{l}{fitting intervals (\AA): 1197.118-1197.304,  1199.330-1199.430,  1201.050-1201.240}   \\
 \multicolumn{4}{l}{$\chi^2$, dof, prob.~of fit:  0.946, 497, 0.800}   \\
      &                &               &                &  &  & \\               
 Kr~I$^C$ &   9.6$\pm$ 0.0  & 4.4$\pm$ 1.0  &  12.26$\pm$0.08   \\
 \multicolumn{4}{l}{fitting intervals (\AA): 1235.830-1235.920}    \\
      &                &               &                &  &  & \\               
Mg~II & -15.9$\pm$ 2.7  &11.3$\pm$ 4.2  &  14.69$\pm$0.15   \\
Mg~II$^A$ &  -2.7$\pm$ 0.0  & 3.7$\pm$ 0.4  &  15.41$\pm$0.08   \\
Mg~II$^B$ &   1.8$\pm$ 0.0  & 2.5$\pm$ 0.4  &  15.41$\pm$0.06  &  &  & \\
Mg~II$^C$ &   9.6$\pm$ 0.0  & 3.7$\pm$ 0.2  &  15.85$\pm$0.03  &  &  & \\
Mg~II$^D$ &  14.4$\pm$ 0.0  & 1.8$\pm$ 0.8  &  15.02$\pm$0.16  &  &  & \\
Mg~II$^E$ &  18.9$\pm$ 0.0  & 6.1$\pm$ 4.3  &  15.01$\pm$0.23  &  &  & \\
Mg~II sum   &          &               &  16.17$\pm$0.03 &  &  & \\ 
 \multicolumn{4}{l}{fitting intervals (\AA):   1239.803-1240.054,  1240.310-1240.510}   \\
      &                &               &                &  &  & \\               
Ni~II$^a$ &  -2.7$\pm$ 0.6  & 2.3$\pm$ 0.4  &  12.92$\pm$0.30   \\
Ni~II$^b$ &   1.8$\pm$ 0.3  & 4.4$\pm$ 1.1  &  13.42$\pm$0.21   \\
Ni~II$^c$ &   9.6$\pm$ 0.0  & 4.9$\pm$ 1.4  &  13.52$\pm$0.18   \\
Ni~II$^d$ &  14.4$\pm$ 0.3  & 2.5$\pm$ 0.4  &  13.07$\pm$0.24   \\
Ni~II$^e$ &  18.9$\pm$ 2.7  & 4.3$\pm$ 3.8  &  12.73$\pm$0.45   \\
Ni~II sum$^\dagger$  &    &               &  13.93$\pm$0.11 &  &  & \\ 
 \multicolumn{4}{l}{fitting intervals (\AA): 1317.150-1317.320, 1393.290-1393.410, 1454.818-1454.960} \\
& 1467.150-1467.400, 1467.650-1467.890 \\
      &                &               &                &  &  & \\               
Ga~II &   0.3$\pm$ 1.2  & 4.5$\pm$ 2.0  &  11.03$\pm$0.11   \\
Ga~II$^C$ &   9.6$\pm$ 0.0  & 3.2$\pm$ 0.7  &  11.36$\pm$0.08  &  &  & \\
Ga~II &  15.6$\pm$ 3.0  & 1.8$\pm$ 5.7  &  10.25$\pm$0.39  &  &  & \\
Ga~II sum   &          &               &  11.55$\pm$0.07 &  &  & \\ 
 \multicolumn{4}{l}{fitting intervals (\AA):  1414.364-1414.495}   \\
      &                &               &                &  &  & \\               
Ge~II$^A$ &  -2.7$\pm$ 0.0  & 3.6$\pm$ 1.1  &  11.41$\pm$0.11  \\
Ge~II$^B$ &   1.8$\pm$ 0.0  & 2.6$\pm$ 0.6  &  11.64$\pm$0.07  &  &  & \\
Ge~II$^C$ &   9.6$\pm$ 0.0  & 2.7$\pm$ 0.2  &  12.05$\pm$0.04  &  &  & \\
Ge~II$^D$ &  14.4$\pm$ 0.0  & 1.9$\pm$ 1.2  &  11.46$\pm$0.21  &  &  & \\
Ge~II$^E$ &  18.9$\pm$ 0.0  & 4.3$\pm$ 4.4  &  11.32$\pm$0.29  &  &  & \\
Ge~II sum   &          &               &  12.37$\pm$0.05 &  &  & \\
 \multicolumn{4}{l}{fitting intervals (\AA):  1237.013-1237.162}   \\
      &                &               &                &  &  & \\               
 O~I$^B$ &   1.8$\pm$ 0.0  & 7.3$\pm$ 1.0  &  17.61$\pm$0.05  \\
  O~I &   9.9$\pm$ 0.3  & 2.0$\pm$ 0.3  &  17.64$\pm$0.05  &  &  & \\
  O~I sum   &          &               &  17.93$\pm$0.04 &  &  & \\
 \multicolumn{4}{l}{fitting intervals (\AA):  1355.546-1355.705}   \\
      &                &               &                &  &  & \\               
 P~II &  -5.1$\pm$ 2.1  & 1.3$\pm$ 1.1  &  12.76$\pm$0.63  \\
 P~II &   0.0$\pm$ 0.6  & 4.1$\pm$ 0.7  &  13.91$\pm$0.04  &  &  & \\
 P~II &   7.8$\pm$ 0.6  & 2.3$\pm$ 0.3  &  13.92$\pm$0.16  &  &  & \\
 P~II &  11.7$\pm$ 0.9  & 3.0$\pm$ 1.1  &  13.81$\pm$0.12  &  &  & \\
 P~II sum   &          &               &  14.37$\pm$0.07 &  &  & \\
 \multicolumn{4}{l}{fitting intervals (\AA):  1301.817-1301.940}   \\
     &                &               &                &  &  & \\               
 Cl~I &  -2.1$\pm$ 1.2  & 2.4$\pm$ 0.6  &  12.90$\pm$0.31   \\
 Cl~I &   2.7$\pm$ 1.2  & 4.2$\pm$ 1.5  &  13.14$\pm$0.13  \\
 Cl~I &   9.6$\pm$ 0.0  & 2.0$\pm$ 0.3  &  13.39$\pm$0.15  &  &  & \\
 Cl~I &  12.9$\pm$ 9.6  & 3.2$\pm$ 9.7  &  12.39$\pm$0.88  &  &  & \\
 Cl~I &  21.0$\pm$ 2.4  & 3.7$\pm$ 3.2  &  12.13$\pm$0.34  &  &  & \\
 Cl~I sum   &          &               &  13.70$\pm$0.12 &  &  & \\
 \multicolumn{4}{l}{fitting intervals (\AA): 1347.200-1347.369}   \\
 \multicolumn{4}{l}{$\chi^2$, dof, prob.~of fit: 1.56, 14, 0.082}    \\ \hline
\enddata
\tablenotetext{}{$^\dagger$~The  {\it f-}values for some of the \NiII\ transitions shown in Table~\ref{tab-fvalues} were
slightly changed after the Ni column densities were computed for this sight line \citep{BB19}.  To be
conservative, we have made a corresponding correction to these column densities on the order of $\leq 0.06$ dex per component, and
increased the errors by 0.03 dex per component and 0.09 dex in the sum. This has no bearing on our determination of \NHI.}
\end{deluxetable}

\begin{deluxetable}{lrrclcc}
\tablecaption{Metal line data for HD~191877. Just as in Table~\ref{table_BD39}, the lower and upper case letters in the ion column denote tied
radial velocities. The fits for the Mn~II, Kr~I, Ge~II, Mg~II, P~II, Ni~II, and O~I ions were done simultaneously.
The fit for Cl~I was done individually.\label{table_HD191877}}
\tablewidth{0pt}
\tablehead{
\colhead{ion} & \colhead{rad. vel. (km~s$^{-1}$)} & \colhead{$b$ (km~s$^{-1}$)} & \colhead{$\log N$ (cm$^{-2}$)}  \\
}
\startdata
Mn~II & -21.3$\pm$ 1.8  & 1.6$\pm$ 1.1  &  11.71$\pm$0.25   \\
Mn~II & -14.1$\pm$ 2.7  & 2.8$\pm$ 0.9  &  12.33$\pm$0.51   \\
Mn~II$^E$ &  -6.6$\pm$ 0.0  & 4.4$\pm$ 1.5  &  13.12$\pm$0.17  \\
Mn~II &  -0.0$\pm$ 1.5  & 3.4$\pm$ 1.9  &  12.59$\pm$0.26   \\
Mn~II sum  &             &               &  13.29$\pm$0.14 &  &  & \\
 \multicolumn{4}{l}{fitting intervals (\AA):  1197.079-1197.219,  1199.296-1199.405,  1201.013-1201.153}  \\
 \multicolumn{4}{l}{$\chi^2$, dof, prob.~of fit: 1.26, 185, 0.010}   \\
      &                &               &                &  &  & \\               
 Kr~I$^E$ &  -6.6$\pm$ 0.0  & 5.7$\pm$ 1.3  &  12.18$\pm$0.08   \\
 \multicolumn{4}{l}{fitting intervals (\AA):  1235.778-1235.848}   \\
      &                &               &                &  &  & \\               
Ge~II & -16.5$\pm$ 5.4  & 6.9$\pm$ 4.6  &  11.27$\pm$0.40  \\
Ge~II$^D$ & -11.1$\pm$ 0.0  & 1.6$\pm$ 0.7  &  11.27$\pm$0.26  &  &  & \\
Ge~II$^E$ &  -6.6$\pm$ 0.0  & 2.1$\pm$ 0.6  &  11.72$\pm$0.27  &  &  & \\
Ge~II$^F$ &  -2.7$\pm$ 0.0  & 4.0$\pm$ 2.6  &  11.78$\pm$0.20  &  &  & \\
Ge~II sum &             &               &  12.18$\pm$0.14 &  &  & \\
 \multicolumn{4}{l}{fitting intervals (\AA): 1236.953-1237.086}    \\
      &                &               &                &  &  & \\               
Mg~II$^a$ & -20.1$\pm$ 0.9  & 3.6$\pm$ 0.5  &  14.75$\pm$0.12   \\
Mg~II$^c$ & -14.4$\pm$ 2.7  & 2.8$\pm$ 1.1  &  14.69$\pm$0.44   \\
Mg~II$^d$ & -11.1$\pm$ 0.9  & 2.0$\pm$ 0.4  &  15.08$\pm$0.26  &  &  & \\
Mg~II$^e$ &  -6.6$\pm$ 0.3  & 2.2$\pm$ 0.6  &  15.44$\pm$0.29  &  &  & \\
Mg~II$^f$ &  -2.7$\pm$ 2.1  & 4.3$\pm$ 2.5  &  15.39$\pm$0.84  &  &  & \\
Mg~II &   0.6$\pm$31.5  & 5.9$\pm$34.0  &  15.03$\pm$2.57  &  &  & \\
Mg~II sum &             &               &  15.93$\pm$0.92 &  &  & \\
 \multicolumn{4}{l}{fitting intervals (\AA): 1239.817-1239.977,  1240.287-1240.447}  \\
      &                &               &                &  &  & \\               
 P~II & -18.3$\pm$ 1.8  & 4.4$\pm$ 1.1  &  13.12$\pm$0.17   \\
 P~II & -12.9$\pm$ 3.0  & 1.7$\pm$ 0.5  &  13.36$\pm$0.58  &  &  & \\
 P~II &  -7.5$\pm$ 0.6  & 3.0$\pm$ 1.1  &  14.00$\pm$0.20  &  &  & \\
 P~II &  -2.1$\pm$ 2.1  & 4.1$\pm$ 2.6  &  13.69$\pm$0.21  &  &  & \\
 P~II sum &             &               &  14.27$\pm$0.15 &  &  & \\
 \multicolumn{4}{l}{fitting intervals (\AA): 1301.758-1301.910}   \\
      &                &               &                &  &  & \\               
Ni~II$^C$ & -14.4$\pm$ 0.0  & 4.6$\pm$ 0.6  &  12.95$\pm$0.30  \\
Ni~II &  -6.3$\pm$ 0.6  & 4.7$\pm$ 1.1  &  13.55$\pm$0.17  &  &  & \\
Ni~II &   0.9$\pm$ 0.9  & 4.2$\pm$ 1.1  &  13.39$\pm$0.13  &  &  & \\
Ni~II sum &             &               &  13.84$\pm$0.11 &  &  & \\
 \multicolumn{4}{l}{fitting intervals (\AA):  1317.110-1317.264}   \\
      &                &               &                &  &  & \\               
  O~I$^A$ & -20.1$\pm$ 0.0  & 4.3$\pm$ 1.1  &  17.23$\pm$0.10  \\
  O~I$^E$ &  -6.6$\pm$ 0.0  & 6.5$\pm$ 0.8  &  17.77$\pm$0.04  &  &  & \\
  O~I sum &             &               &  17.88$\pm$0.04 &  &  & \\ 
 \multicolumn{4}{l}{fitting intervals (\AA): 1355.471-1355.607}   \\
      &                &               &                &  &  & \\               
 Cl~I & -21.9$\pm$ 2.7  & 1.4$\pm$ 0.5  &  11.97$\pm$0.56 \\
 Cl~I & -16.2$\pm$ 1.5  & 3.4$\pm$ 1.5  &  12.69$\pm$0.19   \\
 Cl~I & -11.1$\pm$ 1.2  & 1.7$\pm$ 0.6  &  13.11$\pm$0.37  &  &  & \\
 Cl~I &  -7.2$\pm$ 0.3  & 1.4$\pm$ 1.0  &  13.41$\pm$0.27  &  &  & \\
 Cl~I &  -3.3$\pm$ 0.3  & 3.5$\pm$ 0.3  &  13.58$\pm$0.01  &  &  & \\
 Cl~I sum &             &              &  13.92$\pm$0.11 &  &  & \\
 \multicolumn{4}{l}{fitting intervals (\AA): 1347.121-1347.272}   \\
 \multicolumn{4}{l}{$\chi^2$, dof, prob.~of fit: 2.17, 10, 0.017}   \\ \hline
\enddata
\end{deluxetable}

\clearpage


\section{B. Depletion Correlation Data}

\renewcommand{\thetable}{B\arabic{table}}
\setcounter{table}{0}

In Table~\ref{basic_data} we present column densities and values of the generalized depletion
parameter  $F_*$ for the objects shown in Figure~\ref{fig_depl}. The sources of the data
shown in column 3 are given in Table~\ref{refs}.

\begin{deluxetable}{lcl}
\tablewidth{265pt}
\tablecolumns{3}
\tablecaption{Basic Data in the Correlation
\label{basic_data}}
\tablehead{
\colhead{Item} & \colhead{Value} & \colhead{Source(s)\tablenotemark{a}}
}
\startdata
\cutinhead{HD 3894 (PG 0038+199)}
$\log N({\rm D~I})$&$15.75\pm 0.04$&W++05\\
$\log N({\rm H_{tot}})$&$20.47\pm 0.01$&TP, W++05\\
$F_*({\rm  O})$&$-0.732\pm 0.625$&  W++05\\
$F_*({\rm Fe})$&$ 0.467\pm 0.098$&  W++05\\
$\langle F_*\rangle$&$ 0.438\pm 0.097$\\
\cutinhead{HD 5394 ($\gamma$ Cas)}
$\log N({\rm D~I})$&$15.15^{+0.04}_{-0.05}$&FVY80\\
$\log N({\rm H_{tot}})$&$20.04^{+0.04}_{-0.02}$&FVY80, S98\\
$F_*({\rm  O})$&$ 0.124\pm 0.263$&  MJC98\\
$F_*({\rm Mg})$&$ 0.480\pm 0.060$&  JSS86\\
$F_*({\rm  P})$&$ 0.494\pm 0.099$&  JSS86\\
$F_*({\rm Cl})$&$ 0.616\pm 0.258$&  JSS86\\
$F_*({\rm Ti})$&$ 0.425\pm 0.042$&    S78\\
$F_*({\rm Mn})$&$ 0.250\pm 0.086$&  JSS86\\
$F_*({\rm Fe})$&$ 0.405\pm 0.082$&  JSS86\\
$\langle F_*\rangle$&$ 0.420\pm 0.028$\\
\cutinhead{HD 36486 ($\delta$ Ori A)}
$\log N({\rm D~I})$&$15.06^{+0.07}_{-0.04}$&J++99\\
$\log N({\rm H_{tot}})$&$20.19\pm 0.03$&ST++00, J++00\\
$F_*({\rm  O})$&$ 1.189\pm 0.308$&  MJC98\\
$F_*({\rm Mg})$&$ 0.450\pm 0.049$&  JSS86\\
$F_*({\rm  P})$&$ 0.735\pm 0.215$&   JY78\\
$F_*({\rm Cl})$&$ 0.591\pm 0.091$&  JSS86\\
$F_*({\rm Ti})$&$ 0.471\pm 0.025$&  PTH05\\
$F_*({\rm Cr})$&$ 0.687\pm 0.044$&   RB95\\
$F_*({\rm Mn})$&$ 0.693\pm 0.077$&  JSS86\\
$F_*({\rm Fe})$&$ 0.561\pm 0.055$&  JSS86\\
$F_*({\rm Zn})$&$ 0.548\pm 0.102$&   RB95\\
$\langle F_*\rangle$&$ 0.534\pm 0.018$\\
\cutinhead{HD 37043 ($\iota$ Ori)}
$\log N({\rm D~I})$&$15.30\pm 0.04$&LVY79\\
$\log N({\rm H_{tot}})$&$20.11^{+0.09}_{-0.11}$&LVY79, BSD78\\
$F_*({\rm  O})$&$ 0.429\pm 0.603$& MJHC94\\
$F_*({\rm Mg})$&$ 0.420\pm 0.106$&  JSS86\\
$F_*({\rm  P})$&$ 0.495\pm 0.126$&  JSS86\\
$F_*({\rm Cl})$&$ 0.632\pm 0.127$&  JSS86\\
$F_*({\rm Ti})$&$ 0.353\pm 0.051$&  PTH05\\
$F_*({\rm Mn})$&$ 0.238\pm 0.132$&  JSS86\\
$F_*({\rm Fe})$&$ 0.382\pm 0.104$&  JSS86\\
$\langle F_*\rangle$&$ 0.392\pm 0.037$\\
\cutinhead{HD 37128 ($\epsilon$ Ori)}
$\log N({\rm D~I})$&$15.26^{+0.04}_{-0.06}$&LVY79\\
$\log N({\rm H_{tot}})$&$20.45^{+0.07}_{-0.09}$&LVY79, J++00\\
$F_*({\rm  O})$&$ 0.967\pm 0.443$&  MJC98\\
$F_*({\rm Mg})$&$ 0.530\pm 0.083$&  JSS86\\
$F_*({\rm  P})$&$ 0.538\pm 0.117$&  JSS86\\
$F_*({\rm Cl})$&$ 0.575\pm 0.082$&  JSS86\\
$F_*({\rm Ti})$&$ 0.475\pm 0.042$&  PTH05\\
$F_*({\rm Cr})$&$ 0.618\pm 0.067$&   RB95\\
$F_*({\rm Mn})$&$ 0.471\pm 0.104$&  JSS86\\
$F_*({\rm Fe})$&$ 0.646\pm 0.075$&  JSS86\\
$F_*({\rm Zn})$&$ 0.401\pm 0.160$&   RB95\\
$\langle F_*\rangle$&$ 0.536\pm 0.026$\\
\enddata
\end{deluxetable}

\setcounter{table}{0}

\begin{deluxetable}{lcl}
\tablewidth{265pt}
\tablecolumns{3}
\tablecaption{\textit{(continued)}
\label{basic_data}}
\tablehead{
\colhead{Item} & \colhead{Value} & \colhead{Source(s)\tablenotemark{a}}
}
\startdata
\cutinhead{HD 38666 ($\mu$ Col)}
$\log N({\rm D~I})$&$14.70^{+0.30}_{-0.10}$&YR76\\
$\log N({\rm H_{tot}})$&$19.86\pm 0.02$&HSF99, SCH74\\
$F_*({\rm Mg})$&$ 0.090\pm 0.034$&  HSF99\\
$F_*({\rm Si})$&$ 0.075\pm 0.034$&  HSF99\\
$F_*({\rm  P})$&$ 0.088\pm 0.039$&  HSF99\\
$F_*({\rm Ti})$&$ 0.139\pm 0.018$&  LHW08\\
$F_*({\rm Cr})$&$ 0.086\pm 0.034$&  HSF99\\
$F_*({\rm Mn})$&$ 0.143\pm 0.040$&  HSF99\\
$F_*({\rm Fe})$&$ 0.108\pm 0.023$&  HSF99\\
$F_*({\rm Ni})$&$ 0.170\pm 0.042$&  HSF99\\
$F_*({\rm Zn})$&$ 0.053\pm 0.151$&  HSF99\\
$\langle F_*\rangle$&$ 0.116\pm 0.010$\\
\cutinhead{HD 41161}
$\log N({\rm D~I})$&$16.40\pm 0.05$&OH06\\
$\log N({\rm H_{tot}})$&$21.17\pm 0.02$&TP, SDA21\\
$F_*({\rm  O})$&$ 0.229\pm 0.210$&     TP\\
$F_*({\rm Mg})$&$ 0.443\pm 0.030$&     TP\\
$F_*({\rm  P})$&$ 0.462\pm 0.040$&     TP\\
$F_*({\rm Ti})$&$ 0.467\pm 0.023$&  EPL07\\
$F_*({\rm Mn})$&$ 0.658\pm 0.036$&     TP\\
$F_*({\rm Fe})$&$ 0.600\pm 0.051$&   OH06\\
$F_*({\rm Ni})$&$ 0.507\pm 0.030$&     TP\\
$F_*({\rm Ge})$&$ 0.522\pm 0.066$&     TP\\
$\langle F_*\rangle$&$ 0.503\pm 0.013$\\
\cutinhead{HD 53975}
$\log N({\rm D~I})$&$16.15\pm 0.07$&OH06\\
$\log N({\rm H_{tot}})$&$21.09\pm 0.02$&TP, OH06\\
$F_*({\rm  O})$&$ 0.323\pm 0.297$&     TP\\
$F_*({\rm Mg})$&$ 0.394\pm 0.045$&     TP\\
$F_*({\rm  P})$&$ 0.464\pm 0.037$&     TP\\
$F_*({\rm Ti})$&$ 0.360\pm 0.022$&  EPL07\\
$F_*({\rm Mn})$&$ 0.542\pm 0.047$&     TP\\
$F_*({\rm Fe})$&$ 0.655\pm 0.043$&   OH06\\
$F_*({\rm Ni})$&$ 0.474\pm 0.027$&     TP\\
$F_*({\rm Ge})$&$ 0.606\pm 0.056$&     TP\\
$\langle F_*\rangle$&$ 0.456\pm 0.013$\\
\cutinhead{HD 66811 ($\zeta$ Pup)}
$\log N({\rm D~I})$&$15.11\pm 0.06$&ST++00\\
$\log N({\rm H_{tot}})$&$19.96\pm 0.03$&ST++00, MD76\\
$F_*({\rm Mg})$&$ 0.241\pm 0.052$&    M78\\
$F_*({\rm Si})$&$ 0.102\pm 0.137$&    M78\\
$F_*({\rm  P})$&$ 0.298\pm 0.064$&    M78\\
$F_*({\rm Cl})$&$ 0.223\pm 0.132$&    M78\\
$F_*({\rm Ti})$&$ 0.397\pm 0.025$&  EPL07\\
$F_*({\rm Cr})$&$ 0.425\pm 0.141$&    M78\\
$F_*({\rm Mn})$&$ 0.222\pm 0.062$&    M78\\
$F_*({\rm Fe})$&$ 0.362\pm 0.047$&    M78\\
$F_*({\rm Ni})$&$ 0.410\pm 0.270$&    M78\\
$F_*({\rm Zn})$&$ 0.452\pm 0.335$&    M78\\
$\langle F_*\rangle$&$ 0.343\pm 0.018$\\
\cutinhead{HD 68273 ($\gamma^2$ Vel)}
$\log N({\rm D~I})$&$15.05\pm 0.03$&ST++00\\
$\log N({\rm H_{tot}})$&$19.71\pm 0.03$&ST++00, BSD78\\
$F_*({\rm Mg})$&$ 0.390\pm 0.154$&   FS94\\
$F_*({\rm Si})$&$ 0.302\pm 0.082$&   FS94\\
$F_*({\rm  P})$&$ 0.443\pm 0.220$&   FS94\\
$F_*({\rm Ti})$&$ 0.261\pm 0.020$&  EPL07\\
$F_*({\rm Mn})$&$-0.009\pm 0.104$&   FS94\\
$F_*({\rm Fe})$&$ 0.270\pm 0.075$&   FS94\\
$\langle F_*\rangle$&$ 0.258\pm 0.018$\\
\enddata
\end{deluxetable}

\setcounter{table}{0}

\begin{deluxetable}{lcl}
\tablewidth{265pt}
\tablecolumns{3}
\tablecaption{\textit{(continued)}
\label{basic_data}}
\tablehead{
\colhead{Item} & \colhead{Value} & \colhead{Source(s)\tablenotemark{a}}
}
\startdata
\cutinhead{HD 90087}
$\log N({\rm D~I})$&$16.16\pm 0.06$&H++05\\
$\log N({\rm H_{tot}})$&$21.25\pm 0.02$&TP, SDA21\\
$F_*({\rm  O})$&$ 0.452\pm 0.206$&     TP\\
$F_*({\rm Mg})$&$ 0.433\pm 0.024$&     TP\\
$F_*({\rm  P})$&$ 0.526\pm 0.057$&     TP\\
$F_*({\rm Mn})$&$ 0.436\pm 0.029$&     TP\\
$F_*({\rm Fe})$&$ 0.427\pm 0.050$&  JS07a\\
$F_*({\rm Ni})$&$ 0.480\pm 0.032$&     TP\\
$F_*({\rm Ge})$&$ 0.474\pm 0.062$&     TP\\
$\langle F_*\rangle$&$ 0.451\pm 0.014$\\
\cutinhead{HD 93030 ($\theta$ Car)}
$\log N({\rm D~I})$&$14.98^{+0.18}_{-0.21}$&AJS92\\
$\log N({\rm H_{tot}})$&$20.26\pm 0.08$&DS94, AJS92\\
$F_*({\rm  O})$&$ 2.776\pm 1.126$&  AJS92\\
$F_*({\rm Mg})$&$ 0.417\pm 0.114$&  AJS92\\
$F_*({\rm  P})$&$ 0.477\pm 0.149$&  AJS92\\
$F_*({\rm Cl})$&$ 0.441\pm 0.117$&  AJS92\\
$F_*({\rm Ti})$&$ 0.427\pm 0.040$&  EPL07\\
$F_*({\rm Mn})$&$ 0.453\pm 0.164$&  AJS92\\
$F_*({\rm Fe})$&$ 0.505\pm 0.081$&  AJS92\\
$\langle F_*\rangle$&$ 0.444\pm 0.031$\\
\cutinhead{HD 108248 ($\alpha^1$ Cru)}
$\log N({\rm D~I})$&$14.95\pm 0.05$&YR76\\
$\log N({\rm H_{tot}})$&$19.60\pm 0.10$&YR76, B++83\\
$F_*({\rm Mg})$&$ 0.099\pm 0.105$&  JSS86\\
$F_*({\rm  P})$&$ 0.205\pm 0.134$&  JSS86\\
$F_*({\rm Cl})$&$ 0.285\pm 0.103$&  JSS86\\
$F_*({\rm Ti})$&$ 0.197\pm 0.051$&  EPL07\\
$F_*({\rm Mn})$&$ 0.028\pm 0.129$&  JSS86\\
$F_*({\rm Fe})$&$ 0.154\pm 0.088$&  JSS86\\
$\langle F_*\rangle$&$ 0.177\pm 0.035$\\
\cutinhead{HD 122451 ($\beta$ Cen)}
$\log N({\rm D~I})$&$14.70\pm 0.20$&YR76\\
$\log N({\rm H_{tot}})$&$19.54\pm 0.05$&YR76, Y76\\
$F_*({\rm Si})$&$ 0.366\pm 0.060$& BLWY84\\
$F_*({\rm Ti})$&$ 0.202\pm 0.033$&  EPL07\\
$\langle F_*\rangle$&$ 0.240\pm 0.029$\\
\cutinhead{HD 191877}
$\log N({\rm D~I})$&$15.94^{+0.11}_{-0.06}$&H++03\\
$\log N({\rm H_{tot}})$&$21.12\pm 0.02$&TP, SDA21\\
$F_*({\rm  O})$&$ 0.194\pm 0.434$&     TP\\
$F_*({\rm Mg})$&$ 0.515\pm 0.029$&     TP\\
$F_*({\rm  P})$&$ 0.496\pm 0.040$&     TP\\
$F_*({\rm Ti})$&$ 0.380\pm 0.015$&  PTH05\\
$F_*({\rm Mn})$&$ 0.543\pm 0.034$&     TP\\
$F_*({\rm Ni})$&$ 0.408\pm 0.022$&     TP\\
$F_*({\rm Ge})$&$ 0.510\pm 0.061$&     TP\\
$\langle F_*\rangle$&$ 0.429\pm 0.010$\\
\cutinhead{HD 195965}
$\log N({\rm D~I})$&$15.88\pm 0.07$&H++03\\
$\log N({\rm H_{tot}})$&$21.13\pm 0.02$&H++03, SDA21\\
$F_*({\rm  O})$&$ 0.494\pm 0.162$&  H++03\\
$F_*({\rm Mg})$&$ 0.595\pm 0.051$&  JS07b\\
$F_*({\rm Ti})$&$ 0.506\pm 0.016$&  PTH05\\
$\langle F_*\rangle$&$ 0.514\pm 0.015$\\
\cutinhead{BD +28 4211}
$\log N({\rm D~I})$&$14.95\pm 0.02$&HM03\\
$\log N({\rm H_{tot}})$&$19.85\pm 0.04$&S++02, S++02\\
$F_*({\rm  O})$&$ 1.403\pm 0.295$&   HM03\\
$F_*({\rm Ti})$&$ 0.337\pm 0.043$&  PTH05\\
$F_*({\rm Fe})$&$ 0.260\pm 0.083$&  L++06\\
$\langle F_*\rangle$&$ 0.339\pm 0.038$\\
\enddata
\end{deluxetable}

\setcounter{table}{0}

\begin{deluxetable}{lcl}
\tablewidth{265pt}
\tablecolumns{3}
\tablecaption{\textit{(continued)}
\label{basic_data}}
\tablehead{
\colhead{Item} & \colhead{Value} & \colhead{Source(s)\tablenotemark{a}}
}
\startdata
\cutinhead{WD 2247+583 (Lan 23)}
$\log N({\rm D~I})$&$15.23\pm 0.07$&O++03\\
$\log N({\rm H_{tot}})$&$19.89^{+0.25}_{-0.04}$&WKL99, O++03\\
$F_*({\rm  O})$&$-0.433\pm 1.055$&  L++03\\
$F_*({\rm Fe})$&$ 0.384\pm 0.126$&  O++03\\
$\langle F_*\rangle$&$ 0.373\pm 0.125$\\
\cutinhead{REJ 1738+665}
$\log N({\rm D~I})$&$15.08\pm 0.04$&D++09\\
$\log N({\rm H_{tot}})$&$19.83\pm 0.05$&D++09, D++09\\
$F_*({\rm  O})$&$ 0.977\pm 0.372$&  D++09\\
$F_*({\rm  P})$&$ 0.704\pm 0.102$&  D++09\\
$F_*({\rm Fe})$&$ 0.326\pm 0.043$&  D++09\\
$\langle F_*\rangle$&$ 0.391\pm 0.040$\\
\cutinhead{TD1 32709}
$\log N({\rm D~I})$&$15.30\pm 0.05$&O++06\\
$\log N({\rm H_{tot}})$&$20.08\pm 0.01$&TP, O++06\\
$F_*({\rm  O})$&$ 1.816\pm 0.553$&  O++06\\
$F_*({\rm Fe})$&$ 0.558\pm 0.079$&  O++06\\
$\langle F_*\rangle$&$ 0.584\pm 0.078$\\
\cutinhead{  WD 1034+001}
$\log N({\rm D~I})$&$15.40\pm 0.07$&O++06\\
$\log N({\rm H_{tot}})$&$20.12\pm 0.02$&TP, O++06\\
$F_*({\rm  O})$&$ 1.191\pm 0.487$&  O++06\\
$F_*({\rm Ti})$&$ 0.460\pm 0.098$&  EPL07\\
$F_*({\rm Fe})$&$ 0.472\pm 0.079$&  O++06\\
$\langle F_*\rangle$&$ 0.479\pm 0.061$\\
\cutinhead{BD +39 3226}
$\log N({\rm D~I})$&$15.15\pm 0.05$&O++06\\
$\log N({\rm H_{tot}})$&$20.01\pm 0.01$&TP, O++06\\
$F_*({\rm  O})$&$ 1.601\pm 0.525$&  O++06\\
$F_*({\rm Mg})$&$ 0.502\pm 0.251$&     TP\\
$F_*({\rm  P})$&$ 0.853\pm 0.161$&     TP\\
$F_*({\rm Ti})$&$ 0.217\pm 0.040$&  EPL07\\
$F_*({\rm Mn})$&$ 0.353\pm 0.118$&     TP\\
$F_*({\rm Fe})$&$ 0.350\pm 0.056$&  O++06\\
$F_*({\rm Ni})$&$-0.225\pm 0.086$&     TP\\
$F_*({\rm Ge})$&$-0.276\pm 0.356$&     TP\\
$\langle F_*\rangle$&$ 0.235\pm 0.029$\\
\cutinhead{WD 2317-05 (Feige 110)}
$\log N({\rm D~I})$&$15.47\pm 0.03$&F++02\\
$\log N({\rm H_{tot}})$&$20.26\pm 0.02$&TP, TP\\
$F_*({\rm  O})$&$-0.222\pm 0.392$&  H++05\\
$F_*({\rm Ti})$&$ 0.290\pm 0.019$&  PTH05\\
$\langle F_*\rangle$&$ 0.289\pm 0.019$\\
\enddata
\end{deluxetable}

\setcounter{table}{0}

\begin{deluxetable}{lcl}
\tablewidth{265pt}
\tablecolumns{3}
\tablecaption{\textit{(continued)}
\label{basic_data}}
\tablehead{
\colhead{Item} & \colhead{Value} & \colhead{Source(s)\tablenotemark{a}}
}
\startdata
\cutinhead{  JL 9}
$\log N({\rm D~I})$&$15.78\pm 0.06$&W++04\\
$\log N({\rm H_{tot}})$&$20.71\pm 0.01$&TP, W++04\\
$F_*({\rm  O})$&$-0.171\pm 0.803$&  W++04\\
$F_*({\rm Ti})$&$ 0.369\pm 0.049$&  LHW08\\
$F_*({\rm Fe})$&$ 0.475\pm 0.063$&  W++04\\
$\langle F_*\rangle$&$ 0.408\pm 0.039$\\
\cutinhead{  LSS 1274}
$\log N({\rm D~I})$&$15.86\pm 0.09$&W++04\\
$\log N({\rm H_{tot}})$&$20.99\pm 0.04$&W++04, W++04\\
$F_*({\rm  O})$&$ 0.404\pm 0.412$&  W++04\\
$F_*({\rm Ti})$&$ 0.603\pm 0.029$&  LHW08\\
$F_*({\rm Fe})$&$ 0.599\pm 0.070$&  W++04\\
$\langle F_*\rangle$&$ 0.602\pm 0.026$\\
\cutinhead{  LB 1566}
$\log N({\rm D~I})$&$15.29\pm 0.05$&TP\\
$\log N({\rm H_{tot}})$&$20.21\pm 0.01$&TP, TP\\
$F_*({\rm  O})$&$ 3.926\pm 0.914$&     TP\\
$F_*({\rm Fe})$&$ 0.881\pm 0.050$&     TP\\
$\langle F_*\rangle$&$ 0.890\pm 0.050$\\
\cutinhead{  CPD$-$71 172}
$\log N({\rm D~I})$&$15.63\err{0.08}{0.07}$&TP\\
$\log N({\rm H_{tot}})$&$20.28\pm 0.01$&TP, TP\\
$\langle F_*\rangle = F_*({\rm Fe})$&$ 0.186\pm 0.079$&     TP\\
\enddata
\tablenotetext{a}{Two sources are listed for $N({\rm H_{tot}})$: the first is for the 
determination of $N({\rm H~I})$ and the second refers to $N({\rm H}_2)$.  The 
keys to references are explained in Table~\ref{refs}.  The code TP means the value 
was determined in this paper.}
\end{deluxetable}


\begin{deluxetable}{
l	
l	
}
\tablewidth{300pt}
\tablecaption{References for codes in Table~\ref{tab:d2h} and Appendix 2, Table \ref{basic_data}
\label{refs}}
\tablehead{\colhead{Code\tablenotemark{a}} & \colhead{Reference}}
\startdata
AJS92& \citet{allen92} \\
B++83& \citet{bohlin83} \\
BLWY84& \citet{barker84} \\
BSD78& \citet{bohlin78} \\
D++09& \citet{dupuis09}\\
DS94& \citet{diplas94} \\
EPL07& \citet{ellison07}\\
F++02& \citet{friedman02} \\
F++06& \citet{friedman06} \\
FS94& \citet{fitzpatrick94} \\
FVY80& \citet{ferlet80} \\
H++03& \citet{hoopes03} \\
H++05& \citet{hebrard05} \\
HM03& \citet{hebrard03} \\
HSF99& \citet{howk_etal99} \\
J++00& \citet{jenkins00} \\
J++99& \citet{jenkins99} \\
JS07a& \citet{jensen07a} \\
JS07b& \citet{jensen07b} \\
JSS86& \citet{jenkins86}\tablenotemark{b}\\
JY78& \citet{jura78} \\
L++03& \citet{lehner03} \\
L++06& \citet{linsky06}\tablenotemark{c}\\
LHW08& \citet{lallement08} \\
LVY79& \citet{laurent79} \\
M78& \citet{morton78} \\
MD76& \citet{morton76} \\
MJC98& \citet{meyer98} \\
MJHC94& \citet{meyer94} \\
O++03& \citet{oliveira03} \\
O++06& \citet{oliveira06} \\
OH06& \citet{oliveirahebrard06} \\
PTH05& \citet{prochaska05} \\
RB95& \citet{roth95} \\
S++02& \citet{sonneborn02} \\
SDA21 & \citet{shull21} \\
S78& \citet{stokes78} \\
S98& \citet{sarlin98} \\
SCH74& \citet{spitzer74} \\
SJ98& \citet{sofia98} \\
ST++00& \citet{sonneborn00} \\
TP& This paper\\
W++04& \citet{wood04} \\
W++05& \citet{williger05} \\
WKL99& \citet{wolff99} \\
Y76& \citet{york76a} \\
YR76& \citet{yorkrogerson76} \\
\enddata
\tablenotetext{a}{These codes are identical to the ones given in Table~1 of Jenkins 
(2009), except for a few new sources.}
\tablenotetext{b}{Data from this survey required special treatment; see Section 4.1 
of Jenkins (2009).}
\tablenotetext{c}{Value taken from the listing given in this reference; the original 
reference is unclear.}
\end{deluxetable}


\end{document}